\documentclass[12pt]{article}
\usepackage[T1]{fontenc} 
\usepackage{amsmath}
\usepackage{natbib}
\usepackage{xcolor}
\usepackage{graphicx}
\usepackage[utopia]{mathdesign}
\usepackage{bbm}
\usepackage{microtype}
\usepackage{hyperref}
\usepackage{booktabs}
\usepackage{pdflscape}
\usepackage{caption}
\usepackage{subcaption}
\usepackage{array}   
\newcolumntype{L}{>{$}l<{$}} 

\definecolor{lightblue}{rgb}{0, 0.4, 0.6}
\hypersetup{
    breaklinks=true,
	colorlinks = true,
	citecolor = {lightblue},
	linkcolor = {lightblue},
	urlcolor = {lightblue}
}
\usepackage[a4paper,margin=1.25in]{geometry}

\usepackage{titlesec}

\titleformat*{\section}{\large\bfseries}
\titleformat*{\subsection}{\large\bfseries}

\newtheorem{assumption}{Assumption}
\newtheorem{theorem}{Theorem}
\newtheorem{lemma}{Lemma}

\newtheorem{remark}{Remark}

\title{Inference on LATEs with covariates\footnote{We thank Alberto Abadie, Joshua Angrist, Isaiah Andrews, Michal Koles\'{a}r, James MacKinnon, Anna Mikusheva, Whitney Newey, Tymon S\l{}oczy\'{n}ski, Mikkel S\o{}lvsten, Elie Tamer, Tom Wansbeek, Tiemen Woutersen, Yichong Zhang, and seminar participants at the Aarhus Workshop in Econometrics, the Bristol Econometric Study Group, MIT/Harvard, and Singapore Management University for insightful discussions. We also thank Megan Stevenson for sharing the judge design data. Tom Boot acknowledges financial support by the Dutch Research Council (NWO) as part of grant VI.Veni.201E.11.}}
\date{\vspace{0.5cm}\today}
\author{Tom Boot\\
University of Groningen\\
\urlstyle{rm}\url{t.boot@rug.nl}\and Didier Nibbering\\
Monash University\\
\urlstyle{rm}\url{didier.nibbering@monash.edu}}

\begin{document}

\maketitle
\begin{abstract}
\noindent In theory, two-stage least squares (TSLS) identifies a weighted average of covariate-specific local average treatment effects (LATEs) from a saturated specification, without making parametric assumptions on how available covariates enter the model. In practice, TSLS is severely biased as saturation leads to a large number of control dummies and an equally large number of, arguably weak, instruments. This paper derives asymptotically valid tests and confidence intervals for the weighted average of LATEs that is targeted, yet missed by saturated TSLS. The proposed inference procedure is robust to unobserved treatment effect heterogeneity, covariates with rich support, and weak identification. We find LATEs statistically significantly different from zero in applications in criminology, finance, health, and education.
\end{abstract}

\thispagestyle{empty}
\newpage
\setcounter{page}{1}


\section{Introduction}
With endogenous treatment and a binary instrument, \citet{imbens1994identification} show that the two-stage least squares (TSLS) estimand has a causal interpretation as a local average treatment effect (LATE). Empirically, instrument validity is generally argued conditional on available controls. The causal interpretation of the TSLS estimand is lost if we linearly include these controls \textit{unless} this is a correct parametric assumption \citep*{blandhol2022tsls}. In the absence of a credible justification to linearly include the controls, \citet{angrist1995two} show that TSLS can consistently estimate a weighted average of LATEs provided that the number of possible values of the vector of controls is fixed: a researcher can select a saturated specification that includes (i) a dummy for each unique realized value of the vector of controls and (ii) interactions of the instrument with these dummies. However, in most empirical settings the vector of controls has rich support, and saturated TSLS breaks down. As a result, applied work continues to use the more parsimonious linear specification at the risk of targeting a non-causal estimand.

This paper proposes a new method for inference on the weighted average of LATEs targeted by saturated TSLS. We provide asymptotically valid tests and confidence intervals with sufficient power to detect empirically realistic causal effect sizes, while allowing the number of control dummies and instrument interactions to be of the same order as the sample size. As saturated specifications contain only dummy variables, we can provide low-level assumptions for our results. The most important and rather mild assumption is that each covariate group contains at least two individuals for which the instrument is active and two for which the instrument is inactive. Since the conditional validity of an instrument often relies on categorical controls representing different groups or strata, and our method extends to multivalued treatments and multiple or multivalued instruments, our method applies to a wide variety of empirically relevant settings in applied economics.

Conducting valid inference in a saturated specification is a challenging task \citep{mogstad2024instrumental}. Even with 10 binary controls, there are 1,024 possible values of the covariate vector, and for each possible value a covariate-group dummy and an instrument interaction is included in the regression. When the individuals induced to change treatment status by the instrument and the variation in the instrument are not evenly distributed across the covariate groups, interacting a strong instrument with control dummies may result in instrument interactions that are only weakly related to the treatment. Hence, even in the most favorable saturated specification with a single strong binary instrument, inference has to take many and possibly weak instruments into account. Standard inferential procedures for instrumental variables are known to be biased in the presence of many and weak instruments \citep{bekker1994alternative,staiger1997instrumental}. 

While there is an extensive literature that combines the notion of many and weak instruments, this either excludes unobserved heterogeneity in treatment effects or a large number of covariate groups. For instance, \citet*{bekker2003finite, chao2005consistent,hausman2012instrumental, crudu2021inference,mikusheva2021inference,matsushita2020jackknife,lim2022conditional}, among others, focus on the linear instrumental variable model with a homogeneous slope coefficient. \citet{lee2018consistent} develops a TSLS variance estimator that allows for heterogeneous treatment effects, which however is not robust to the large number of instrument interactions in typical saturated specifications. \citet{kolesar2013estimation} proposes estimators free of many and weak instrument bias while allowing for heterogeneous treatment effects. However, inferential procedures based on these estimators have only been developed under the assumption that the number of control dummies is a negligible fraction of the sample size \citep{evdokimov2018inference}.

This paper conducts inference on the weighted average of LATEs that is robust against the key features of saturated economic data: unobserved treatment effect heterogeneity, many control dummies and instrument interactions, weak identification strength, and conditional heteroskedasticity. We first show that the recently proposed estimator from \citet*{chao2023jackknife}, designed to estimate a homogeneous slope coefficient in a panel data setting, consistently estimates a weighted average of LATEs in the setting of saturated instrumental variable estimation (SIVE). We then show that in the present context, it is crucial to take treatment effect heterogeneity into account in the inference stage, as it has a nontrivial effect on the variance that standard variance estimators do not capture. Throughout, the analysis allows the number of control dummies and instrument interactions to be asymptotically non-negligible relative to the sample size. For consistency, the condition that we impose on the identification strength has been shown by \citet{mikusheva2021inference} to be the weakest possible. Our results also provide a fully identification-robust procedure.

We conduct a series of Monte Carlo simulations which set-up mimics key features of the data used in \citet{card1995using}, and has also been used by \citet{blandhol2022tsls}. The results illustrate that our proposed estimator is median unbiased, even with a large number of covariate groups and weak instrument strength. Our inference method yields nominal size control regardless of the instrument strength when the number of covariate groups is small, and becomes progressively more conservative when the number of covariate groups increases under weak identification. At the same time, our method has sufficient statistical power to detect empirically realistic causal effect sizes. In contrast, linear and fully saturated TSLS and various jackknife estimators incur a bias that increases with the number of covariate groups and as the strength of the instrument decreases, and the corresponding tests show large size distortions.

The empirical relevance of our method for treatment evaluation is illustrated with empirical applications from different fields of applied economics: a judge design study from \citet{stevenson2018distortion}, an experiment on banking access by \citet{dupas2018banking}, and in the supplementary material, we further study the effect of schooling \citep{card1995using}, health insurance \citep*{finkelstein2012oregon}, and an educational program \citep*{muralidharan2019disrupting}. We find that our method produces realistic LATE estimates in settings where the literature has found issues with existing methods for inferring LATEs. In settings in which assumptions underlying TSLS are harmless, our method produces very similar results with a negligible loss in precision. Hence, our inference method is able to detect statistically significant LATEs, while the applied researcher is assured that results are not driven by the often implicitly made assumptions underlying existing methods.

To the best of our knowledge, we are the first to tackle the statistical challenges in inferring the weighted average of covariate-specific LATEs as derived by \citet{angrist1995two} that explicitly allows for weak monotonicity: the instrument may induce treatment take-up for some covariate groups, while it may discourage take-up for other groups. Weak monotonicity has been shown to be empirically relevant in judge designs \citep{aizer2015juvenile,mueller2015criminal,sigstad2023monotonicity}, information provision experiments \citep{vilfort2023interpreting}, and a wide range of other applied economic settings \citep{de2017tolerating}. \citet{sloczynski2020should} provides a review of 25 articles published in AEA journals between 2009 and 2015 and finds that 22 papers report at least one TSLS regression with individuals induced in and out of treatment. 

A strand of related literature aims to infer the unconditional LATE with semiparametric estimators \citep{abadie2003semiparametric}, non-parametric estimators \citep{frolich2007nonparametric}, and machine learning techniques \citep*{chernozhukov2018double}. The unconditional LATE weighs the covariate-specific reduced forms and first stages with the probability density function of the covariates. This parameter only has a causal interpretation under the notion of strong monotonicity: The instrument moves treatment take up in the same direction for all covariate groups. Under this assumption, the instrument interactions are not required for identification, which removes the inferential challenges with many and weak instruments. 

The remainder of this article is organized as follows. Section~\ref{sec:LATEs} explains the current practice of inferring LATEs with covariates from the data and its challenges. Section~\ref{sec:SIVE} introduces our proposed causal estimand and its inference procedure, supported by large sample theoretical results. Section~\ref{sec:mc} discusses the Monte Carlo simulations, Section~\ref{sec:application} the empirical applications, and Section~\ref{sec:conclusion} concludes.

\section{LATEs with covariates}\label{sec:LATEs}
Suppose we are interested in the causal effect of a binary treatment $T_i$ on an outcome $Y_i$, for individuals $i=1,\dots,n$. For each individual, define the potential outcomes $Y_i(1)$ and $Y_i(0)$ corresponding to the values of $Y_i$ if individual $i$ is treated or not treated, respectively. Hence, the treatment effect is defined as $Y_i(1)-Y_i(0)$. The treatment is potentially endogenous, and a binary instrument $Q_i$ and a vector of covariates $X_i$ are available to help to identify a causal effect. The developed theory applies equally well to the extensions to multivalued treatments and instruments in \citet{angrist1995two}.

Define $\mathbb{X}=\{x_1,\dots,x_G\}$ as the set of all possible $G$ unique realizations of $X_i$. Denote the potential treatment statuses $T_i(1)$ and $T_i(0)$ corresponding to the values of $T_i$ if individual $i$'s treatment assignment is given by $Q_i=1$ and $Q_i=0$, respectively. In case the outcome $Y_i$ also directly depends on $Q_i$, its corresponding potential outcomes are given by $Y_i(Q_i,T_i)$. If we condition on the covariates, the four instrumental variable (IV) assumptions in the \citet{imbens1994identification} framework are the following.

\begin{assumption}\label{ass:IV}
\,    With $\mathbb{E}[Y_{i}(q,t)]$ for $q\in\{0,1\}$ and $t\in\{0,1\}$ bounded, we have:
    \begin{enumerate}
    \item \textbf{Independence}: $(Y_i(q,t),T_i(q))\perp Q_i | X_i$ for $q\in\{0,1\}$ and $t\in\{0,1\}$,
\item \textbf{Exclusion}: $\mathbb{P}(Y_i(1,t)=Y_i(0,t)|X_i)=1$ $a.s.$ for $t\in\{0,1\}$,
    \item \textbf{Relevance}: $\mathbb{P}[T_i(1)\neq T_i(0)| X_i]>0$ $a.s.$,
    \item \textbf{Monotonicity}: $\mathbb{P}[T_i(1)\geq T_i(0) | X_i]=1$ $a.s.$, or $\mathbb{P}[T_i(1)\leq T_i(0) | X_i]=1$ $a.s.$
\end{enumerate}
\end{assumption}
These assumptions allow for complete treatment effect heterogeneity across all individuals, and do not impose any parametric assumptions. Following the exclusion restriction, we write $Y_{i}(T_{i})$ for the potential outcomes. The relevance assumption holds if there are individuals with $T_i(1)\neq T_i(0)$ for each $X_i$. These individuals are either compliers ($T_i(1) > T_i(0)$) or defiers ($T_i(1) < T_i(0)$).
We weaken the relevance assumption in Section~\ref{sec:estimation_challenges} to allow for subpopulations with no compliers and no defiers. The monotonicity assumption is referred to in \citet{blandhol2022tsls} as \textit{weak} monotonicity, because it allows the effect of the instrument on the treatment to have a different direction for each $X_i$. In other words, we can have compliers for some values of $X_i$ and defiers for other values of $X_i$. Strong monotonicity requires $\mathbb{P}[T_{i}(1)\geq T_{i}(0)]=1$ or $\mathbb{P}[T_{i}(1)\leq T_{i}(0)]=1$, and therefore assumes that the effect of switching on the instrument on potential treatment status is (weakly) in the same direction for all individuals. 

Within the LATE framework, causal effects are estimated of the form
\begin{align}\label{eq:param_interest}
   \tau =& \sum_g \omega(x_g) \tau(x_g) \text{ with }  \sum_g \omega(x_g)=1, \,\omega(x_g) \geq 0 \text{ for } g=1,\dots,G, \text{ and }\\
   \tau(x_g) =& \mathbb{E}[Y_i(1)-Y_i(0)|T_i(1)\neq T_i(0),X_i=x_g].
\end{align}
The causal effect is then a positively weighted average of covariate-specific LATEs. The following well-known result, see for instance \citet{angrist2009mostly}, shows that the covariate-specific LATEs $\tau(x_g)$ are indeed identified.
\begin{lemma}\label{lemm:latex}
Under Assumption~\ref{ass:IV}, 
\begin{align}
    \frac{\mathbb{E}[Y_i|Q_i=1,X_i={x}_g]-\mathbb{E}[Y_i|Q_i=0,X_i={x}_g]}{\mathbb{E}[T_i|Q_i=1,X_i={x}_g]-\mathbb{E}[T_i|Q_i=0,X_i={x}_g]} = \tau(x_g).
\end{align}
\end{lemma}

In practice, the number of observations corresponding to each $X_i$ is usually small, and the moments in Lemma~\ref{lemm:latex} cannot be accurately estimated. Therefore, practitioners commonly rely on regression to estimate a weighted average of the covariate-specific LATEs as the parameter of interest, where the weights are automatically selected through the regression model that is specified. The default regression specification includes the covariates linearly in the first and second stage. However, \citet{blandhol2022tsls} show that if this parametric assumption is incorrect, then TSLS has no causal interpretation.

\subsection{Saturating the TSLS specification}
\citet{angrist1995two} show that $\tau$ can be estimated by TSLS in saturated specifications: the first stage includes dummies for each possible value of $X_i$ and a full set of interactions between these dummies and the instrument, and the second stage includes the treatment variable and the control dummies. 
We define the $G\times1$ vector $W_i$ with elements $W_{ig}={1}[X_i=x_g]$, indicating the covariate group of individual $i$. The $G\times1$ vector $Z_i$ contains the instrument interactions $Z_{ig}=Q_i{1}[X_i=x_g]$. Note that $\sum_g W_{ig}=1$ and $\sum_g Z_{ig} = Q_i$. Define $n_g = \sum_i W_{ig}$ as the number of individuals in covariate group $g$, and $m_{1,g} = \sum_i Z_{ig}$ and $m_{0,g} = n_g-m_{1,g}$ as the number of individuals in covariate group $g$ with an active and inactive instrument, respectively.

Both the full set of covariate group indicators and the full set of instrument interactions are required for nonparametric estimation of $\tau$ by TSLS. In this specification, the conditional expectation of the instrument given the covariate groups is linear in the covariate group indicators. This allows for correctly partialling out the covariates, which otherwise may induce negative weights into $\tau$. Without the instrument interactions, the first stage does not necessarily reproduce the direction of the monotonicity assumption in all covariate groups. Hence, omission of the instrument interactions requires the direction of the monotonicity to be invariant to the covariate group.

It is clear that saturation can only work if we observe each covariate value more than once. Moreover, to achieve identification, each covariate group has to include both individuals with an active and an inactive instrument. The setup we consider, with many control dummies and many instrument interactions, requires the number of observations in each group to satisfy the following assumption.

\begin{assumption}\label{ass:group_size} $m_{1,g}\geq 2$ and $m_{0,g}\geq 2$ $a.s.$ for $g=1,\dots,G$.
\end{assumption}
This assumption requires that both the number of individuals with an active instrument and nonactive instrument has to be larger than one in each covariate group. Hence, without additional assumptions, our setup focuses on discrete controls. This applies to a wide range of applications, as the conditional validity of instrumental variables is most commonly argued on the basis of categories or strata. Assuming that both values of the instrument can be observed in each covariate group, is similar to an overlap or common support assumption. 

\subsection{Estimation challenges}\label{sec:estimation_challenges}
While saturation results in a causal TSLS estimand if the number of possible covariate values is small, it is not straightforward to find a causal estimand in the empirically more common setting in which the controls have rich support. Since each group requires an indicator, and the instrument is interacted with these indicators, this automatically results into a large set of control dummies and a large set of instruments. In this setting, TSLS is known to be biased, see e.g.\ \citet{bekker1994alternative}. 

Moreover, the instrument interactions may weaken the instrument strength. Instrument strength is measured by the first stage signal, which can be written as $\text{FS}=\sum_g \mathbb{P}[X_i=x_g]\pi(x_g)^2 \mathbb{V}[Q_i|X_i=x_g]$ with complier or defier shares $\pi(x_g)=\mathbb{P}[T_i(1) \neq T_i(0)|X_i=x_g]$ and treatment assignment variation $\mathbb{V}[Q_i|X_i=x_g]$. If the complier or defier share and the variation in treatment assignment are homogeneous across covariate groups, that is $\pi(x_g)=\pi$ and $\mathbb{V}[Q_i|X_i=x_g]=\mathbb{V}[Q_i]$, the strength of the instrument interactions $Z$ equals the strength of the instrument $Q$. However, in settings where $\pi(x_g)$ is large in groups with a low number of treated or untreated units, while $\pi(x_g)$ is small in groups with a number of treated units close to half of the number of group members, the first stage is likely weak. A large number of potentially weak instruments can exacerbate the bias in TSLS, see e.g.\ \citet{chao2005consistent}. We impose the following assumption on the identification strength, which ensures existence of a causal estimand and weakens Assumption~\ref{ass:IV}.3.
\begin{assumption}\label{ass:relevance}
Replace Assumption~\ref{ass:IV}.3 (Relevance) with: there exists a $g\in \{1,\ldots,G\}$ such that $\sum_{i=1}^{n}\mathbb{P}[T_i(1) \neq T_i(0)|X_i=x_g]> 0$.
\end{assumption}
Under Assumption~\ref{ass:relevance}, we need to have a nonzero probability to find compliers or defiers in at least one covariate group. Even for that covariate group, the number of compliers or defiers may be small, as Assumption~\ref{ass:relevance} allows $\mathbb{P}[T_i(1) \neq T_i(0)|X_i=x_g]$ to converge to zero as the sample size increases. 

Estimators that have been proposed in settings with many or weak instruments often assume the first stage and reduced form errors to be independent across observations:
\begin{assumption}\label{ass:indep}
Define $\varepsilon_{i} = Y_{i}-\mathbb{E}[Y_{i}|Q,X]$ and $u_{i} = T_{i}-\mathbb{E}[T_{i}|Q,X]$. Then, $(\varepsilon_{i},u_{i})$ is independent across $i$ conditional on $(Q,X)$.
\end{assumption}
We follow suit, and maintain this assumption throughout this section to illustrate the estimation challenges with existing methods. In Supplemental Appendix~\ref{A:cluster}, we discuss how to relax this assumption to allow for more general correlation structures in our proposed method. 
 
\subsection{Two-stage least squares estimand}
Define the $n$-dimensional vectors $ Y = (Y_{1},\ldots,Y_{n})'$ and $ T = (T_{1},\ldots,T_{n})'$, and the $n \times G$-dimensional matrices $ W = (W_{1}',\ldots, W_{n}')'$ and $Z = ( Z_{1}',\ldots, Z_{n}')'$. Define the residual maker matrix $M_{W} = I_n - W(W'W)^{-1}W'$ with $I_n$ the $n$-dimensional identity matrix. The TSLS estimand is commonly defined as
\begin{align}
    \beta^{\text{TSLS}}=\frac{\mathbb{E}[T'P Y|Q,X]}{\mathbb{E}[T'PT|Q,X]},
\end{align}
where $P=M_{W}Z(Z'M_{W}Z)^{-1}Z'M_{W}$ partials out the controls $W$ from the first stage and the second stage. However, this estimand does not identify $\tau$ when the number of covariate groups is large as the following result makes precise.
\begin{lemma}\label{lemm:2sls}
Under Assumption~\ref{ass:IV} and \ref{ass:indep} it holds that
\begin{align}
    \beta^{\emph{TSLS}}=\frac{\sum\nolimits_g \tilde{\emph{P}}[X_i=x_g]\pi(x_g)^2 \tilde{\emph{V}}[Q_i|X_i=x_g] \tau(x_g)+\frac{1}{n}\sum\nolimits_i\mathbb{E}[u_i\varepsilon_i|Q,X]P_{ii}}{\sum\nolimits_g \tilde{\emph{P}}[X_i=x_g]\pi(x_g)^2 \tilde{\emph{V}}[Q_i|X_i=x_g] +\frac{1}{n}\sum\nolimits_i\mathbb{E}[u_i^2|Q,X]P_{ii}},
\end{align}
where $\tilde{\emph{P}}[X_i=x_g]=\frac{n_g}{n}$, $\pi(x_g)=\emph{P}[T_i(1) \neq T_i(0)|X_i=x_g]$, $\tilde{\emph{V}}[Q_i|X_i=x_g]=\frac{m_{1,g}}{n_g}\frac{m_{0,g}}{n_g}$, and $P_{ii}=\sum_g \frac{1}{n_g}W_{ig}\frac{(Z_{ig}-\tilde{\emph{P}}[Q_i=1|X_i=x_g])^2}{\tilde{\emph{V}}[Q_i|X_i=x_g]}$ with $\tilde{\emph{P}}[Q_i=1|X_i=x_g]=\frac{m_{1,g}}{n_g}$.
\end{lemma}
This result can be found in \citet{evdokimov2018inference}. We provide a short proof in Appendix~\ref{A:2sls} as some of the steps are used to derive subsequent results. Note that $\tilde{\text{P}}[X_i=x_g]$, $\tilde{\text{V}}[Q_i|X_i=x_g]$, and $\tilde{\text{P}}[Q_i=1|X_i=x_g]$ are the sample analogues of ${\mathbb{P}}[X_i=x_g]$, ${\mathbb{V}}[Q_i|X_i=x_g]$, and ${\mathbb{P}}[Q_i=1|X_i=x_g]$. 

Lemma~\ref{lemm:2sls} shows that when both the second term in the numerator and the second term in the denominator go to zero faster than the first term in the numerator and denominator for $\beta^{\text{TSLS}}$, we identify $\tau$ with the weights in \eqref{eq:param_interest} equal to $\omega^{\text{TSLS}}(x_g) = \frac{\tilde{\emph{P}}[X_i=x_g]\pi(x_g)^2 \tilde{\emph{V}}[Q_i|X_i=x_g]}{\sum_g \tilde{\emph{P}}[X_i=x_g]\pi(x_g)^2 \tilde{\emph{V}}[Q_i|X_i=x_g]}$. It is clear that this is the case when $P_{ii}$ is small. Since $
   P_{ii}=\sum_g W_{ig}\left(Z_{ig}\frac{1}{n_g}\frac{m_{0,g}}{m_{1,g}} + (1-Z_{ig})\frac{1}{n_{g}}\frac{m_{1,g}}{m_{0,g}}\right)$,
TSLS has a causal interpretation if in each covariate group the number of individuals for which the instrument is active ($m_{1,g}$) and the number of individuals for which the instrument is inactive ($m_{0,g}$) is large. This requirement becomes more stringent as the strength of the instrument interactions measured by $\sum_g \tilde{\emph{P}}[X_i=x_g]\pi(x_g)^2 \tilde{\emph{V}}[Q_i|X_i=x_g]$ decreases.

In empirical applications, the number of covariate groups is generally large and nonnegligible relative to the number of individuals. In this case, at least a number of covariate groups has to contain a small number of observations, and the additional terms in Lemma~\ref{lemm:2sls} do not go to zero. This bias is known as the many instrument bias of TSLS. Under homogenous treatment effects, and assuming homoskedastic errors, one can follow  \citet{bekker1994alternative} in using LIML to avoid this bias. However, \citet{kolesar2013estimation} points out that with treatment effect heterogeneity, the estimand of LIML is generally not causal. 

\begin{remark}
In Lemma~\ref{lemm:2sls}, we condition both on the instrument and the covariates. This estimand can generally be more accurately inferred from the data relative to the unconditional counterpart. This point is made by \citet{crump2009dealing} in a regression context. In the IV context, \citet{evdokimov2018inference} show that both the conditional and unconditional estimands are a weighted combination of covariate specific LATEs, where the unconditional estimand integrates out sampling uncertainty in the combination weights. As such, confidence intervals for the unconditional estimand are wider. We focus on the conditional estimand in the subsequent analysis. 
\end{remark}

\subsection{Jackknife instrumental variables estimands}\label{subsec:jive}
It follows from Lemma~\ref{lemm:2sls} that the bias in TSLS is due to the diagonal elements $P_{ii}$. A frequently used approach to reduce many instrument bias is to employ a jackknife-style correction \citep*{angrist1999jackknife,ackerberg2009improved}. In the current setting, we could remove the diagonal of $P$ using the diagonal matrix $D_P$ with $[D_{P}]_{ii}=P_{ii}$, to obtain the estimand 
\begin{align}\label{eq:jive}
    \beta^{\text{JIVE}}=\frac{\mathbb{E}[T'(P-D_P) Y|Q,X]}{\mathbb{E}[T'(P-D_P)T|Q,X]}.   
\end{align}
This diagonal removal has been the basis of recent papers in the literature on identification-robust inference under many instrument sequences \citep*{mikusheva2021inference,crudu2021inference,matsushita2020jackknife}. It ensures that the many instrument bias present in TSLS disappears. However, 
in the presence of a potentially large set of control variables, the consequence of removing the diagonal elements $P_{ii}$ is that the controls are no longer projected out. %

An alternative jackknife approach is to first partial out the controls before removing the diagonal. This gives the following estimand
\begin{align}\label{eq:jive2}
    \beta^{\text{IJIVE}}=\frac{\mathbb{E}[T'M_W (P-D_P) M_W Y|Q,X]}{\mathbb{E}[T'M_W (P-D_P)M_W T|Q,X]}.   
\end{align}
\citet{ackerberg2009improved} refer to this estimand as the improved JIVE (IJIVE), where they additionally scale the matrix $(P-D_P)$ with $(I-D_P)^{-1}$ in \eqref{eq:jive2}. Although IJIVE appropriately handles the controls, it reintroduces a many instrument bias as $[M_{W}(P-D_{P})M_{W}]_{ii}=[P-M_{W}D_{P}M_{W}]_{ii}\neq 0$. Supplemental Appendix~\ref{A:jive} derives the bias expressions for JIVE and IJIVE. %

\section{Saturated instrumental variable estimation}\label{sec:SIVE}

\subsection{A causal estimand}
We now consider an estimand identical to that of TSLS when the number of possible values of the vector of controls is fixed, but that does not suffer from the many instrument bias when the number of covariate values grows. Define the matrix $V=[W,Z]$ consisting of the covariate group indicators and the instrument interactions, and the residual maker matrix ${M}_{W,Z}=I-V(V'V)^{-1}V'$. The SIVE estimand is specified as
\begin{align}\label{eq:sive}
    \beta^{\text{SIVE}}=\frac{\mathbb{E}[T'(P-{M}_{W,Z}D{M}_{W,Z}) Y|Q,X]}{\mathbb{E}[T'(P-{M}_{W,Z}D{M}_{W,Z})T|Q,X]},
\end{align}
where $D$ is a diagonal matrix with diagonal elements such that $P_{ii}=[{M}_{W,Z}D{M}_{W,Z}]_{ii}$. It follows that \eqref{eq:sive} is a jackknife estimand, which removes the diagonal of $P$ and hence the many instrument bias in the TSLS estimand. At the same time, by pre- and post-multiplying $D$ by ${M}_{W,Z}$, the controls are projected out correctly and the bias in JIVE is prevented. In addition to the controls, ${M}_{W,Z}$ also projects out the instrument interactions, and therefore does not affect the causal estimand that TSLS is targeting. 

The SIVE estimator has been proposed by \citet*{chao2023jackknife} as Fixed Effect Jackknife IV (FEJIV). They show that the diagonal elements of $D$ can be obtained by solving a system of linear equations with a unique solution, and derive consistency results in a linear instrumental variable panel data model with fixed effects, where the number of instruments increases slowly with the sample size. Within our saturated setting, we can derive a closed-form expression for $D$, and show that the estimand identifies a weighted average of covariate-specific LATEs. These results require a weaker set of assumptions as we allow the number of control dummies and instrument interactions to be asymptotically non-negligible relative to the sample size. Furthermore, Assumption~\ref{ass:group_size} relaxes Assumption 6 in \citet{chao2023jackknife} that requires $m_{1,g}\geq 3$ and $m_{0,g}\geq 3$.

The following result shows that the diagonal matrix $D$ in \eqref{eq:sive} exists under Assumption~\ref{ass:group_size}.  
\begin{lemma}\label{lemm:diag}
 Suppose Assumption~\ref{ass:group_size} holds and let $D$ be a diagonal matrix with elements
    \begin{align}
    D_{ii}=  \sum_g \frac{1}{n_g} W_{ig} \left[\frac{m_{0,g}}{m_{1,g}-1}Z_{ig}+\frac{m_{1,g}}{m_{0,g}-1}(1-Z_{ig})\right].
\end{align}
    Then, $P_{ii}=[{M}_{W,Z}D{M}_{W,Z}]_{ii}$. 
\end{lemma}
The proof is deferred to Appendix~\ref{A:diag}. The result shows that when the number of covariate values is small and both $m_{1,g}$ and $m_{0,g}$ are large, the diagonal elements $D_{ii}$ are small and the estimator reduces to the TSLS estimator. 

Because the term that is subtracted in the numerator and denominator of \eqref{eq:sive} is orthogonal to the instruments and controls, it is straightforward to establish that the SIVE estimand has a causal interpretation that is identical to the unbiased TSLS estimand, without requiring the number of control dummies or instrument interactions to be small. 
\begin{theorem}\label{lemm:sive}
Under Assumptions~\ref{ass:IV} to \ref{ass:indep},
    \begin{align}
        \beta^{\emph{SIVE}}=\frac{\sum\nolimits_{g} \tilde{\emph{P}}[X_i=x_g]\pi(x_g)^2 {\tilde{\emph{V}}}[Q_i|X_i=x_g] \tau(x_g)}{\sum\nolimits_{g} \tilde{\emph{P}}[X_i=x_g]\pi(x_g)^2 {\tilde{\emph{V}}}[Q_i|X_i=x_g]}=\sum_{g} \omega(x_g)\tau(x_g).
    \end{align}
\end{theorem}
The proof is deferred to Appendix~\ref{A:sive}. 
\begin{remark}
Under the same assumptions, the Unbiased JIVE (UJIVE) proposed by \citet{kolesar2013estimation} has the estimand:  $\beta^{\text{UJIVE}}=\sum_{g} \frac{n_g}{n_g-1}\omega(x_g)\tau(x_g)$. The difference between the SIVE and UJIVE estimand can be substantial with small covariate groups, with UJIVE placing 4/3 times the SIVE weight on groups with only four observations. The robust inference procedure we develop in this paper also applies to the UJIVE estimand. 
\end{remark}

\subsection{Inference on the estimand}
While having a causal estimand is a crucial first step, we also need to be able to infer the estimand from the data. In this section, we therefore consider the testing problem
\begin{align}
    H_0: \beta^{\text{SIVE}}=\beta_{0} \text{ against } H_1: \beta^{\text{SIVE}}\neq\beta_{0},
\end{align}
for a given $\beta_{0}$. The testing procedure is standard and based on the fact that under $H_{0}$, 
\begin{align}\label{eq:test_statistic}
\frac{\hat{\beta}^{\text{SIVE}}-\beta_{0}}{\sqrt{{\mathbb{V}}[\hat{\beta}^{\text{SIVE}}|Q,X]}}\rightarrow_{d} N(0,\lambda),
\end{align}
where we provide conditions on the strength of the instrument interactions under which $\lambda=1$ or $\lambda\leq 1$. 
In \eqref{eq:test_statistic}, $\hat{\beta}^{\text{SIVE}}$ is the sample analogue of \eqref{eq:sive} and given by,
\begin{align}\label{eq:sivehat}
    \hat{\beta}^{\text{SIVE}}=\frac{T'(P-{M}_{W,Z}D{M}_{W,Z}) Y}{T'(P-{M}_{W,Z}D{M}_{W,Z})T}.
\end{align}
The crucial part to make the test operational is to find an estimator for the variance of $\hat{\beta}^{\text{SIVE}}$ that appropriately accounts for treatment effect heterogeneity, as well as the presence of many control dummies and many instrument interactions. Let $v_{i} = \varepsilon_{i}-u_{i}\beta^{\text{SIVE}}$ and denote by $\sigma_{u,i}^2 = \mathbb{E}[u_{i}^2|Q,X]$, $\sigma_{v,i}^2 = \mathbb{E}[v_{i}^2|Q,X]$, and $\sigma_{uv,i} = \mathbb{E}[u_{i}v_{i}|Q,X]$. We propose the following variance estimator for $\hat{\beta}^{\text{SIVE}}$ that is decomposed into two parts, 
\begin{equation}\label{eq:Vhat}
\hat{\mathbb{V}}[\hat{\beta}^{\text{SIVE}}|Q,X] =\hat{\mathbb{V}}_{1}-\hat{\mathbb{V}}_{2}.
\end{equation}
The first part of \eqref{eq:Vhat} is given by
\begin{equation}\label{eq:V1hat}
\begin{split}
    \hat{\mathbb{V}}_1&= \frac{(Y-T\hat{\beta}^{\text{SIVE}})'A D_{\hat{\sigma}_{u}^2} A( Y- T\hat{\beta}^{\text{SIVE}}) +  T' A D_{\hat{\sigma}_{v}^2} A T + 2( Y- T\hat{\beta}^{\text{SIVE}})' A D_{\hat{\sigma}_{uv}} A T}{(T' A T)^2},
    \end{split}
\end{equation}
where $A=P-M_{W,Z}DM_{W,Z}$ with $D$ as defined in Lemma~\ref{lemm:diag}, and $D_{\hat{\sigma}_{u}^2}$, $D_{\hat{\sigma}_{v}^2}$ and $D_{\hat{\sigma}_{uv}}$ are diagonal matrices with $[D_{\hat{\sigma}_{u}^2}]_{ii}=\hat{\sigma}_{u,i}^2$, $[D_{\hat{\sigma}_{v}^2}]_{ii}=\hat{\sigma}_{v,i}^2$, and $[D_{\hat{\sigma}_{uv}}]_{ii}=\hat{\sigma}_{uv,i}$ on their respective diagonals.
In the presence of many instruments, standard heteroskedasticity robust Eicker-Huber-White variance estimators are inconsistent \citep{cattaneo2018inference}.  We therefore consider the \citet{hartley1969variance} (HRK) variance estimators that are given by,
\begin{equation}\label{eq:unbvar}
\begin{split}
\hat{\sigma}_{u,i}^2 &= e_{i}'( M_{W,Z}\odot  M_{W,Z})^{-1}( M_{W,Z} T\odot  M_{W,Z} T),\\
\hat{\sigma}_{v,i}^2 &=  e_{i}'( M_{W,Z}\odot  M_{W,Z})^{-1}( M_{W,Z}( Y- T\hat{\beta}^{\text{SIVE}})\odot  M_{W,Z}( Y- T\hat{\beta}^{\text{SIVE}})),\\
\hat{\sigma}_{uv,i} &=  e_{i}'( M_{W,Z}\odot  M_{W,Z})^{-1}( M_{W,Z}( Y- T\hat{\beta}^{\text{SIVE}})\odot  M_{W,Z} T).
\end{split}
\end{equation}
When $\hat{\beta}^{\text{SIVE}}$ is replaced by its population counterpart $\beta^{\text{SIVE}}$, the estimators are unbiased conditional on the instrument and covariates. This removes the main driver of the inconsistency of standard heteroskedasticity robust variance estimators. 

The second part of \eqref{eq:Vhat} corrects for an upward bias in \eqref{eq:V1hat} stemming from the use of the HRK variance estimators instead of the corresponding population quantities. In Appendix~\ref{A:bias} we show how to estimate these interactions to form the bias-correction term $\hat{\mathbb{V}}_{2}$. This term eliminates the bias if $m_{1,g}\geq 4$ and $m_{0,g}\geq 4$ and partially removes the bias when $m_{1,g}\geq 3$ and $m_{0,g}\geq 3$. 

\begin{remark}\label{remark:varest_small}
The estimators in \eqref{eq:unbvar} require a strengthening of Assumption~\ref{ass:group_size} to $m_{1,g}\geq 3$ and $m_{0,g}\geq 3$ for the inverse of $M_{W,Z}\odot M_{W,Z}$ to exist. Instead, we can use the following estimators on the individuals with an instrument status that is only shared with one other individual in the same covariate group:
\begin{equation}\label{eq:unbvar_smallg}
\begin{split}
\hat{\sigma}_{u,i}^2 &= 4e_{i}'( M_{W,Z} T\odot  M_{W,Z} T),\\
\hat{\sigma}_{v,i}^2 &= 4e_{i}'( M_{W,Z}( Y- T\hat{\beta}^{\emph{SIVE}})\odot  M_{W,Z}( Y- T\hat{\beta}^{\emph{SIVE}})),\\
\hat{\sigma}_{uv,i} &= 4e_{i}'( M_{W,Z}( Y- T\hat{\beta}^{\emph{SIVE}})\odot  M_{W,Z} T).
\end{split}
\end{equation}
These estimators generate a positive bias in the variance estimator \eqref{eq:Vhat} as we show in Supplemental Appendix~\ref{A:addvarest}. Hence, the presence of many small groups will make the inference procedure more conservative.
\end{remark}

\begin{remark}\label{remark: computation}
The saturated specification allows for the estimation of $\beta^{\emph{SIVE}}$, the variance of its estimator, and the error variances separately for each covariate group. It follows that no large matrix operations are needed, even when the number of groups is large. For instance, the estimator in \eqref{eq:sivehat} is constructed as follows
\begin{align}
    \hat{\beta}^{\emph{SIVE}}= \sum_g w_g \hat{\beta}^{\emph{SIVE}}_g, \quad w_g=\frac{T_{(g)}'A_{(g)}T_{(g)}}{\sum_g T_{(g)}'A_{(g)}T_{(g)}},\quad \hat{\beta}^{\emph{SIVE}}_g=\frac{T_{(g)}'A_{(g)}Y_{(g)}}{T_{(g)}'A_{(g)}T_{(g)}},
\end{align}
where $T_{(g)}$ and $Y_{(g)}$ are the $n_g$-dimensional vectors with the observations in respectively $T$ and $Y$ corresponding to covariate group $g$, while $A_{(g)}$ represents the $n_g \times n_g$ matrix with the rows and columns in $A$ that correspond to covariate group $g$. The variance estimator in \eqref{eq:Vhat} and the error variance estimators in \eqref{eq:unbvar} are also constructed in a groupwise fashion. 
\end{remark}

\subsection{Assumptions}\label{subsec:assumptions}
To study the asymptotic properties of the test statistic in \eqref{eq:test_statistic} with variance estimator \eqref{eq:Vhat}, we impose the following assumptions. Throughout, $c$ denotes a generic positive constant that can differ between occurrences. 
 \begin{assumption}\label{ass:large_sample}
 \, \\
 For $i=1,\ldots,n$, it holds almost surely that 
$\mathbb{E}[{u}_{i}^2|Q, X]\geq c>0$, $\mathbb{E}[\varepsilon_{i}^2|Q, X]\geq c>0$, $|\text{corr}[u_i,\varepsilon_i|Q, X]|\leq c<1$, $\mathbb{E}[{u}_{i}^8|Q, X]\leq c<\infty$, and $\mathbb{E}[{\varepsilon}_{i}^8| Q, X]\leq c<\infty$.
 \end{assumption}
Assumption~\ref{ass:large_sample} ensures that the distribution of the test statistic is nondegenerate and controls the behavior of the estimators for the conditional variances of $u_{i}$ and $\varepsilon_{i}$.

Assumptions \ref{ass:IV} to \ref{ass:large_sample} allow for the inference on the weighted average of LATEs in \eqref{eq:param_interest} in a wide range of empirically relevant settings. First, the treatment effects $Y_i(1)-Y_i(0)$ are allowed to be heterogeneous across all $i$. It follows that treatment effects may vary across the covariate groups defined by the elements of $\mathbb{X}$, and hence SIVE identifies a weighted average of potentially heterogeneous conditional treatment effects. This is in line with the LATE identification literature. Note that the literature on inference in IV models often assumes that the data satisfies a model along the lines of 
\begin{align}\label{eq:linearIV}
  T_i = \pi'Z_i + \delta_1'W_i+\varepsilon_i, \quad   Y_i = T_i\beta + \delta_2'W_i +u_i, 
\end{align}
in which the treatment effect of $T_i$ on $Y_i$ is modelled with $\beta$ which is specified to be homogeneous. Since we conduct inference with saturated instruments and covariates, no parametric assumptions or a model specification is required.

Second, the first stage and reduced form errors $u_i$ and $\varepsilon_i$ are allowed to be heteroskedastic.
Although heteroskedasticity has been accounted for in existing methods
with many instruments (e.g.\ \cite{hausman2012instrumental}, \cite{mikusheva2021inference}, \cite{crudu2021inference}), these consider IV models as in \eqref{eq:linearIV} with homogeneous effects.
In our setting, assuming homoskedastic errors would also implicitly restrict treatment effect heterogeneity.

Third, inference based on our test statistic in \eqref{eq:test_statistic} is asymptotically valid under a minimal assumption on the identification strength. That is, the test works with sets of instrument interactions that have a strong or a semi-strong signal. The instrument interactions are referred to as strong when the concentration parameter
\begin{equation}\label{eq:mun}
   \frac{n}{G}\text{FS}\rightarrow \infty \text{ with FS}=\sum_g \tilde{\mathbb{P}}[X_i=x_g] \pi(x_g)^2 \tilde{\text{V}}[Q_i|X_i=x_g].
\end{equation}
Since $|\pi(x_g)|\leq 1$ and $\text{Var}[Q_i|X_i=x_g]\leq \frac{1}{4}$, in the scenario with the strongest identification possible within each covariate group, we require $\frac{n}{4} \gg G$ for the instrument interaction set to be considered as strong. In this case, we show that the SIVE estimator is consistent, the variance estimator \eqref{eq:Vhat} is consistent, and the asymptotic confidence intervals constructed from the test statistic attain nominal coverage. Under semi-strong identification, that is if$\frac{n}{\sqrt{G}}\text{FS}\rightarrow\infty$ and $G\rightarrow\infty$, the SIVE estimator remains consistent and inference may be conservative if there exist covariate groups with $m_{1,g}\leq 3$ or $m_{0,g}\leq 3$. \citet{mikusheva2021inference} show that this is the weakest identification strength under which a consistent estimator exists. In the case that one is worried that the identification is even weaker so that $\frac{n}{\sqrt{G}}\text{FS} \rightarrow C<\infty$ for some constant $C\in[0,\infty)$ and $G \rightarrow \infty$, the results we provide allow for the construction of fully identification robust confidence intervals by inverting a score statistic as in \citet{kleibergen2005testing}.

\begin{remark}
Considering the use of clustered standard errors in many applied IV analyses \citep{abadie2023should,mackinnon2023cluster}, it is useful to relax Assumption~\ref{ass:indep}. Supplemental Appendix~\ref{A:cluster} shows that SIVE remains consistent and proposes a variance estimator for settings with clustered errors. More specifically, our results allow for clustering at, for instance, the household level provided that each person in a household has a different covariate vector.
\end{remark}

\subsection{Large sample theory}\label{sec:large_sample}
We first provide a consistency result for $\hat{\beta}^{\text{SIVE}}$. 

\begin{lemma}\label{lemm:consistency}
Under Assumptions \ref{ass:IV} to \ref{ass:large_sample} with $\frac{n}{\sqrt{G}}\emph{FS}\rightarrow_{p}\infty$, it holds that $\hat{\beta}^{\emph{SIVE}}-\beta^{\emph{SIVE}}\rightarrow_{p} 0$.
\end{lemma}
The proof is deferred to Appendix~\ref{subsec:consistency}. The condition on the identification strength covers strong identification and semi-strong identification as defined in Section~\ref{subsec:assumptions}. The most stringent case occurs when $G\propto n$, in which case we require that $\sqrt{n}\text{FS}\rightarrow_{p}\infty$. Even then, $\text{FS}$ is allowed to {decrease} to zero asymptotically. 

The following result shows the asymptotic validity of a standard $t$-test that uses the variance estimator from \eqref{eq:Vhat}. What is particularly important to note is that the theorem does not limit the rate at which $G$ can grow with $n$. In particular, we allow $G/n\rightarrow c\in (0,1)$ which are the many-instrument sequences by \citet{bekker1994alternative}. 

\begin{theorem}\label{corr:dist2} Under Assumptions~\ref{ass:IV} to \ref{ass:large_sample} and if $\frac{n}{\sqrt{G}}\text{FS}\rightarrow_{a.s.}\infty$,
\begin{equation}\label{eq:corr:dist2}
\frac{\hat{\beta}^{\emph{SIVE}}-\beta^{\emph{SIVE}}}{\sqrt{\hat{\mathbb{V}}[\hat{\beta}^{\emph{SIVE}}|Q,X]}}\rightarrow_{d} N(0,\lambda),
\end{equation}
with $\hat{\mathbb{V}}[\hat{\beta}^{\emph{SIVE}}|Q,X]$ from \eqref{eq:Vhat} and $\lambda\leq 1$. Moreover, (i) $\lambda = 1$ if $\frac{n}{G}\emph{FS}\rightarrow_{a.s.}\infty$, (ii) $\lambda=1$ if $m_{1,g}\geq 4$ and $m_{0,g}\geq 4$, (iii) $\lambda \in[1/4,1]$ if $m_{1,g}\geq 3$ and $m_{0,g}\geq 3$.
\end{theorem}

The proof is deferred to Appendix~\ref{app:dist}. Theorem~\ref{corr:dist2} shows that under strong identification or with sufficiently large groups through a strengthening of Assumption~\ref{ass:group_size}, we achieve asymptotically nominal coverage. In the presence of smaller groups and weaker identification, the procedure becomes conservative. If $m_{1,g}\geq 3$ and $m_{0,g}\geq 3$, Theorem~\ref{corr:dist2} shows that we can precisely quantify how conservative the procedure maximally is. Technically, the most challenging part to establish Theorem~\ref{corr:dist2} is that the variance estimator converges to a quantity that is equal to, or at least as large as the population variance, conditional on the covariates $X$ and instrument $Q$.

\begin{remark}\label{rem:robust} A natural question regarding Theorem~\ref{corr:dist2} concerns settings in which the identification may be even weaker than that required by the theorem. Our results can be used to construct an identification-robust procedure. Consider the null hypothesis $H_{0}\colon \beta^{\text{SIVE}}=\beta_{0}$. By replacing the estimator $\hat{\beta}^{\text{SIVE}}$ in the variance estimator \eqref{eq:Vhat} by $\beta_{0}$,  Theorem~\ref{corr:dist2} holds under $H_{0}$ without the requirement that $\frac{n}{\sqrt{G}}\text{FS}\rightarrow_{a.s.}\infty$ and only requires  $G\rightarrow\infty$. Hence, confidence intervals can be constructed using test inversion.
\end{remark}

\section{Monte Carlo Study}\label{sec:mc}
We illustrate the finite sample properties of the proposed inference method relative to existing methods in an empirically relevant setting. We build on the Monte Carlo experiment considered in \citet{blandhol2022tsls} that is designed to match some key features of the data used by \citet{card1995using}. We consider the inference method proposed in Section~\ref{sec:SIVE} and the estimators discussed in Section~\ref{sec:LATEs}: TSLS including the controls linearly (TSLS-L), TSLS with a saturated specification (TSLS-S), and the jackknife estimators JIVE, IJIVE, and UJIVE.

\textbf{Set-up:} Consider a sample with $n=3,000$ observations. The endogenous treatment and the outcome variable are generated as
\begin{equation}\label{mc:dgp}
\begin{split} 
    T_{i}&= \Phi(u_{i})\leq p_{0}1[Q_{i}=0] + p_{1}1[Q_{i}=1], \\
    Y_{i} &= \log(129.7 + 1247.7X_{i} -2149.0 X_{i}^2 + 1515.7 X_{i}^3) + \gamma_i T_{i} + \varepsilon_{i},
    \end{split}
\end{equation}
where $\Phi(.)$ denotes the cumulative distribution function of a standard normal random variable. We generate $(u_{i},\varepsilon_{i})'\sim N( 0,\Sigma)$ where $[\Sigma]_{11}=[\Sigma]_{22}=1$ and $[\Sigma]_{12}=0.527$. We set $\gamma_i=1.2$ across all $i$, which corresponds to $\tau=1.2$. We extend this to heterogeneous treatment effects in Supplemental Appendix~\ref{A:mc}. We set $p_{0}=0.22$ and vary the identification strength through $p_{1}\in\{0.39,0.49,0.59,0.69\}$. The binary instrument satisfies 
\begin{equation}\label{eq:EQ}
    \mathbb{E}[Q_{i}|X_{i}] = 0.119 + 1.785 X_{i} - 1.534X_i^2 + 0.597 X_i^3.
\end{equation}
We have a single control variable $X_{i}$ that can take on $G=\{1,50, 200\}$ values taken from a one-dimensional Halton sequence. 

\begin{figure}[t]
\centering
\caption{Absolute median bias}\label{mc:bias} \includegraphics[width=1\textwidth,trim=1.8cm 0 2cm 0,clip]{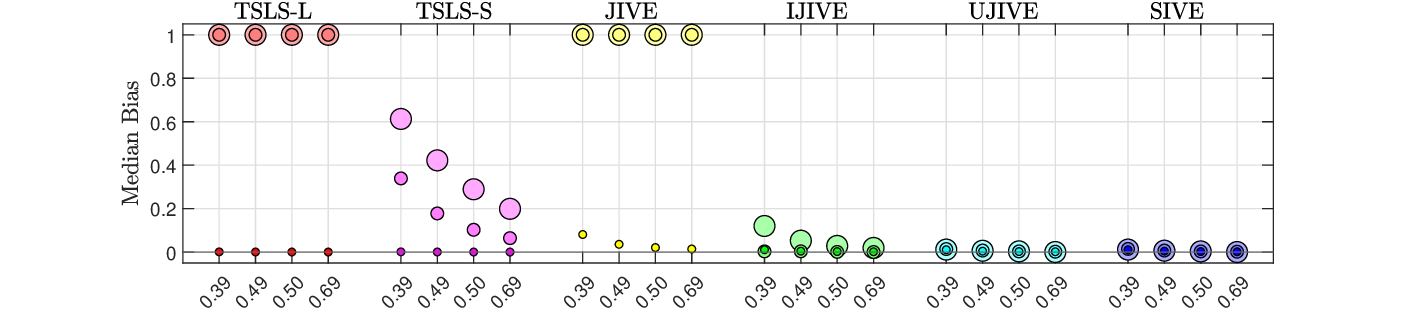}\vspace{0.2cm}
     \begin{minipage}{0.9\textwidth}
        \footnotesize{
\textit{Note}: the figure shows the absolute median bias with the causal estimand in a setting without treatment effect heterogeneity. The size of the circles indicates the number of covariate groups with the small circle corresponding to $G=1$, the medium circle corresponding to $G=50$ and the large circle corresponding to $G=200$. The $x$-axis is the instrument strength $p_1-p_0$, with $p_0=0.22$ and $p_1 = \{0.39,0.49,0.59,0.69\}$. Because TSLS-L and JIVE show large biases, the reported bias values are truncated at one.}
\end{minipage}
\end{figure}

\textbf{Bias:} We first study the bias of the estimators. Figure~\ref{mc:bias} shows the absolute median bias as a function of the instrument strength $p_1$, where values for the absolute bias are truncated at one. The small circles correspond to $G=1$ covariate groups, the medium-sized circles to $G=50$ groups, and the large circles to $G=200$ groups. 

We find that SIVE is unbiased across all settings under consideration. For estimators commonly used in applied work, TSLS-L and TSLS-S, the bias becomes substantial when the number of covariate groups increases. In this setting, TSLS-L does not correspond to a causal estimand and TSLS-S suffers from many instrument bias, which is more pronounced in settings where the instruments are weak. From the jackknife estimators, JIVE is not performing much better. Since the controls are not appropriately projected out in this estimator, it incurs a large omitted variable bias. IJIVE and UJIVE are performing considerably better, with the former only incurring bias when there are many weak instruments and the latter is unbiased in all settings.

\begin{figure}[t]
\centering
\caption{Test size}\label{mc:size}
 \includegraphics[width=1\textwidth,trim=1.8cm 0 2cm 0,clip]{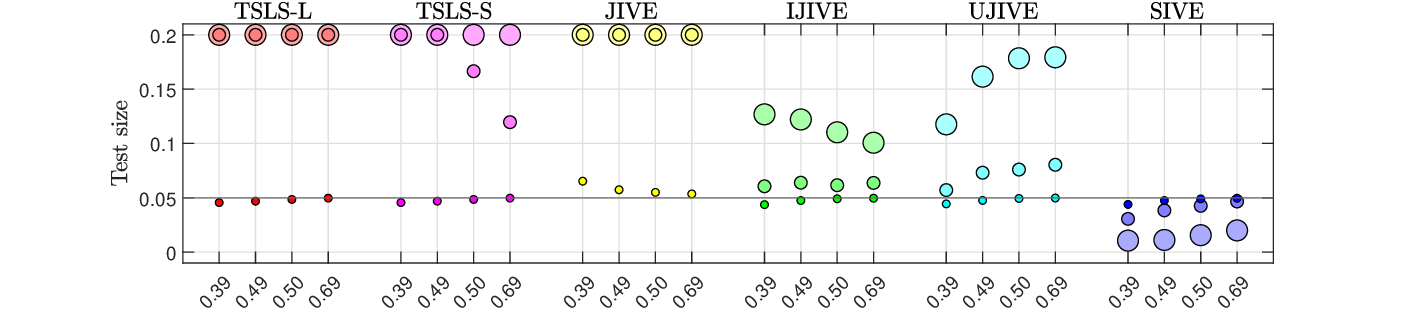}
    \begin{minipage}{0.95\textwidth}
    \vspace{0.2cm}
        \footnotesize{
\textit{Note}: 
the figure shows the size of testing $H_{0}\colon \beta=1.2$ at a nominal level of $5\%$ in a setting without treatment effect heterogeneity. The size of the circles indicates the number of covariate groups, with the small circle corresponding to $G=1$, the medium circle corresponding to $G=50$, and the large circle corresponding to $G=200$. The $x$-axis is the instrument strength $p_1-p_0$, with $p_0=0.22$ and $p_1 = \{0.39,0.49,0.59,0.69\}$. Because TSLS-L, TSLS-S, and JIVE show large size distortions, the reported size values are truncated at 0.2. TSLS-L and TSLS-S use heteroskedasticity-robust (HC0) standard errors, JIVE and IJIVE use variance estimators proposed in \citet{ackerberg2009improved}, UJIVE as in \citet{evdokimov2018inference}, and SIVE uses \eqref{eq:Vhat}.}
\end{minipage}
\end{figure}

\textbf{Size:} Second, we examine the empirical size of a test of $H_{0}\colon \beta=\tau$ with $\tau=1.2$ and a nominal test size of 5\%. Figure~\ref{mc:size} shows the test size as a function of the instrument strength, where values for the size are truncated at 0.2.

SIVE obtains close to nominal size control for $G=1$ for all values of the instrument strength. As follows from Theorem~\ref{corr:dist2}, the test becomes progressively more conservative for a larger number of covariate groups and increasing the instrument strength makes the test less conservative. For TSLS-L, TSLS-S, and JIVE we observe substantial size distortions for all instrument strengths. This can be explained by the biases observed in Figure~\ref{mc:bias}. IJIVE and UJIVE also show substantial deviations from nominal size. This can be explained by the assumptions under which the variance estimators for these estimators are proposed: They require the number of control dummies to be a negligible fraction of the sample size.

In settings with heterogeneous treatment effects, we find similar or slightly less conservative sizes with SIVE compared to the homogeneous settings, but the size of the other methods further moves away from the nominal size. Instead of $\gamma_i=1.2$ in \eqref{mc:dgp}, we set $\gamma_i=6$ for observations in covariate groups with fewer observations than the median sized covariate group, and $\gamma_i=1.2$ for the remaining observations. Detailed results are reported in Supplemental Appendix~\ref{A:mc}. This appendix also shows that potential alternative variance estimators, such as the ones proposed by \citet{lee2018consistent} and \citet{chao2023jackknife}, do not control size in settings with treatment effect heterogeneity and/or many instrument interactions.

\begin{figure}[t]
\centering
\caption{Statistical power curves SIVE}\label{mc:power}
    \includegraphics[width=1\textwidth,trim=1.8cm 0 2cm 0,clip]{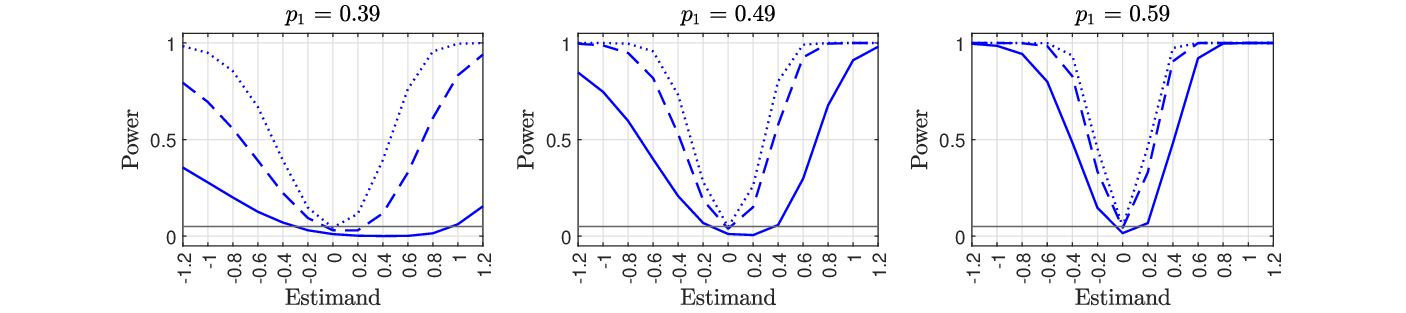}
    \vspace{0.2cm}
     \begin{minipage}{0.9\textwidth}
        \footnotesize{
\textit{Note}: the figure shows the statistical power of testing $H_{0}\colon \beta=0$ at a nominal level of $5\%$ in a setting without treatment heterogeneity. The panels correspond to different instrument strengths $p_1-p_0$, with $p_0=0.22$ and $p_1 = \{0.39,0.49,0.59\}$. The dotted line corresponds to $G=1$, the dashed line to $G=50$, and the solid line to $G=200$. The $x$-axis shows the value of $\gamma_i$ in \eqref{mc:dgp}, which equals $\beta$ in this setting.}
\end{minipage}
\end{figure}

\textbf{Power:}
Third, we consider the statistical power of a test of $H_{0}\colon \beta=0$  with a nominal test size of 5\%. Since our proposed method is the only one that accurately controls size, we focus on the power of SIVE. Figure~\ref{mc:power} shows the power curves of SIVE with different numbers of covariate groups and varying instrument strength. We consider values of $\beta$ between -1.2 and 1.2, from which the latter is set as an empirically realistic effect in this setting by \citet{blandhol2022tsls}. Hence, we consider power curves in an empirically relevant range of treatment effects. Although Figure~\ref{mc:size} shows that SIVE can be conservative, we find that the test still has power even when the number of covariate groups is large and the instruments are weak. For modest or strong instruments, power is close to one when $\beta=1.2$, even with a large number of covariate groups.

\section{Empirical applications}\label{sec:application}
This section illustrates the empirical relevance of the proposed method for treatment evaluation. We consider the LATE of pretrial detention using a judge design and the LATE of banking access using a field experiment. Supplemental Appendix~\ref{A:application} discusses empirical applications to the LATEs of schooling, health insurance, and an educational program. 

\subsection{The impact of pretrial detention}
First, we discuss the LATE of pretrial detention on conviction using the judge design data studied by \citet{stevenson2018distortion}. The data has also been recently analyzed by \citet{mogstad2024instrumental} and \citet{sloczynski2020should}. The initial study already discusses violations of strong monotonicity caused by judges being relatively strict for certain types of defendants but relatively lenient for other types of defendants. 

The sample consists of 331,971 arrests between 2006 and 2013 from the Philadelphia court records. Each case is randomly assigned to one of seven judges. For ease of exposition, we select the two judges with the lowest and highest proportion of cases in which
the defendant is detained pretrial in 2006 and 2007. To ensure the exogeneity of the judge selection, we then consider the sample of 86,765 cases assigned to one of these two judges between 2008 and 2013. 
The binary outcome variable equals one if the case outcome is conviction, and the binary treatment variable equals one if the defendant is detained pretrial. The binary instrument equals one if the case was assigned to the more lenient judge in 2006-2007.
Following one of the control specifications in \citet{sloczynski2020should}, our analysis includes 19 controls: 17 offense type dummies and two race dummies.

The first three columns of Table~\ref{tab:app} show the TSLS-L, TSLS-S, and SIVE estimates, with their corresponding standard errors. TSLS-L relies on the parametric assumption of a linear relation between the controls and the outcome and treatment, and a strong monotonicity assumption on the instrument.  The estimator is inconsistent if either of those assumptions is violated, which probably explains the implausible magnitude of the estimate. The TSLS-S estimate is more realistic, but may suffer from a many and potentially weak instrument bias. The SIVE estimate is slightly larger than the TSLS-S estimate, but substantially smaller than the TSLS-L estimate. Using the appropriate variance estimator, the standard error is considerably smaller than using the standard variance estimator for TSLS-L and slightly larger than the standard error of TSLS-S. 
We conclude that our proposed method is able to correct for issues in LATE estimation in the commonly used judge design. The SIVE estimate is in line with the literature, corrects for the bias in TSLS estimates caused by violations of the linearity assumption on the controls or the strong monotonicity assumption on the instruments, and is statistically significant. 

\begin{table}[t]
  \centering
  \caption{Estimated effects of pretrail detention and banking access}
    \begin{tabular}{lrrrrrr}
    
          & \multicolumn{3}{c}{\citet{stevenson2018distortion}} & \multicolumn{3}{c}{\citet{dupas2018banking}} \\
          \cline{2-4}  \cline{5-7}
          & \multicolumn{1}{l}{TSLS-L} & \multicolumn{1}{l}{TSLS-S} & \multicolumn{1}{l}{SIVE} & \multicolumn{1}{l}{TSLS-L} & \multicolumn{1}{l}{TSLS-S} & \multicolumn{1}{l}{SIVE} \\
          \hline
    Estimate & 3.580 & 0.204 & 0.275 & 20.117 & 20.511 & 20.726 \\
    Std. Error & 0.734 & 0.036 & 0.111 & 2.795 & 2.790 & 3.001 \\
    \hline
    Variables & 19    & 2228  & 1548   & 43    & 246   & 224 \\
    Observations & 86765 & 86108 & 85573 & 6007  & 6007  & 5960 \\
    \hline
    \end{tabular}%
  \label{tab:app}\\%
\begin{minipage}{0.85\textwidth}
  	\vspace{0.2cm}
  	\footnotesize{\textit{Note:} the total number of variables includes the controls and instrument in TSLS-L, and the covariate groups and instrument interactions with all covariate groups in TSLS-S and SIVE. Due to lack of variation in the instrument for some covariate groups, the final two estimators may use less observations, denoted in the fourth row. Standard errors are heteroskedasticity-robust in the first three columns, and clustered at the household level in the final three columns.}
    \end{minipage}
\end{table}%

\subsection{The impact of banking access}
Second, we consider a field experiment for which we have no a priori indications that the TSLS estimates are problematic. \citet{dupas2018banking} study a randomized opportunity to open a free bank account to estimate the LATE of opening a bank account on savings balance. 
We focus on the LATE on savings in formal financial institutions in Uganda. The outcome variable is savings stocks in 2010 US dollars at formal financial institutions including commercial banks, microfinance banks, and savings and credit cooperatives. The binary treatment variable equals one for households who actually opened the accounts and made at least one deposit. The binary instrument equals one for households offered a voucher to open bank accounts with no financial costs. The analysis includes 43 controls; 41 strata dummies cover the stratified randomization by gender, occupation, and bank branch. Two wave dummies indicate the three rounds of follow-up surveys, administered approximately 6, 12, and 18 months after accounts were opened. The sample size is 6,007, standard errors are clustered at the household level, and our sample includes 2,085 households.

The final three columns of Table~\ref{tab:app} show the estimates for the LATE of banking access. The estimate in the first column is identical to the estimate in \citet[Table 4, Column (1), Panel A]{dupas2018banking}. The differences between the estimates of TSLS-L, TSLS-S, and SIVE are small. The standard error of SIVE is slightly larger, but this does not change the qualitative results.
Supplemental Appendix~\ref{A:application} finds that SIVE also produces similar results to TSLS in other carefully conducted field experiments. Using SIVE in these settings, the practitioner can be assured that the results are not driven by any implicit assumptions. The SIVE confidence intervals respect the full nonparametric nature of the LATE framework, and this does not necessarily imply a substantial loss in precision. 

\section{Conclusion}\label{sec:conclusion}
Although TSLS with covariates is known to require saturation to obtain a causal interpretation, one of the main concerns of applied economists are the statistical challenges involved in inference with a saturated specification. We propose a method that tackles these challenges. This method requires a valid instrument that satisfies the standard \citet{imbens1994identification} assumptions, and minimal assumptions on the number of observed individuals in each covariate group and the identification strength. Unlike existing procedures, we allow arbitrary treatment effect heterogeneity and covariates with rich support, while still being able to detect empirically relevant effect sizes.
\newpage
\begin{appendix}

\section{Definitions and notation}\label{A:defsandnotation}
Throughout the appendix, $\iota_{n}$ and $0_{n}$ are $n\times 1$ vectors of ones/zeros, $I_{n}$ is the $n\times n$ identity matrix, $O_{n_{1},n_{2}}$ is an $n_{1}\times n_{2}$ matrix of zeros, $\lambda_{\max}(A)$ and $\text{tr}(A)$ denote the maximum eigenvalue and trace of the matrix $A$, $A\odot B$ is the elementwise product between matrices $A$ and $B$ of the same dimension, and $\beta$ refers to $\beta^{\text{SIVE}}$ unless noted otherwise. Define  $r_{n} = \left\{\begin{array}{cl}
    n\text{FS} & \text{if }    n\text{FS}\rightarrow\infty,\\
    G & \text{otherwise}.
    \end{array}\right.$ with $\frac{n}{G}\text{FS}=\frac{n}{G}\pi'Z'M_W Z\pi/n$ from \eqref{eq:mun}.

For any $n\times n$ matrix $M$, $M_{(g)}$ corresponds to the $n_{g}\times n_{g}$ submatrix of $M$ composed of individuals with $W_{ig}=1$. Without loss of generality, we assume that the observations are ordered according to the dummies in the matrix $W$ indicating the values of the covariates. As a subsequent ordering, we assume again without loss of generation that the observations are ordered based on the value of the instrument. That is, 
\begin{equation}
    W = \left(\begin{array}{cccc}
    \iota_{n_{1}} &  0_{n_{1}}&\ldots &  0_{n_{1}}\\
     0_{n_{2}} &  \iota_{n_{2}} & \ldots &  0_{n_{2}}\\
    \vdots & \vdots & \ddots &\vdots\\
     0_{n_{G}} &  0_{n_{G}} &\ldots & \iota_{n_{G}}
    \end{array}\right), \quad  Z = \left(\begin{array}{llll}
    \iota_{m_{1,1}} &  0_{m_{1,1}}&\ldots &  0_{m_{1,1}}\\
     0_{m_{0,1}} &  0_{m_{0,1}}&\ldots &  0_{m_{0,1}}\\
     0_{m_{1,2}} &  \iota_{m_{1,2}} & \ldots &  0_{m_{1,2}}\\
     0_{m_{0,2}} &  0_{m_{0,2}} & \ldots &  0_{m_{0,2}}\\
    \vdots & \vdots & \ddots &\vdots\\
     0_{m_{1,G}} &  0_{m_{1,G}} &\ldots & \iota_{m_{1,G}}\\
     0_{m_{0,G}} &  0_{m_{0,G}} &\ldots &  0_ {m_{0,G}}
    \end{array}\right).
\end{equation}
The TSLS projection matrix $P=M_{W}Z(Z'M_{W}Z)^{-1}M_{W}Z$ is block diagonal with blocks
\begin{equation}\label{eq:Pg}
P_{(g)}=\left(\begin{array}{cc}
\frac{1}{n_{g}}\frac{m_{0,g}}{m_{1,g}}\iota_{m_{1,g}}\iota_{m_{1,g}}'& 
-\frac{1}{n_{g}}\iota_{m_{1,g}}\iota_{m_{0,g}}'\\
-\frac{1}{n_{g}}\iota_{m_{0,g}}\iota_{m_{1,g}}' &\frac{1}{n_{g}}\frac{m_{1,g}}{m_{0,g}}\iota_{m_{0,g}}\iota_{m_{0,g}}'\end{array}\right).
\end{equation}
Define $L=V(V'V)^{-1}V'$, which is a block diagonal matrix with blocks
\begin{equation}\label{eq:L}
\begin{split}
   L_{(g)} &=\left(\begin{array}{cc}
   \frac{1}{m_{1,g}}\iota_{m_{1,g}}\iota_{m_{1,g}}'  &  O_{m_{1,g},m_{0,g}}\\
       O_{m_{0,g},m_{1,g}} &  \frac{1}{m_{0,g}}\iota_{m_{0,g}}\iota_{m_{0,g}}'
    \end{array}\right).
\end{split}\end{equation}
The matrix $M_{W,Z} = I_{n}-L$, used to bias correct TSLS and to estimate the variance, is then block diagonal with blocks 
\begin{equation}\label{eq:Mtil_g}
\begin{split}
    M_{W,Z,(g)} &= I_{n_{g}}-L_{(g)} = I_{n_{g}}-\left(\begin{array}{cc}
   \frac{1}{m_{1,g}}\iota_{m_{1,g}}\iota_{m_{1,g}}'  &  O_{m_{1,g},m_{0,g}}\\
       O_{m_{0,g},m_{1,g}} &  \frac{1}{m_{0,g}}\iota_{m_{0,g}}\iota_{m_{0,g}}'
    \end{array}\right).
\end{split}\end{equation}
The matrix $A = P-M_{W,Z}DM_{W,Z}$ is block diagonal with diagonal blocks
\begin{equation}\label{eq:Ag}
\begin{split}
    {A}_{(g)} &= \left(\begin{array}{cc}
\frac{m_{0,g}}{n_g(m_{1,g}-1)}(\iota_{m_{1,g}}\iota_{m_{1,g}}'-I_{m_{1,g}}) & -\frac{1}{n_{g}}\iota_{m_{1,g}}\iota_{m_{0,g}}'\\
-\frac{1}{n_{g}}\iota_{m_{0,g}}\iota_{m_{1,g}}' &\frac{m_{1,g}}{n_g(m_{0,g}-1)}(\iota_{m_{0,g}}\iota_{m_{0,g}}'-I_{m_{0,g}}) 
\end{array}\right).
\end{split}
\end{equation}
The eigenvalues of $A$ are $0$ with multiplicity 1, $1$ with multiplicity $1$, $-m_{0,g}/(n_g(m_{1,g}-1))$ with multiplicity $m_{1,g}-1$ and $-m_{1,g}/(n_g(m_{0,g}-1))$ with multiplicity $m_{0,g}-1$. Since $m_{1,g}\geq 2$ and $m_{0,g}\geq 2$, the eigenvalues are on the [-1,1] interval. 

When estimating the variance of the SIVE estimator, we define $C = (A\odot L)J(A\odot L)-((A\odot L)J(A\odot L))\odot I$ with $J$ a diagonal matrix with $[J]_{ii}=(1-L_{ii})^{-1}(1-2L_{ii})^{-2}$. The matrix $C$ is block diagonal with blocks
\begin{equation}\label{eq:C}
     C_{(g)} = \left(\begin{array}{cc}
    \frac{m_{0,g}^2}{n_g^2(m_{1,g}-1)^3}\big(\iota_{m_{1,g}}\iota_{m_{1,g}}'- I_{m_{1,g}}\big) &  O_{m_{1,g},m_{0,g}}\\
       O_{m_{0,g},m_{1,g}}  & \frac{m_{1,g}^2}{n_g^2(m_{0,g}-1)^3}\big(\iota_{m_{0,g}}\iota_{m_{0,g}}'- I_{m_{0,g}}\big)
    \end{array}\right),
\end{equation}
from which it follows that $C_{ij}\leq A_{ij}^2$.

In the proofs we use the elementwise bound  $A\leq B$ where $B$ is block diagonal with blocks 
\begin{equation}\label{eq:tildeAg}
B_{(g)} = 2\frac{m_{1,g}m_{0,g}}{n_{g}}\left(\begin{array}{cc}
\frac{1}{m_{1,g}^2}\iota_{m_{1,g}}\iota_{m_{1,g}}'& \frac{1}{m_{1,g}m_{0,g}}\iota_{m_{1,g}}\iota_{m_{0,g}}'\\
\frac{1}{m_{1,g}m_{0,g}}\iota_{m_{0,g}}\iota_{m_{1,g}}' &\frac{1}{m_{0,g}^2}\iota_{m_{0,g}}\iota_{m_{0,g}}'  \end{array}\right).
\end{equation}
Several properties of this matrix are established in Supplemental Appendix~\ref{A:Aresults}.

Finally, $D_{{u}^2}$ is a diagonal matrix with elements $[D_{{u}^2}]_{ii}=u_{i}^2$. We similarly define $D_{{v}}^2$ and $D_{{uv}}$. Define $D_{\sigma_{u}^2} = \mathbb{E}[D_{u^2}|Q,X]$ and analogous definitions are used for $D_{\sigma_{v}^2}$ and $D_{\sigma_{uv}}$. 

\section{Proofs - estimands and consistency}

\subsection{Proof \texorpdfstring{Lemma~\ref{lemm:2sls}}{Lemma \ref{lemm:2sls}}}\label{A:2sls}
In a saturated specification, we can write
\begin{align}
    T_i = \pi'Z_{i} + \psi'W_{i} +u_i,
    \qquad
    Y_i = \theta' Z_{i} + \phi'W_{i} +\varepsilon_i,\label{eq:2slseq}
\end{align}
where $\pi=(\pi(x_1),\dots,\pi(x_g))'$ with $\pi(x_g)=\mathbb{P}[T_i(1) > T_i(0)|X_i=x_g]-\mathbb{P}[T_i(1) < T_i(0)|X_i=x_g]$, $\psi=(\psi_1,\dots,\psi_g)'$ with $\psi_g=\mathbb{E}[T_i|Z_{ig}=0,W_{ig}=1]$, $\theta=(\theta(x_1),\dots,\theta(x_g))'$ with $\theta(x_g)=\tau(x_g)\pi(x_g)$, and $\phi=(\phi_1,\dots,\phi_g)'$ with $\phi_g=\mathbb{E}[Y_i|Z_{ig}=0,W_{ig}=1]$.

For the TSLS estimand, we now obtain the following.
\begin{equation}\label{eq:tslsestimand}
\begin{split}
    \beta^{\text{TSLS}}
    &=\frac{\pi'Z'M_{W}Z\theta+\mathbb{E}[u'P\varepsilon|Q,X]}{\pi'Z'M_{W} Z\pi+\mathbb{E}[u'P u|Q,X]}\\
     &= \frac{\sum_g  \frac{n_g}{n}\pi(x_g)^2 \frac{m_{1,g}}{n_g}\frac{m_{0,g}}{n_g}\tau(x_g) + \frac{1}{n}\sum_i \mathbb{E}[u_i \varepsilon_i|Q,X] P_{ii}}{\sum_g \frac{n_g}{n} \pi(x_g)^2 \frac{m_{1,g}}{n_g}\frac{m_{0,g}}{n_g} + \frac{1}{n}\sum_i \mathbb{E}[u_i^2|Q,X] P_{ii}},
\end{split}
\end{equation}
where we use that $Z'M_{W}Z$ is a $G\times G$ diagonal matrix with elements $n_g\frac{m_{1,g}}{n_g}\frac{m_{0,g}}{n_g}$.

Note that $e_i'M_{W}Z$ is a $1 \times G$ vector with as elements the residual of observation $i$ in the regression of $Z_{ig}$ on $W_{ig}$, so $[e_i'M_{W}Z]_g= Z_{ig}-W_{ig}\tilde{\text{P}}[Z_i=1|X_i=x_g]$, where $\tilde{\text{P}}[Z_i=1|X_i=x_g]=\frac{m_{1,g}}{n_g}$. Hence, $P_{ii}=e_i'M_{W}Z(Z'M_{W}Z)^{-1}Z'M_{W}e_i = \sum_g \frac{1}{n_g}W_{ig}\frac{(Z_{ig}-\tilde{\text{P}}[Q_i=1|X_i=x_g])^2}{\text{Var}[Q_i|X_i=x_g]}$.

\subsection{Proof \texorpdfstring{Lemma~\ref{lemm:diag}}{Lemma \ref{lemm:diag}}}\label{A:diag}
First, we show that $P_{ii}=[M_{W,Z}DM_{W,Z}]_{ii}$ if the elements of the diagonal matrix $D$ are set equal to $D_{ii} = \sum_{k=1}^{n}[(M_{W,Z}\odot M_{W,Z})^{-1}]_{ik}P_{kk}$. Using that $D$ is diagonal and $M_{W,Z}$ symmetric,
\begin{equation}
    \begin{split}
    [M_{W,Z}DM_{W,Z}]_{ii} 
    &= \sum_{j}[M_{W,Z}\odot M_{W,Z}]_{ij}\sum_{k=1}^{n}P_{kk}[(M_{W,Z}\odot M_{W,Z})^{-1}]_{kj}\\
    &= \sum_{k=1}^{n}P_{kk}\sum_{j}[M_{W,Z}\odot M_{W,Z}]_{ij}[(M_{W,Z}\odot M_{W,Z})^{-1}]_{kj} =P_{ii}.
    \end{split}
\end{equation}
Next, we derive an expression for $D_{ii}$. Starting with $
    M_{W,Z,(g)}$ as in \eqref{eq:Mtil_g}, it follows that 
\begin{align}
    &[(M_{W,Z} \odot M_{W,Z})^{-1}]_{(g)}\label{eq:invMtilMtil}\\
    &= \left(\begin{array}{cc}
   \frac{m_{1,g}}{m_{1,g}-2}\left( I_{m_{1,g}} - \frac{1}{m_{1,g}(m_{1,g}-1)}\iota_{m_{1,g}}\iota_{m_{1,g}}'\right) &  O_{m_{1,g},m_{0,g}}\\
       O_{m_{0,g},m_{1,g}} &   \frac{m_{0,g}}{m_{0,g}-2}\left( I_{m_{0,g}} - \frac{1}{m_{0,g}(m_{0,g}-1)}\iota_{m_{0,g}}\iota_{m_{0,g}}'\right)
      \end{array}\right)\nonumber.
\end{align}
Using this expression, and with $P_{ii}$ as in \eqref{eq:Pg},
\begin{equation}\label{eq:Diiapp}\begin{split}
    D_{ii}
    &= \sum_g W_{ig} \left[\frac{m_{1,g}}{m_{1,g}-2}\left(1-\frac{m_{1,g}}{m_{1,g}(m_{1,g}-1)}\right)\frac{1}{n_g}\frac{m_{0,g}}{m_{1,g}}Z_{ig} \right.\\
    &\qquad \left.+ \frac{m_{0,g}}{m_{0,g}-2}\left(1-\frac{m_{0,g}}{m_{0,g}(m_{0,g}-1)}\right)\frac{1}{n_g}\frac{m_{1,g}}{m_{0,g}}(1-Z_{ig})\right]\\
    &=\sum_g \frac{1}{n_g} W_{ig} \left[\frac{m_{0,g}}{m_{1,g}-1}Z_{ig}+\frac{m_{1,g}}{m_{0,g}-1}(1-Z_{ig})\right].
\end{split}\end{equation}
While the derivation above implicitly assumes that $m_{1,g}\geq 3$ and $m_{0,g}\geq 3$ for the inverse in \eqref{eq:invMtilMtil} to be well-defined, it is straightforward to verify that $D_{ii}$ as given on the final line of \eqref{eq:Diiapp} yields a zero diagonal for $P-M_{W,Z}DM_{W,Z}$ even if $m_{1,g}=2$ and/or $m_{0,g}=2$. 

\subsection{Proof \texorpdfstring{Theorem~\ref{lemm:sive}}{Theorem~ \ref{lemm:sive}}}\label{A:sive}
We write the SIVE estimand as
\begin{equation}
    \beta=
    \frac{\mathbb{E}[T'PY|Q,X]-\mathbb{E}[T'M_{W,Z}DM_{W,Z} Y|Q,X]}{\mathbb{E}[T'PT|Q,X]-\mathbb{E}[T'M_{W,Z}DM_{W,Z} T|Q,X]}.
\end{equation}
Using \eqref{eq:2slseq}, as well as the fact that by construction $[M_{W,Z}DM_{W,Z}]_{ii}=P_{ii}$, we have
\begin{equation}\begin{split}
    \mathbb{E}[T'M_{W,Z}DM_{W,Z} Y|Q,X] &= \mathbb{E}[u'M_{W,Z}DM_{W,Z} \varepsilon|Q,X]=\sum_i \mathbb{E}[u_i \varepsilon_i |Q,X]P_{ii},\\
    \mathbb{E}[T'M_{W,Z}DM_{W,Z} Y|Q,X]& =\mathbb{E}[u'M_{W,Z}DM_{W,Z} u|Q,X]\sum_i \mathbb{E}[u_i^2  |Q,X]P_{ii}.
\end{split}\end{equation}
Combining this result with \eqref{eq:tslsestimand},
\begin{equation}\begin{split}
   \beta &= \frac{\sum\nolimits_g  \frac{n_g}{n}\pi(x_g)^2 \frac{m_{1,g}}{n_g}\frac{m_{0,g}}{n_{g}} \tau(x_g) }{\sum\nolimits_g \frac{n_g}{n} \pi(x_g)^2 \frac{m_{1,g}}{n_g}\frac{m_{0,g}}{n_{g}}  },
\end{split}\end{equation}
which is well-defined since $\tau(x_g)\leq c$ $a.s.$ by Assumption~\ref{ass:IV}, while Assumption~\ref{ass:group_size} and \ref{ass:relevance} ensure that the denominator is not zero.

\subsection{Proof \texorpdfstring{Lemma~\ref{lemm:consistency}}{Lemma \ref{lemm:consistency}}}\label{subsec:consistency}
First, we establish a bound on $\text{tr}(A^2)$, where we use that the elements of $A$ corresponding to individuals in covariate group $g$ are given in \eqref{eq:Ag}.
\begin{equation}\label{eq:trA2upper}
\begin{split}
    \text{tr}(A^2) &= G + \text{tr}(M_{W,Z}DM_{W,Z}D)\\
    & = G+\sum_{g=1}^{G}\frac{1}{n_{g}^2}\left[\frac{m_{0,g}^2}{m_{1,g}-1} + \frac{m_{1,g}^2}{m_{0,g}-1}\right]
    \leq 3G,
    \end{split}
\end{equation}
where the last line uses that $m_{1,g}\geq 2$ and $m_{0,g}\geq 2$. 

Second, let $\hat{\beta}-\beta = S_{n}/D_{n}$ with
\begin{equation}\label{eq:Sn}
    \begin{split}
    S_{n}=r_{n}^{-1/2}T'A(Y-T\beta),\quad
    D_{n}=r_{n}^{-1/2}T'AT.
    \end{split}
\end{equation} 
We refer to $S_{n}$ as the score. 
Let $\zeta = \theta-\beta\pi$ and $v_{i} = \varepsilon_{i}-u_{i}\beta$. Denote $\sigma_{u,i}^2=\mathbb{E}[u_{i}^2|Q,X]$, $\sigma_{v,i}^2=\mathbb{E}[v_{i}^2|Q,X]$, and $\sigma_{uv,i}=\mathbb{E}[u_{i}v_{i}|Q,X]$. We have $\mathbb{E}[S_{n}|Q,X]=0$ and
\begin{equation}\label{eq:boundvar}
\begin{split}
   \mathbb{V}(S_{n}| Q, X) &= r_{n}^{-1}\sum_{i=1}^{n}\sigma_{u,i}^2[M_{W}Z\zeta]_{i}^2 + \sigma_{v,i}^2[M_{W}Z\pi]_{i}^2 + 2\sigma_{uv,i}[M_{W}Z\pi]_{i}[M_{W}Z\zeta]_{i}\\
   &\qquad +r_{n}^{-1}\sum_{i=1}^{n}\sum_{j\neq i}A_{ij}^2(\sigma_{uv,i}\sigma_{uv,j} + \sigma_{v,i}^2\sigma_{u,j}^2)\\
   &\leq c[r_{n}^{-1}\zeta'Z'M_{W}Z\zeta+r_{n}^{-1}\pi'Z'M_{W}Z\pi) + r_{n}^{-1}\text{tr}(A^2)]\leq c,\quad a.s.,
   \end{split}
\end{equation}
where the first inequality follows from Assumption~\ref{ass:large_sample} and the second inequality from Assumption~\ref{ass:IV}, the definition of $r_{n}$, and the bound in \eqref{eq:trA2upper}. For $D_{n}$, we have $
    \mathbb{E}[D_{n}|Q,X]= r_{n}^{-1/2} \pi' Z' M_{W}Z\pi \equiv r_{n}^{1/2}H_{n}$ and
\begin{equation}\label{eq:boundhess}
\begin{split}
    \mathbb{V}(D_{n}| Q, X) &\leq 4 r_{n}^{-1}\mathbb{E}[( u' M_{W} Z\pi)^2|Q,X] + r_{n}^{-1}\mathbb{E}[( u' A u)^2|Q,X]\\
    & \leq 4\max_{i}\sigma_{u,i}^2r_{n}^{-1}\pi' Z' M_{W} Z\pi + 2r_{n}^{-1}\max_{i}\sigma_{u,i}^4\text{tr}( A^2)\leq c,\quad a.s.
    \end{split}
\end{equation}
We can now write
\begin{equation}
\begin{split}
    \hat{\beta}-\beta& =\frac{r_{n}^{-1/2}T'A(Y-T\beta)/(r_{n}^{1/2}H_{n})}{1 + r_{n}^{1/2}(r_{n}^{-1}T'AT-H_{n})/(r_{n}^{1/2}H_{n})} \equiv \frac{A_{n}}{B_{n}}.
    \end{split}
\end{equation}
Assume that $r_{n}^{-1/2}H_{n}^{-1}\rightarrow_{p}0$. From \eqref{eq:boundvar} it follows that $A_{n}\rightarrow_{p}0$. From \eqref{eq:boundhess} we have that $r_{n}^{1/2}(r_{n}^{-1}T'AT-H_{n})/(r_{n}^{1/2}H_{n}) \rightarrow_{p}0$, and hence $B_{n}\rightarrow_{p}1$. We conclude that  when $r_{n}^{-1/2}H_{n}^{-1}=\big(r_{n}^{-1/2}\pi'Z'M_{W}Z\pi\big)^{-1}\rightarrow_{p}0$, the SIVE estimator is consistent: $\hat{\beta}-\beta\rightarrow_{p}0$. Finally, under weak identification, $r_{n}^{1/2}H_{n} = \sqrt{G}\frac{\pi'Z'M_{W}Z\pi}{G} = \sqrt{G}\frac{n\text{FS}}{G}$ so that requiring that $r_{n}^{-1/2}H_{n}^{-1}\rightarrow_{p}0$ is the same as $\frac{n\text{FS}}{\sqrt{G}}\rightarrow_{p}\infty$. 

\section{Proof \texorpdfstring{Theorem~\ref{corr:dist2}}{Lemma \ref{thm:dist}}}\label{app:dist}

\subsection{Asymptotic normality}\label{proof:asynorm}
First, we establish asymptotic normality of the score $S_{n}$ defined in \eqref{eq:Sn} after normalizing it with the square root of its conditional variance. This step relies on a central limit theorem for quadratic forms of growing rank. Consider the self-normalized version of $S_{n}$,
\begin{equation}  S_{n}^{*}=\frac{r_{n}^{-1/2} T' A( Y- T\beta)}{\mathbb{V}(S_n|Q,X)^{1/2}}.
\end{equation}
To show that $S_{n}^{*}\rightarrow_{d} N(0,1)$, we use Lemma D.2 in \citet{evdokimov2018inference}, which in turn is based on Lemma A.2 in \citet{chao2012asymptotic}.
\begin{lemma}\label{lemCGT} \emph{[\citet{evdokimov2018inference}, Lemma D.2]}  Define $R =  t' u +  s' v +  u' \bar{A} v$, where $[ \bar{A}]_{ii}=0$.
    Suppose that almost surely: (1) $\mathbb{E}[u_{i}| Q, X]=\mathbb{E}[v_{i}| Q, X]=0$, $\mathbb{E}[u_{i}^4| Q, X]\leq c<\infty$, and $\mathbb{E}[v_{i}^4| Q,X]\leq c<\infty$; (2) $\mathbb{V}(R| Q, X)^{-1}\leq c<\infty$; (3) $\sum_{i=1}^{n}(t_{i}^4 + s_{i}^4)\rightarrow 0$; (4) $\emph{tr}( \bar{A}^4)\rightarrow 0$.
    Then, $
        \mathbb{V}(R| Q, X)^{-1/2} R\rightarrow_{d} N(0,1).$
\end{lemma}
Let $R = S_{n}$, so that $ t = r_{n}^{-1/2} M_{W}Z\zeta$ and $ s = r_{n}^{-1/2} M_{W}Z\pi$ and $ \bar{A} =r_{n}^{-1/2}A$. Condition 1 holds by the definition of $u_i$ and $\varepsilon_i$ in Assumption~\ref{ass:indep}. For Condition 3, we consider
\begin{equation}\label{eq:boundeiAr}
\begin{split}
    |[ M_{W} Z\pi]_{i}|
    & \leq \max_{g} \pi(x_g)\left\{\begin{array}{ll}
    \frac{m_{0,g}}{n_{g}} & \text{if } Z_{ig}=1, W_{ig}=1,\\
    \frac{m_{1,g}}{n_{g}} & \text{if } Z_{ig}=0, W_{ig}=1,\\
    0&\text{otherwise}.
    \end{array}\right. \leq c<\infty, \quad a.s.
\end{split}
\end{equation}
The same result follows for $ M_{W}Z\zeta$ by noting that $\theta(x_g)\leq c<\infty$ $a.s.$ by Assumption~\ref{ass:IV}. From \eqref{eq:boundeiAr} now follows that $| e_{i}' M_{W}Z\pi|\leq c<\infty$ $a.s.$ uniformly over $i$. Similarly, $| e_{i}'M_{W}Z\zeta|\leq c<\infty$ $a.s.$ uniformly over $i$. We then obtain
\begin{equation}
    \sum_{i=1}^{n}(t_{i}^4+s_{i}^4)\leq cr_{n}^{-2}\sum_{i=1}^{n}(t_{i}^2+s_{i}^2) =cr_{n}^{-1}(\pi'Z'M_{W}Z\pi/r_{n} + \zeta'Z'M_{W}Z\zeta/r_{n}).
\end{equation}
The term in parentheses is almost surely bounded as shown in Appendix~\ref{subsec:consistency}. Since $r_{n}^{-1}\rightarrow_{a.s.}0$, Condition 3 of Lemma~\ref{lemCGT} holds. 
To verify Condition 2 and 4 we separately consider strong and semi-strong/weak identification.

\textit{Strong identification}$\quad$ We start by showing that $u'\bar{A}v=o_{p}(1)$. Note that 
\begin{equation}
    \begin{split}
        \mathbb{V}(G^{-1/2} u' A v|Q,X)& = G^{-1}\sum_{j\neq i}\sum_{k\neq m}A_{ij}A_{mk}\mathbb{E}[u_{i}u_{k}v_{j}v_{m}|Q,X]\\
        & = G^{-1}\sum_{j\neq i}A_{ij}^2(\mathbb{E}[u_{i}v_{i}u_{j}v_{j}|Q,X]+\mathbb{E}[u_{i}^2v_{j}^2|Q,X])\\
        &\leq cG^{-1}\text{tr}( A^2)\leq c\quad a.s.,
    \end{split}
\end{equation}
using that $u_{i}$ and $v_{i}$ have bounded fourth conditional moments by Assumption~\ref{ass:large_sample} and the bound established in \eqref{eq:trA2upper}. Then, by the definition of $r_{n}$, $\mathbb{V}( u' A v|Q,X)/r_{n}\leq c G/r_{n}\rightarrow_{a.s.}0$.
We can then conclude that $r_{n}^{-1/2} u' A v = o_{p}(1)$. It now suffices to consider $\tilde{R}= t' u +  s' v$.

Consider Condition 2. As in \eqref{eq:boundvar}, 
\begin{equation}
\begin{split}
   \mathbb{V}(\tilde{R}|Q,X) &= r_{n}^{-1}\sum_{i=1}^{n}\sigma_{u,i}^2[M_{W}Z\zeta]_{i}^2 + \sigma_{v,i}^2[M_{W}Z\pi]_{i}^2 + 2\sigma_{uv,i}[M_{W}Z\pi]_{i}[M_{W}Z\zeta]_{i}.
   \end{split}
\end{equation}
Define $\Sigma_{i} = \left(\begin{array}{cc}\sigma_{u,i}^2 &\sigma_{u\varepsilon,i}\\ \sigma_{u\varepsilon,i}&\sigma_{\varepsilon_{i}}^2\end{array}\right)$ and $
h_{i}=\sum_{g}Z_{ig}\frac{m_{0,g}}{n_g}
      - (W_{ig}-Z_{ig})  \frac{m_{1,g}}{n_{g}}$. By Assumption~\ref{ass:large_sample}, using \eqref{eq:boundeiAr},
\begin{align}
       \mathbb{V}(\tilde{R}|Q,X) &=r_{n}^{-1}\sum_{i=1}^{n}h_{i}^2\sum_{g=1}^{G}W_{ig}\bigg(\zeta(x_{g})^2\sigma_{u,i}^2 + \pi(x_{g})^2\sigma_{v,i}^2 + 2\pi(x_{g})\zeta(x_{g})\sigma_{uv,i}\bigg)\nonumber\\&=r_{n}^{-1}\sum_{i=1}^{n}h_{i}^2\sum_{g=1}^{G}W_{ig}[\theta(x_{g}),\pi(x_{g})]\left(\begin{array}{cc}1 & 0 \\ -\beta & 1\end{array}\right)\Sigma_{i}\left(\begin{array}{cc}1 & -\beta \\ 0 & 1\end{array}\right) [\theta(x_{g}),\pi(x_{g})]'\nonumber\\
        &\geq cr_{n}^{-1}\sum_{i=1}^{n}h_{i}^2\sum_{g=1}^{G}W_{ig}\left(\theta(x_{g})^2 + \pi(x_{g})^2\right).
    \end{align}
Using the definition or $r_n$, we see that $r_{n}^{-1}\sum_{i=1}^{n}h_{i}^2\sum_{g=1}^{G}\pi(x_{g})^2 
=1$. It follows that $\mathbb{V}(\tilde{R}|Q,X)\geq c>0$ $a.s.$ and Condition 2 of Lemma~\ref{lemCGT} holds.

\textit{Semi-strong/weak identification and $G\rightarrow\infty$}$\quad$
We start with verifying Condition 2 in Lemma~\ref{lemCGT}. By Assumption~\ref{ass:large_sample}, $\min_{i}\sigma_{v,i}^2\geq c>0$ and $\min_{i}\sigma_{u,i}^2\geq c>0$. The first line of \eqref{eq:trA2upper} shows that $\text{tr}(A^2)\geq G$. It now follows that $
    \mathbb{V}(S_{n}|Q,X)\geq G^{-1}\text{tr}( D_{\sigma_{v}^2} A D_{\sigma_{u}^2} A)
    \geq cG^{-1}\text{tr}( A^2) 
    \geq c>0\quad a.s.$ 
For Condition 4, note that $
G^{-1}\text{tr}( A^2) \leq 3$ $a.s.$ by \eqref{eq:trA2upper} and $\lambda_{\max}(A^2)= 1$.
Then , $\text{tr}(\bar{A}_{n}^4)=r_{n}^{-2}\text{tr}( A^4) \leq \frac{G^2}{r_{n}^2}\frac{\text{tr}(A^2)}{G}\frac{\lambda_{\max}(A^2)}{G}\rightarrow_{a.s.}0.$

\subsection{Consistency of the variance estimator}\label{app:consistentvar}
We now analyze the variance estimator for the score $S_{n}$, given by $(T'AT)^2\hat{\mathbb{V}}[\hat{\beta}|Q,X]$ with $\hat{\mathbb{V}}[\hat{\beta}|Q,X]$ from \eqref{eq:Vhat}. Section~\ref{subsec:unbvarest} provides initial results on the \citet{hartley1969variance} variance estimators. Section~\ref{app:varscore} shows that $\hat{\mathbb{V}}_{1}$ defined in \eqref{eq:V1hat} converges to a quantity at least as large as $\mathbb{V}(S_{n}|Q,X)$. Section~\ref{A:bias} provides the precise expression for $\hat{\mathbb{V}}_{2}$ in \eqref{eq:Vhat} and shows under what conditions this term removes the upward bias. Throughout, we assume that $m_{1,g}\geq 3$ and $m_{0,g}\geq 3$ so that we can use \eqref{eq:unbvar} for the regression error variances and covariances. We relax this assumption in Supplemental Appendix~\ref{sec:smallgroups}. 

\subsubsection{HRK (co)variance estimators}\label{subsec:unbvarest}
 
From the first stage and reduced form equations \eqref{eq:2slseq}, we see that $ M_{W,Z}T =  M_{W,Z} u$ and $ M_{W,Z}( Y- T\beta) =  M_{W,Z} v$. We now have the following estimators,
\begin{equation}\label{eq:sigmabetahattobeta}
\begin{split}
\hat{\sigma}_{u,i}^2 &=  e_{i}'( M_{W,Z}\odot  M_{W,Z})^{-1}( M_{W,Z} T\odot  M_{W,Z} T)\\
     &=  e_{i}'( M_{W,Z}\odot  M_{W,Z})^{-1}( M_{W,Z} u\odot  M_{W,Z} u)\\
\hat{\sigma}_{uv,i}&=  e_{i}'( M_{W,Z}\odot  M_{W,Z})^{-1}( M_{W,Z}( Y-     T\hat{\beta})\odot  M_{W,Z} T),\\
    & = \hat{\sigma}_{uv,i}(\beta) -(\hat{\beta}-\beta) \hat{\sigma}_{u,i}^2,\\   
\hat{\sigma}_{v,i}^2& =  e_{i}'( M_{W,Z}\odot  M_{W,Z})^{-1}( M_{W,Z}( Y- T\hat{\beta})\odot  M_{W,Z}( Y- T\hat{\beta}))\\
    & = \hat{\sigma}_{v,i}^2(\beta) -2(\hat{\beta}-\beta)\hat{\sigma}_{uv,i}(\beta) + (\hat{\beta}-\beta)^2\hat{\sigma}_{u,i}^2,
    \end{split}
\end{equation}
where we write $\hat{\sigma}_{v,i}^2(\beta)$ and $\hat{\sigma}_{uv,i}({\beta})$ as the infeasible analogues of $\hat{\sigma}_{v,i}^2$ and $\hat{\sigma}_{uv,i}$. Since $\hat{\beta}-\beta=o_{p}(1)$, the contribution of the corresponding terms in $\hat{\sigma}_{uv,i}$ and $\hat{\sigma}_{v,i}^2$ to the variance estimator for the score is $o_{p}(1)$ as well.
Using \eqref{eq:invMtilMtil} and the definition of $L$ in \eqref{eq:L} we can write
\begin{equation}\label{eq:unbvarest_decomp}
\begin{split}
    \hat{\sigma}_{u,i}^2 &= u_{i}^2-u_{i}\frac{2}{1-2L_{ii}}\sum_{j\neq i}L_{ij}u_{j}  + \frac{1}{(1-L_{ii})(1-2L_{ii})}\sum_{j=1}^{n}\sum_{k\neq j}L_{ij}L_{ik}u_{j}u_{k},\\
    \hat{\sigma}_{v,i}^2(\beta)
    &= v_{i}^2-v_{i}\frac{2}{1-2L_{ii}}\sum_{j\neq i}L_{ij}v_{j}  + \frac{1}{(1-L_{ii})(1-2L_{ii})}\sum_{j=1}^{n}\sum_{k\neq j}L_{ij}L_{ik}v_{j}v_{k},\\
    \hat{\sigma}_{uv,i}(\beta) &= u_{i}v_{i}-v_{i}\frac{1}{1-2L_{ii}}\sum_{j\neq i}L_{ij}u_{j}-u_{i}\frac{1}{1-2L_{ii}}\sum_{j\neq i}L_{ij}v_{j} \\
    &\qquad + \frac{1}{(1-L_{ii})(1-2L_{ii})}\sum_{j=1}^{n}\sum_{k\neq j}L_{ij}L_{ik}u_{j}v_{k}.
    \end{split}
\end{equation}
Taking the conditional expectation and using independence across $i$, we see that the estimators $\hat{\sigma}_{u,i}^2$, $\hat{\sigma}_{uv,i}(\beta)$ and $\hat{\sigma}_{v,i}^2(\beta)$ are (conditionally) unbiased for $\sigma_{u,i}^2$, $\sigma_{uv,i}$ and $\sigma_{v,i}^2$, respectively. The estimators only exist if $m_{1,g}>2$ and $m_{0,g}>2$ for all $g=1,\ldots,G$. We discuss how we handle groups with $m_{1,g}=2$ or $m_{0,g}=2$ in Supplemental Appendix~\ref{sec:smallgroups}.

\subsubsection{The variance of the score and its estimator}\label{app:varscore}
We can rewrite the variance of the score given in \eqref{eq:boundvar} as
\begin{align}
   \mathbb{V}(S_{n}|Q,X)&= r_{n}^{-1}\bigg[ \zeta'Z'M_{W}  D_{\sigma_{u}^2} M_{W}Z\zeta +  \pi'Z'M_{W} D_{\sigma_{v}^2}M_{W}Z\pi+ 2  \zeta'Z'M_{W} D_{\sigma_{uv}} M_{W}Z\pi \nonumber\\
    &\quad\quad + \underbrace{\iota' D_{\sigma_{uv}}( A\odot A) D_{\sigma_{uv}}\iota + \iota' D_{\sigma_{v}^2}( A\odot A) D_{\sigma_{u}^2}\iota}_{({B}.0)}\bigg],\label{eq:B0}
    \end{align}
with $D_{\sigma_{u}^2}$, $D_{\sigma_{v}^2}$ and $D_{\sigma_{uv}}$ defined in Appendix~\ref{A:defsandnotation}.
\eqref{eq:V1hat} estimates $\mathbb{V}(S_{n}|Q,X)$ by
\begin{equation}\label{eq:estimatorvariance}
\begin{split}
    \hat{\mathbb{V}}(S_{n}|Q,X) &= r_{n}^{-1}\left[( Y- T\hat{\beta})' A  D_{\hat{\sigma}_{u}^2} A( Y- T\hat{\beta}) +  T' A D_{\hat{\sigma}_{v}^2} A T + 2( Y- T\hat{\beta})' A D_{\hat{\sigma}_{uv}} A T\right]. 
    \end{split}
    \end{equation}
Consider the infeasible variance estimator,
\begin{align}
   \mathbb{V}_{\inf}(S_n|Q,X)&= r_{n}^{-1}\bigg[ \zeta'Z'M_{W}  D_{u^2} M_{W}Z\zeta +  \pi'Z'M_{W} D_{{v}^2}M_{W}Z\pi+ 2  \zeta'Z'M_{W} D_{{uv}} M_{W}Z\pi\nonumber \\
    &\quad\quad + \iota' D_{{uv}}( A\odot A) D_{{uv}}\iota + \iota' D_{{v}^2}( A\odot A) D_{{u}^2}\iota\bigg].\label{eq:estimatorvarianceinf}
    \end{align}
The variance estimator in \eqref{eq:estimatorvariance} can be decomposed as
\begin{equation}\label{eq:estvariancedecomposed}
\begin{split}
        & \hat{\mathbb{V}}(S_{n}|Q,X)   
        \\&=\mathbb{V}(S_n|Q,X) + \mathbb{V}_{\inf}(S_n|Q,X)-\mathbb{V}(S_n|Q,X) \\
        &\quad + \underbrace{r_{n}^{-1}\iota'D_{v^2}(A\odot A)D_{u^2}\iota + r_{n}^{-1}\iota'D_{uv}(A\odot A)D_{uv}\iota}_{(B.1)} + \underbrace{ r_{n}^{-1}(v' A D_{\hat{\sigma}_{u}^2} A v-\iota'D_{v^2}(A\odot A)D_{u^2}\iota)}_{(B.2)} \\
        &\quad +\underbrace{r_{n}^{-1} (u' A D_{\hat{\sigma}_{v}^2(\beta)} A u -\iota'D_{u^2}(A\odot A)D_{v^2}\iota)}_{(B.3)} +  \underbrace{2r_{n}^{-1}(v' A D_{\hat{\sigma}_{uv}(\beta)}A u-\iota'D_{uv}(A\odot A)D_{uv}\iota)}_{(B.4)} \\
        &\quad +\underbrace{r_{n}^{-1} \zeta'Z'M_W (D_{\hat{\sigma}_{u}^2}-D_{u^2}) M_W Z\zeta}_{(Z.1)} +2\underbrace{r_{n}^{-1} v' A D_{\hat{\sigma}_{u}^2}M_W Z\zeta}_{(Z.2)} \\
        &\quad + r_{n}^{-1} \pi'Z'M_W (D_{\hat{\sigma}_{v}^2(\beta)}-D_{v^2}) M_W Z\pi+ 2 r_{n}^{-1}\zeta'Z'M_W (D_{\hat{\sigma}_{uv}(\beta)} -D_{uv})M_W  Z\pi \\
        &\quad + 2 r_{n}^{-1}u' A D_{\hat{\sigma}_{v}^2(\beta)} M_W Z\pi + 2r_{n}^{-1}\zeta'Z'M_W D_{\hat{\sigma}_{uv}(\beta)} A u + 2r_{n}^{-1}\pi'Z'M_W D_{\hat{\sigma}_{uv}(\beta)} A v +R_{1},
        \end{split}
    \end{equation}
    where we use \eqref{eq:sigmabetahattobeta} and with the remainder term
    \begin{align}
         R_{1}&=4r_{n}^{-1}\bigg[(\hat{\beta}-\beta)^2(Z\pi+u)'AD_{\hat{\sigma}_{u}^2}A(Z\pi+u) -(\hat{\beta}-\beta)(Z\pi+u)'AD_{\hat{\sigma}_{uv}(\beta)}A(Z\pi+u)\nonumber\\
        &\qquad\quad -(\hat{\beta}-\beta)(Z\pi+u)'AD_{\hat{\sigma}_{u}^2}A(Z\zeta+v)\bigg].
        \end{align}
For identification-robust inference, we replace $\hat{\beta}=\beta$ so that under the null $R_{1}=0$, and Remark~\ref{rem:robust} follows from the proof presented here for Theorem~\ref{corr:dist2}. 
    
We now prove the following. First, $\mathbb{V}_{\text{inf}}(S_n|Q,X)-\mathbb{V}(S_n|Q,X)\rightarrow_{p}0$. Second, we show that under weak identification $(B.1)$, $(B.2)$ and $(B.3)$ from \eqref{eq:estvariancedecomposed} converge to a positive bias term. Finally, we show that $(Z.1)$ and $(Z.2)$ in \eqref{eq:estvariancedecomposed} converge to zero in probability. The remaining terms in \eqref{eq:estvariancedecomposed} converge to zero by the same arguments.

\textbf{Part I: $\mathbb{V}_{\inf}(S_n|Q,X)-\mathbb{V}(S_n|Q,X)\rightarrow_{p}0.$} It is sufficient to show that
\begin{align}
    r_{n}^{-1} \zeta'Z'M_W( D_{u^2}- D_{\sigma_{u}^2}) M_W Z\zeta&\rightarrow_{p} 0,\label{eq:part1term1}\\
    r_{n}^{-1} (\iota'D_{v^2} (A\odot A)D_{u^2}\iota-\iota'D_{\sigma_v}^2(A\odot A)D_{\sigma_u}^2\iota)&\rightarrow_{p} 0.\label{eq:part1term2}
    \end{align}
    The other terms follow by the same arguments.
For \eqref{eq:part1term1},  the conditional expectation $\mathbb{E}[r_{n}^{-1}\zeta'Z'M_W( D_{u^2}- D_{\sigma_{u}^2})M_WZ\zeta|Q,X]=0$, and the variance satisfies
\begin{align}
    \mathbb{V}(r_{n}^{-1} \zeta'Z'M_{W}( D_{u^2}- D_{\sigma_{u}^2}) M_{W}Z\zeta|Q,X) &= r_{n}^{-2}\sum_{i=1}^{n}\mathbb{E}[(u_{i}^2-\sigma_{u,i}^2)^2|Q,X][M_WZ\zeta]_{i}^4\nonumber\\
    & \leq cr_{n}^{-2}\sum_{i=1}^{n}[M_WZ\zeta]_{i}^2 \rightarrow_{a.s.}0,
    \end{align}
where the inequality uses Assumption~\ref{ass:large_sample} and the arguments following \eqref{eq:boundeiAr}.
For \eqref{eq:part1term2}, $\mathbb{E}[ r_{n}^{-1} (\iota'D_{v^2} (A\odot A)D_{u^2}\iota-\iota'D_{\sigma_v}^2(A\odot A)D_{\sigma_u}^2\iota)|Q,X]=0$. The variance satisfies
\begin{align}
      &  \mathbb{V}( r_{n}^{-1} (\iota'D_{v^2} (A\odot A)D_{u^2}\iota-\iota'D_{\sigma_v}^2(A\odot A)D_{\sigma_u}^2\iota)|Q,X)\nonumber\\
      &=r_{n}^{-2}\sum_{i=1}^{n}\sum_{j\neq i}\sum_{k=1}^{n}\sum_{l\neq k}\mathbb{E}[(v_{i}^2u_{j}^2-\sigma_{v,i}^2\sigma_{u,i}^2)(v_{k}^2 u_{l}^2-\sigma_{v,k}^2\sigma_{u,l}^2)|Q,X]A_{ij}^2A_{kl}^2\nonumber\\
      &=r_{n}^{-2}\sum_{i=1}^{n}\sum_{j\neq i}\mathbb{E}[(v_{i}^2u_{j}^2-\sigma_{v,i}^2\sigma_{u,i}^2)^2+(v_{i}^2u_{j}^2-\sigma_{v,i}^2\sigma_{u,i}^2)(v_{j}^2 u_{i}^2-\sigma_{v,j}^2\sigma_{u,i}^2)|Q,X]A_{ij}^4\nonumber\\
      &\leq cr_{n}^{-2}\sum_{i=1}^{n}\sum_{j\neq i}A_{ij}^4\leq cr_{n}^{-2}\text{tr}(A^2)\rightarrow_{a.s.}0,
    \end{align}
where we use Assumption~\ref{ass:large_sample} and \eqref{eq:trA2upper}.

\textbf{Part II: Show that under weak identification $(B.1)$, $(B.2)$ and $(B.3)$ from \eqref{eq:estvariancedecomposed} converge to a positive bias term.} It follows from the results in Part I that $(B.1)-(B.0)\rightarrow_{p}0$. For $(B.2)$, we decompose it further as $(B.2) = (B.2a) + (B.2b)$ with
\begin{equation}
\begin{split}
(B.2a) &= r_{n}^{-1}(v'AD_{\hat{\sigma}_{u}^2}Av-v'AD_{{u}^2}Av),\\
(B.2b) &= r_{n}^{-1}(v'AD_{{u}^2}Av-\iota D_{v^2}(A\odot A)D_{u^2}\iota).
\end{split}
\end{equation}
For $(B.2b)$, $\mathbb{E}[(B.2b)|Q,X] =r_{n}^{-1}\sum_{i=1}^{n}\sum_{k\neq i}\sum_{j\neq \{i,k\}}\mathbb{E}[v_{i}v_{k}u_{j}^2|Q,X]A_{ij}A_{kj}= 0.$
The variance can be upper bounded as
\begin{align}
&\mathbb{V}((B.2b)| Q,  X) \\
&= r_{n}^{-2}\mathbb{E}\big[\big(\sum_{i=1}^{n}\sum_{k\neq i}\sum_{j\neq \{i,k\}}v_{i}v_{k}u_{j}^2A_{ij}A_{kj}\big)^2\big|Q,X\big]\nonumber\\
&=r_{n}^{-2}\sum_{i_{1},i_{2}}\sum_{i_3\neq i_1}\sum_{i_4\neq i_2}\sum_{i_{5}\neq \{i_{3},i_{1}\}}\sum_{i_6\neq\{i_{2},i_{4}\}}\big(\mathbb{E}[v_{i_{1}}v_{i_{2}}v_{i_{3}}v_{i_{4}}|Q,X]\sigma_{u,i_5}^2\sigma_{u,i_6}^2A_{i_{1}i_{3}}A_{i_{1}i_{5}}A_{i_{2}i_{4}}A_{i_{2}i_{6}}\nonumber\\
&\quad +\mathbb{E}[v_{i_{1}}v_{i_{2}}v_{i_{3}}v_{i_{4}}(u_{i_5}^2-\sigma_{u,i_5}^2)(u_{i_6}^2-\sigma_{u,i_6}^2)|Q,X]A_{i_{1}i_{3}}A_{i_{1}i_{5}}A_{i_{2}i_{4}}A_{i_{2}i_{6}}\big)\nonumber\\
&\leq cr_{n}^{-2}\big[\sum_{i_{1}..i_{3}}( B_{i_{1}i_{2}}^2 B_{i_{1}i_{3}}^2 +  B_{i_{1}i_{2}}^2 B_{i_{1}i_{3}} B_{i_{2}i_{3}})  + \sum_{i_{1}..i_{4}}( B_{i_{1}i_{2}}^2 B_{i_{1}i_{3}} B_{i_{1}i_{4}}+ B_{i_{1}i_{2}}^2 B_{i_{1}i_{3}} B_{i_{2}i_{4}})\big]\nonumber\\
&=cr_{n}^{-2}\big[\iota'( B\odot B)^2\iota + \text{tr}( B( B\odot B) B) +\sum_{i_{1}}\iota'( B\odot B)e_{i_{1}}(e_{i_{1}} B\iota)^2 + \iota' B( B\odot B) B\iota\big]\nonumber\\
&\leq cr_{n}^{-2}G\rightarrow_{a.s.} 0,\label{var:b2b}
    \end{align}
with $B$ defined in \eqref{eq:tildeAg} and the last inequality follows by results $(R9)-(R11)$ from Supplemental Appendix~\ref{A:Aresults}.

For $(B.2a)$, using the results from Section~\ref{subsec:unbvarest}, $(B.2a) = (B.2a.1) + (B.2a.2)$ with
\begin{equation}
\begin{split}
   (B.2a.1)
   &=-r_{n}^{-1}\sum_{i=1}^{n}\sum_{j\neq i}\frac{2}{1-2L_{ii}}( e_{i}' A v)^2 u_{i}L_{ij}u_{j}, \\
  (B.2a.2) &=r_{n}^{-1}\sum_{i=1}^{n}\sum_{j=1}^{n}\sum_{k\neq j}\frac{1}{(1-L_{ii})(1-2L_{ii})}( e_{i}' A v)^2L_{ij}L_{ik}u_{j}u_{k}.
\end{split}
\end{equation}
Since $[ A]_{ii}=0$, we have that $\mathbb{E}[(B.2a.1)|Q,X] = 0$. For $(B.2a.2)$, we have
\begin{equation}
    \begin{split}
       \mathbb{E}[(B.2a.2)|Q,X] = r_{n}^{-1} \sum_{i=1}^{n}\frac{1}{(1-L_{ii})(1-2L_{ii})}\sum_{j=1}^{n}\sum_{k\neq j}L_{ij}L_{ik}A_{ij}A_{ik}\sigma_{uv,j}\sigma_{uv,k}.
    \end{split}
\end{equation}
With the matrix $C$ defined in \eqref{eq:C}, the bias arising from $(B.2a.2)$ is
     $\mathbb{E}[(B.2a.2)|Q,X] =  r_{n}^{-1}\iota'D_{\sigma_{uv}} CD_{\sigma_{uv}}\iota.$ Similarly, $(B.3)$ and $(B.4)$will yield a bias totaling up to
\begin{equation}
\begin{split}
r_{n}^{-1}\mathbb{E}[ (B.2)+(B.3)+(B.4)|Q,X] &= 2r_{n}^{-1}\iota'D_{\sigma_{uv}} CD_{\sigma_{uv}}\iota +2r_{n}^{-1}\iota'D_{\sigma_{u}^2} C D_{\sigma_{v}^2}\iota\\
     &=2r_{n}^{-1}\sum_{i=1}^{n}\sum_{j>i}C_{ij}\mathbb{E}[(u_{i}v_{j}+u_{j}v_{i})^2|Q,X],
 \end{split}
\end{equation}
which is nonnegative as $C_{ij}\geq 0$ for $i\neq j$. The bias is bounded: $
       r_{n}^{-1}\sum_{i=1}^{n}\sum_{j>i}C_{ij}\mathbb{E}[(u_{i}v_{j}+u_{j}v_{i})^2|Q,X]\leq cr_{n}^{-1}\sum_{i=1}^{n}\sum_{j>i}C_{ij}
         \leq c r_{n}^{-1}G
        \leq c.$
Aggregating the expectations $(B.1)-(B.4)$, the upward bias in the variance estimator for the score equals
\begin{equation}\label{eq:vbias_2}
\begin{split}
    \Delta_{b}&=r_{n}^{-1}\big(\iota'D_{\sigma_{uv}}[(A\odot A)+2C]D_{\sigma_{uv}}\iota + \iota'D_{\sigma_{v}^2}[(A\odot A)+2C]D_{\sigma_{u}^2}\iota\big)\\
    & \leq 3r_{n}^{-1}\big(\iota'D_{\sigma_{uv}}(A\odot A)D_{\sigma_{uv}}\iota + \iota'D_{\sigma_{v}^2}(A\odot A)D_{\sigma_{u}^2}\iota\big).
    \end{split}
\end{equation}
This indicates that under semi-strong or weak identification, we may overestimate the variance term $(B.0)$ by a factor 4.

It remains to be shown that the terms $(B.2)$ and $(B.3)$ converge to their expectation. We show this for $(B.2)$. The result for $(B.3)$ follows analogously. The result for $(B.2b)$ is already established in \eqref{var:b2b}. For the variance of $(B.2a.1)$ and $(B.2a.2)$, we have
\begin{align}
    \mathbb{V}(B.2a.1|Q,X) &=4r_{n}^{-2}\sum_{i_{1},i_3}\sum_{i_{2}\neq i_{1}}\sum_{i_{4}\neq i_{3}}\frac{1}{(1-2L_{i_{1}i_{1}})(1-2L_{i_{3}i_{3}})}\nonumber\\
    &\qquad \times \mathbb{E}[( e_{i_{1}}' A v)^2( e_{i_{3}} A v)^2L_{i_{1}i_{2}}L_{i_{3}i_{4}}u_{i_{1}}u_{i_{2}}u_{i_{3}}u_{i_{4}}|Q,X]\nonumber\\
   &\leq cr_{n}^{-2}\sum_{i_{1},i_{3}}\sum_{i_{2}\neq i_{1}}\sum_{i_{4}\neq i_{3}}\sum_{i_{5}\neq i_{1}}\sum_{i_{6}\neq i_{1}}\sum_{i_{7}\neq i_{3}}\sum_{i_{8}\neq i_{3}}|A_{i_{1}i_{5}}||A_{i_{1}i_{6}}||A_{i_{3}i_{7}}||A_{i_{3}i_{8}}|\nonumber\\
   &\qquad\times L_{i_{1}i_{2}}L_{i_{3}i_{4}}|\mathbb{E}[u_{i_{1}}u_{i_{2}}u_{i_{3}}u_{i_{4}}v_{i_{5}}v_{i_{6}}v_{i_{7}}v_{i_{8}}|Q,X]|,\label{eq:IIa1}
   \end{align}
   \begin{align}
    \mathbb{V}(B.2a.2|Q,X) &=r_{n}^{-2}\sum_{i_{1},i_2,i_6,i_7}\sum_{i_{3}\neq i_{2}}\sum_{i_{4}\neq i_{1}}\sum_{i_{5}\neq i_{1}}\sum_{i_{8}\neq i_{7}}\sum_{i_{9}\neq i_{6}}\sum_{i_{10}\neq i_{6}}c_{i_{1}}c_{i_{6}}A_{i_{1}i_{4}}A_{i_{6}i_{9}}A_{i_{1}i_{5}}A_{i_{6}i_{10}}\nonumber\\
    &\qquad\times L_{i_{1}i_{2}}L_{i_{1}i_{3}}L_{i_{6}i_{7}}L_{i_{6}i_{8}}\mathbb{E}[u_{i_{2}}u_{i_{3}}v_{i_{4}}v_{i_{5}}u_{i_{7}}u_{i_{8}}v_{i_{9}}v_{i_{10}}|Q,X]\nonumber\\
    &-4r_{n}^{-2}\sum_{i_{1},i_4}\sum_{i_{2}\neq i_{1}}\sum_{i_{3}\neq \{i_{2},i_{1}\}}\sum_{i_{5}\neq i_{4}}\sum_{i_{6}\neq \{i_{5},i_{4}\}}c_{i_{1}}c_{i_{4}}A_{i_{1}i_{2}}A_{i_{4}i_{5}}A_{i_{1}i_{3}}A_{i_{4}i_{6}}\nonumber\\
    &\qquad\times L_{i_{1}i_{2}}L_{i_{4}i_{5}}L_{i_{1}i_{3}}L_{i_{4}i_{6}}\mathbb{E}[v_{i_{2}}v_{i_{3}}u_{i_{2}}u_{i_{3}}v_{i_{5}}v_{i_{6}}u_{i_{5}}u_{i_{6}}|Q,X],\label{eq:IIa2}
\end{align}
where $c_{i} = (1-L_{ii})^{-1}(1-2L_{ii})^{-1}$. 
To bound these expressions, we use independence of $u_{i}$ and $v_{i}$ across $i$ and take into account the restrictions on the indices as indicated under the summations signs. The conditional expectations of products of $u_{i}$ and $v_{i}$ in these terms are all almost surely bounded by Assumption~\ref{ass:large_sample}, so we can take these out of the summations. Finding the nonzero terms in \eqref{eq:IIa1} and \eqref{eq:IIa2} is now a combinatorial exercise that can be executed using symbolic programming. The nonzero terms are bounded in Supplemental Appendix~\ref{A:varbounds}, showing that both \eqref{eq:IIa1} and \eqref{eq:IIa2} converge to zero almost surely.

\textbf{Part III: Show that $(Z.1)$ and $(Z.2)$ in \eqref{eq:estvariancedecomposed} converge to zero in probability.} Starting with $(Z.1)$, and using the expressions in \eqref{eq:unbvarest_decomp}, $(Z.1)  = (Z.1a) + (Z.1b)$, where
\begin{equation}
\begin{split}
    (Z.1a)
 &=-r_{n}^{-1}\sum_{i=1}^{n}\sum_{j\neq i}\frac{2}{1-2L_{ii}}[M_{W}Z\zeta]_{i}^2u_{i}L_{ij}u_{j}, \\
(Z.1b)&=r_{n}^{-1}\sum_{i=1}^{n}\sum_{j=1}^{n}\sum_{k\neq j}\frac{1}{(1-L_{ii})(1-2L_{ii})}[M_{W}Z\zeta]_{i}^2L_{ij}L_{ik}u_{j}u_{k}.
    \end{split}
\end{equation}
Using independence across $i$, we find that $(Z.1a)$ and $(Z.1b)$ have expectation zero, conditional on $Q$ and $X$. For the variance,
\begin{equation}\label{eq:VI1}
    \begin{split}
    \mathbb{V}(Z.1a|Q,X) &= r_{n}^{-2}\sum_{i=1}^{n}\sum_{k=1}^{n}\frac{8}{(1-2L_{ii})(1-2L_{kk})}\sigma_{u,i}^2\sigma_{u,k}^2L_{ik}^2[M_{W}Z\zeta]_{i}^2[M_{W}Z\zeta]_{k}^2\\
    &\leq c\max_{k}[M_{W}Z\zeta]_{k}^2r_{n}^{-2}\sum_{i=1}^{n}[M_{W}Z\zeta]_{i}^2\rightarrow_{a.s.}0,
    \end{split}
    \end{equation}
where we use that $[M_{W}Z\zeta]_{i}^2\leq c<\infty$ $a.s.$ and $\sum_{k=1}^{n}L_{ik}^2 = L_{ii}<1/2$. Likewise,
\begin{equation}\label{eq:VI2}
\begin{split}
    \mathbb{V}(Z.1b|Q,X) & = r_{n}^{-2}\sum_{i=1}^{n}\sum_{l=1}^{n}\frac{2}{(1-L_{ii})(1-2L_{ii})(1-L_{ll})(1-2L_{ll})}\\
    &\qquad \times \sum_{j=1}^{n}\sum_{k\neq j}L_{ij}L_{ik}L_{lj}L_{lk}\sigma_{u,j}^2\sigma_{u,k}^2[M_{W}Z\zeta]_{i}^2[M_{W}Z\zeta]_{l}^2\\
    &\leq cr_{n}^{-2}\sum_{i=1}^{n}\sum_{l=1}^{n}[M_{W}Z\zeta]_{i}^2[M_{W}Z\zeta]_{l}^2\sum_{j=1}^{n}\sum_{k\neq j}L_{ij}L_{ik}L_{lj}L_{lk}\\
    &\leq c\max_{l}[M_{W}Z\zeta]_{l}^2r_{n}^{-2}\sum_{i=1}^{n}[M_{W}Z\zeta]_{i}^2\rightarrow_{a.s.}0,
    \end{split}
\end{equation}
using that $\sum_{l,j}\sum_{k\neq j}L_{ij}L_{ik}L_{lj}L_{lk} = \sum_{j=1}^{n}\sum_{k\neq j}L_{ij}L_{ik}L_{jk}=\sum_{j=1}^{n}L_{ij}^2-\sum_{j=1}^{n}L_{ij}^2L_{jj}\leq L_{ii}<1/2$. 
By \eqref{eq:VI1} and \eqref{eq:VI2}, $
    r_{n}^{-1} \zeta'Z'M_{W}( D_{\hat{\sigma}_u^2}- D_{{u}^2})M_{W}Z\zeta\rightarrow_{p}0.$
Similarly, we have $
    r_{n}^{-1}\big(\pi'Z'M_{W}( D_{\hat{\sigma}_{v}^2(\beta)}- D_{{v}^2}) M_{W}Z\pi,\zeta'Z'M_{W}( D_{\hat{\sigma}_{uv}(\beta)}- D_{uv}) M_{W}Z\pi\big)\rightarrow_{p}0.$

We now turn to the final variance term $(Z.2)=(Z.2a) + (Z.2b)$, where
\begin{equation}
\begin{split}
(Z.2a) & =r_{n}^{-1}\sum_{k=1}^{n}\sum_{i\neq k}v_{k}A_{ik} [M_{W}Z\zeta]_{i} u_{i}\frac{2}{1-2L_{ii}}\sum_{j\neq i}L_{ij}u_{j}, \\
(Z.2b)&=r_{n}^{-1}\sum_{k=1}^{n}\sum_{i\neq k}v_{k}A_{ik} [M_{W}Z\zeta]_{i} \frac{1}{(1-L_{ii})(1-2L_{ii})}\sum_{j=1}^{n}\sum_{s\neq j}L_{ij}L_{is}u_{j}u_{s}.
\end{split}
\end{equation}
We have $\mathbb{E}[(Z.2)|Q,X]=0$. For the variance,
\begin{equation}\label{eq:varZ2a}
    \begin{split}
        \mathbb{V}(Z.2a|Q,X) & = r_{n}^{-2}\sum_{k=1}^{n}\sum_{l=1}^{n}\sum_{i\neq k}\sum_{m\neq l}A_{ik}A_{ml} [M_{W}Z\zeta]_{i}[M_{W}Z\zeta]_{m} \\
        &\quad \times \frac{2}{1-2L_{ii}}\frac{2}{1-2L_{mm}}\sum_{j\neq i}L_{ij}\sum_{s\neq m}L_{ms}\mathbb{E}[v_{k}v_{l}u_{i}u_{m}u_{j}u_{s}|Q,X]\\
        &\leq cr_{n}^{-2}\sum_{i,j,k}\bigg[ B_{ij}^2L_{ij}L_{jk} +  B_{ij} B_{ik}L_{ik}L_{jk}+B_{ij}^2L_{jk}^2\\
        &\qquad\qquad +  B_{ij} B_{ik}L_{jk}^2+ B_{ij} B_{jk}L_{ik}L_{jk}+ B_{ik} B_{jk}L_{ik}L_{jk}\bigg]\rightarrow_{a.s.}0.
    \end{split}
\end{equation}
The required bounds are provided in Supplemental Appendix~\ref{A:varbounds}.


For the variance of $(Z.2b)$, we have
\begin{align}
     &   \mathbb{V}(Z.2b|Q,X) \nonumber\\
        & \leq cr_{n}^{-2}\sum_{k=1}^{n}\sum_{l=1}^{n}\sum_{i\neq k}\sum_{m\neq l}\sum_{j=1}^{n}\sum_{s\neq j}\sum_{r=1}^{n}\sum_{t\neq r}|A_{ik}||A_{ml}|L_{ij}L_{is}L_{mr}L_{ms}\mathbb{E}[v_{k}v_{l}u_{j}u_{s}u_{r}u_{t}|Q,X]\nonumber\\
        &\leq  cr_{n}^{-2}\sum_{i_{1},\ldots,i_{4}}
 B_{i_{1}i_{3}} B_{i_{2}i_{4}}L_{i_{1}i_{3}}L_{i_{1}i_{4}}L_{i_{2}i_{3}}L_{i_{2}i_{4}}
+cr_{n}^{-2}\sum_{i_{1},\ldots,i_{5}}\big[
 B_{i_{1}i_{4}} B_{i_{1}i_{5}}L_{i_{2}i_{4}}L_{i_{2}i_{5}}L_{i_{3}i_{4}}L_{i_{3}i_{5}} \nonumber\\
&\quad+
 B_{i_{1}i_{4}} B_{i_{2}i_{5}}L_{i_{1}i_{4}}L_{i_{2}i_{5}}L_{i_{3}i_{4}}L_{i_{3}i_{5}}+
 B_{i_{1}i_{4}} B_{i_{2}i_{5}}L_{i_{1}i_{5}}L_{i_{2}i_{4}}L_{i_{3}i_{4}}L_{i_{3}i_{5}}\big]\rightarrow_{a.s.}0.\label{eq:varZ2b}
    \end{align}
Again, the required bounds are provided in Supplemental Appendix~\ref{A:varbounds}.

We conclude that $\hat{V}_{1}$ converges to a quantity at least as large as its target. We now consider how to remove the upward bias in this variance estimator.

\subsubsection{Variance bias correction}\label{A:bias}
In this section, following Lemma~\ref{lemm:consistency}, we replace $\hat{\beta}$ by $\beta$ in the HRK variance estimators \eqref{eq:unbvar} throughout. In \eqref{eq:vbias_2} we find that the numerator in the variance estimator \eqref{eq:V1hat} incurs an upward bias under weak identification. This bias arises because of the presence of the following terms,
\begin{equation}\label{eq:vbias}
    \Delta_{b} = \iota'D_{u^2}(A\odot A + 2C)D_{v^2}\iota+ \iota'D_{uv}(A\odot A + 2C)D_{uv}\iota,
    \end{equation}
where $C$ is given in \eqref{eq:C}. Estimating \eqref{eq:vbias} by replacing $D_{u^2}$ with $D_{\hat{\sigma}_{u}^2}$, $D_{v^2}$ with $D_{\hat{\sigma}_{v}^2}$ and $D_{uv}$ with $D_{\hat{\sigma}_{uv}}$ does not resolve this bias: the products $\hat{\sigma}_{u,i}\hat{\sigma}_{v,j}$ and $\hat{\sigma}_{uv,i}\hat{\sigma}_{uv,j}$ are themselves biased estimators for $\mathbb{E}[u_{i}^2v_{j}^2|Q,X]$ and $\mathbb{E}[u_{i}v_{i}u_{j}v_{j}^2|Q,X]$ if $i$ and $j$ are in the same covariate group and have the same instrument status. We can correct this bias as follows. 

Let $\alpha_{1,g}=m_{1,g}^2-3m_{1,g}+2$, $\alpha_{0,g}=m_{0,g}^2-3m_{0,g}+2$, and $Z_{ig}^{c}=1-Z_{ig}$, and define
\begin{equation}\label{eq:vbias_est}
    \hat{\Delta}_{b} = \iota'D_{\hat{\sigma}_{u}^2}B_{1}\odot(A\odot A + 2C)D_{\hat{\sigma}_{v}^2}\iota +  \iota' D_{\hat{\sigma}_{uv}}B_{2}\odot(A\odot A + 2C)D_{\hat{\sigma}_{uv}}\iota,
 \end{equation}
where 
\begin{equation}\label{eq:Bij}
\begin{split}
 B_{1,ij} &= 
\sum_{g=1}^{G}\frac{\alpha_{1,g}^2}{(\alpha_{1,g}+1)(\alpha_{1,g}-2)}Z_{ig}Z_{jg}1[m_{1,g}\geq 4] \\
&\quad + \frac{\alpha_{0,g}^2}{(\alpha_{0,g}+1)(\alpha_{0,g}-2)}W_{ig}W_{jg}Z_{ig}^c Z_{jg}^c 1[m_{0,g}\geq 4]+ W_{ig}Z_{ig}^c Z_{jg} +Z_{ig}W_{jg}Z_{jg}^c,\\ 
 B_{2,ij} &= 
\sum_{g=1}^{G}\frac{\alpha_{1,g}(\alpha_{1,g}+2)}{(\alpha_{1,g}+1)(\alpha_{1,g}-2)}Z_{ig}Z_{jg}1[m_{1,g}\geq 4]\\
&\quad + \frac{\alpha_{0,g}(\alpha_{0,g}+2)}{(\alpha_{0,g}+1)(\alpha_{0,g}-2)}W_{ig}W_{jg}Z_{ig}^c Z_{jg}^c 1[m_{0,g}\geq 4]+ W_{ig}Z_{ig}^c Z_{jg} +Z_{ig}W_{jg}Z_{jg}^c. 
 \end{split}
\end{equation}
With $\hat{\Delta}_{b}$ from \eqref{eq:vbias_est}, we then define $\hat{\mathbb{V}}_{2}$ in the variance estimator \eqref{eq:Vhat} as
\begin{equation}\label{eq:exprV2hat}
\hat{\mathbb{V}}_{2} = \hat{\Delta}_{b}/(T'AT)^2.
\end{equation}
This removes the bias from pairs of observations with opposite treatment status if they belong to groups with $m_{1,g}\geq 3$ or $m_{0,g}\geq 3$ and from pairs with the same treatment status for $m_{1,g}\geq 4$ and $m_{0,g}\geq 4$. The following result underlies part 2 of Theorem~\ref{corr:dist2} that states that we asymptotically attain nominal coverage if $m_{1,g}\geq 4$ and $m_{0,g}\geq 4$. 

\begin{lemma}\label{lem:debias} Suppose the conditions of Theorem~\ref{corr:dist2} hold with Assumption~\ref{ass:group_size} strengthened to $m_{1,g}\geq 4$, $m_{0,g}\geq 4$. With $\Delta_{b}$ as in \eqref{eq:vbias}, $\hat{\Delta}_{b}$ as in \eqref{eq:vbias_est}, $r_{n}^{-1}\big(\hat{\Delta}_{b} -\Delta_{b}\big)\rightarrow_{p}0$.
\end{lemma}
To prove the result, we first show unbiasedness of $\hat{\Delta}_{b}$ under the stated assumptions. The remainder of the proof is deferred to the Supplemental Appendix~\ref{supp:consistencycorrection} as it relies on the same techniques used to show consistency of the variance estimator in Appendix~\ref{app:varscore}. 

Using the block structure of $A$ and $C$ shown in \eqref{eq:Ag} and \eqref{eq:C}, we can write
\begin{equation}
\begin{split}
\Delta_{b}
&=\sum_{g=1}^{G}\big(A_{11,(g)}^2 + 2C_{11,(g)}\big)\sum_{i}\sum_{j\neq i}Z_{ig}Z_{jg}\rho_{ij}\\
&\qquad+A_{10,(g)}^2\sum_{i}\sum_{j\neq i}W_{jg}Z_{ig}Z_{jg}^c \rho_{ij} +A_{01,(g)}^2\sum_{i}\sum_{j\neq i}W_{ig}Z_{ig}^c Z_{jg}\rho_{ij} \\
&\qquad+\big(A_{00,(g)}^2 + 2C_{00,(g)}\big)\sum_{i}\sum_{j\neq i}W_{ig}W_{jg}Z_{ig}^{c}Z_{jg}^{c}\rho_{ij},
\end{split}
\end{equation}
with $A_{kl,(g)}$ and $C_{kl,(g)}$ denoting the value of the entries of \eqref{eq:Ag} and \eqref{eq:C} corresponding to two individuals from covariate group $g$ with instrument values $k$ and $l$, and
with $\rho_{ij}=u_{i}v_{i}u_{j}v_{j} + u_{i}^2v_{j}^2$. To estimate $\sum_{i\neq j}W_{jg}Z_{ig}Z_{jg}^{c}\rho_{ij}$ and $\sum_{i\neq j}W_{ig}Z_{ig}^{c}Z_{jg}\rho_{ij}$, we can multiply the HRK variance estimators for the individual components as these are evaluated on independent data. To estimate $\sum_{i\neq j}Z_{ig}Z_{jg}\rho_{ij}$, it is helpful to observe that we do not require estimators for individual products $\sigma_{u,i}\sigma_{v,j}$ and $\sigma_{uv,i}\sigma_{uv,j}$, but only for sums over these products in the same covariate group with the same instrument status.

Take indices $i$ and $l$ such that $Z_{ig}=1$ and $Z_{lg}=1$. Using the decomposition in \eqref{eq:unbvarest_decomp}, 
\begin{align}
&\hat{\sigma}_{uv,i}\hat{\sigma}_{uv,l}\\
&=\bigg(u_{i}v_{i} + \frac{1}{m_{1,g}-1}\big(u_{i}\sum_{j\neq i}Z_{jg}v_{j} + v_{i}\sum_{j\neq i}Z_{jg}u_{j} + \frac{1}{m_{1,g}-2}\sum_{j}\sum_{k\neq j}Z_{jg}Z_{kg}u_{j}v_{k}\big)\bigg)\nonumber\\
&\times\bigg(u_{l}v_{l} + \frac{1}{m_{1,g}-1}\big(u_{l}\sum_{j\neq l}Z_{jg}v_{j} + v_{l}\sum_{j\neq l}Z_{jg}u_{j}  + \frac{1}{m_{1,g}-2}\sum_{j}\sum_{k\neq j}Z_{jg}Z_{kg}u_{j}v_{k}\big)\bigg).\nonumber
\end{align}
Taking the expectation conditional on the covariates and instruments, and summing over all $\{i,l\}$ for which $Z_{ig}=1$, $Z_{lg}=1$ and $i\neq l$, we get
\begin{equation}\label{eqvar1}
\begin{split}
   & \sum_{i}\sum_{l\neq i}Z_{ig}Z_{lg}\mathbb{E}[\hat{\sigma}_{uv,i}\hat{\sigma}_{uv,l}|Q,X]\\
   &=\mathbb{E}\bigg[\frac{\alpha_{1,g}-1}{\alpha_{1,g}}\sum_{i}\sum_{l\neq i}Z_{ig}Z_{lg}u_{i}v_{i}u_{l}v_{l}-\frac{1}{\alpha_{1,g}}\sum_{i}\sum_{l\neq i}Z_{ig}Z_{lg}u_{i}^2v_{l}^2\big|Q,X\bigg].
\end{split}
\end{equation}
    Similar calculations show that,  \begin{equation}\label{eqvar2}
    \begin{split}         &\sum_{i}\sum_{l\neq i}Z_{ig}Z_{lg}\mathbb{E}[\hat{\sigma}_{u,i}\hat{\sigma}_{v,l}|Q,X] \\
    &= \mathbb{E}\bigg[\sum_{i}\sum_{l\neq i}Z_{ig}Z_{lg}u_{i}^2v_{l}^2 -\frac{2}{\alpha_{1,g}}\sum_{i}\sum_{l\neq i}Z_{ig}Z_{lg}u_{i}v_{i}u_{l}v_{l}\big|Q,X\bigg].
    \end{split}
\end{equation}
The expressions in \eqref{eqvar1} and \eqref{eqvar2} can be solved to find the unbiased estimators
\begin{equation}\label{eq:unbvares}
\begin{split}\widehat{\sum_{i}\sum_{l\neq i}Z_{ig}Z_{lg}u_{i}v_{i}u_{l}v_{l}} &= \bigg(\frac{\alpha_{1,g}-1}{\alpha_{1,g}}-\frac{2}{\alpha_{1,g}^2}\bigg)^{-1}\sum_{i}\sum_{l\neq i}Z_{ig}Z_{lg}(\hat{\sigma}_{uv,i}\hat{\sigma}_{uv,l}+\frac{1}{\alpha_{1,g}}\hat{\sigma}_{u,i}\hat{\sigma}_{v,l} )\\
      \widehat{\sum_{i}\sum_{l\neq i}Z_{ig}Z_{lg}u_{i}^2v_{l}^2} &=  \sum_{i}\sum_{l\neq i}Z_{ig}Z_{lg}\hat{\sigma}_{u,i}\hat{\sigma}_{v,l}+\frac{2}{\alpha_{1,g}} \widehat{\sum_{i}\sum_{l\neq i}Z_{ig}Z_{lg}u_{i}v_{i}u_{l}v_{l}}.      \end{split}    
\end{equation}
To complete the proof, we sum the estimators in \eqref{eq:unbvares} to obtain
\begin{equation}
\begin{split}
   & \widehat{\sum_{i}\sum_{l\neq i}Z_{ig}Z_{lg}u_{i}v_{i}u_{l}v_{l}}+  \widehat{\sum_{i}\sum_{l\neq i}Z_{ig}Z_{lg}u_{i}^2v_{l}^2}\\
    &=\frac{\alpha_{1,g}^2}{(\alpha_{1,g}+1)(\alpha_{1,g}-2)}\sum_{i}\sum_{l\neq i}\hat{\sigma}_{u,i}\hat{\sigma}_{v,l} + \frac{\alpha_{1,g}(\alpha_{1,g}+2)}{(\alpha_{1,g}+1)(\alpha_{1,g}-2)}\sum_{i}\sum_{l\neq i}\hat{\sigma}_{uv,i}\hat{\sigma}_{uv,l},
   \end{split}
\end{equation}
which coincides with the result in \eqref{eq:Bij}. Analogous calculations with $m_{1,g}$ replaced by $m_{0,g}$ yield the estimator for $\sum_{i\neq j}W_{ig}W_{jg}Z_{ig}^{c}Z_{jg}^{c}\rho_{ij}$. In total, this yields \eqref{eq:vbias_est}.

We note that these estimators only exist if 
\begin{equation}
\frac{\alpha_{1,g}-1}{\alpha_{1,g}}-\frac{2}{\alpha_{1,g}^2}
= \frac{m_g(m_{g}^3-6m_{g}^2+12m_{g}-9)}{(m_{g}^2-3m_{g}+2)^2}\neq 0.
\end{equation}
This rules out $m_{1,g}=3$ and $m_{0,g}=3$. The quantity should also be finite, which rules out $m_{1,g}=2$ and $m_{0,g}=2$. We conclude that we require $m_{1,g}\geq 4$ and $m_{0,g}\geq 4$. 

\end{appendix}


\setlength{\bibsep}{2pt plus 0.3ex}
\bibliographystyle{chicagoa}
\bibliography{library}  



\newpage
\noindent
\setcounter{page}{1}
\begin{center}
	{\bf \Large{Supplemental Appendix for\\ ``Inference on LATEs with covariates''}}
\end{center}

\vspace{10pt}
\setcounter{equation}{0}
\setcounter{figure}{0}
\setcounter{table}{0}
\setcounter{section}{0}
\renewcommand{\thetable}{S\arabic{table}}
\renewcommand{\thefigure}{S\arabic{figure}}
\renewcommand{\thesection}{S\Alph{section}}
\renewcommand{\thesubsection}{S\Alph{section}.\arabic{subsection}}
\renewcommand{\theequation}{S\arabic{equation}}

\noindent
This supplemental appendix has five parts:

\begin{itemize}
    \item[] {\bf Part~A}: Extension to clustering.
	\item[] {\bf Part~B}: Jackknife instrumental variables estimation.
    \item[] {\bf Part~C}: Additional results on variance estimation.
	\item[] {\bf Part~D}: Additional numerical experiments.
	\item[] {\bf Part~E}: Additional empirical applications.
\end{itemize}

\section{Extension to clustering}\label{A:cluster}

Consider a setting where the projection errors $u_{i}$ and $\varepsilon_{i}$ are clustered in $K$ groups indexed by $k=1,\ldots,K$ and $n_{k}$ observations per cluster. Denote by $L_{K}$ an $n\times K$ matrix with $L_{K,ik}=1$ if observation $i$ is a member of cluster $k$. For any vector $v$, denote by $v_{(k)}$ the subvector of individuals from cluster $k$. For any matrix $V$, denote by $V_{(kl)}$ the submatrix with the rows (columns) corresponding to individuals from cluster $k$ ($l$). We can accommodate the following clustering structure. 
\begin{assumption}\label{ass:cluster} 
(1) Let $e_{(k)} = (\varepsilon_{(k)}',u_{(k)}')'$ and $\Sigma{(k)} = \mathbb{E}[e_{(k)}e_{(k)}'|Q,X]$. Then, $c_{1}\leq \lambda_{\min}(\Sigma_{(k)})\leq \lambda_{\max}(\Sigma_{(k)})\leq c_{2}$ $a.s.$ for positive constants $\{c_{1},c_{2}\}$. 
(2) For $g=1,\ldots, G$, $
\max_{k=1,\ldots,K}\sum_{i=1}^{n}W_{ig}L_{K,ik}=1.$
\end{assumption}
Part 1 is the analogue of the lower bounds on the variances and upper bound on the correlation in Assumption~\ref{ass:large_sample} in a clustered setting. To illustrate Part 2, consider the case where clustering takes place at the household level. Assumption~\ref{ass:cluster} states that each person in a household has a different covariate vector, i.e.\ each person is unique within their household. This implies that we always have at least as many clusters as we have covariate groups.
The following identification result extends Theorem~\ref{lemm:sive} to a clustered setting.
\begin{theorem}\label{lemm:siveclust}
Under Assumption~\ref{ass:IV}, \ref{ass:group_size}, \ref{ass:relevance} and \ref{ass:cluster} it holds that
    \begin{align}
        \beta^{\emph{SIVE}}=\frac{\sum\nolimits_g \tilde{\emph{P}}[X_i=x_g]\pi(x_g)^2 {\tilde{\emph{V}}}[Q_i|X_i=x_g] \tau(x_g)}{\sum\nolimits_g \tilde{\emph{P}}[X_i=x_g]\pi(x_g)^2 {\tilde{\emph{V}}}[Q_i|X_i=x_g]}=\sum_g \omega(x_g)\tau(x_g).
    \end{align}
\end{theorem}
The proof is an immediate consequence of Assumption~\ref{ass:cluster}. Bias under clustering could be caused by the term $u'A \varepsilon$. Although by construction $\mathbb{E}[u_i\varepsilon_{i}|Q,X]A_{ii}=0$, potentially $\mathbb{E}[u_i\varepsilon_{j}|Q,X]A_{ij}\neq 0$ if $i$ and $j$ are in the same cluster. However, for any pair of indices $\{i,j\}$ that are in the same cluster, $A_{ij}=0$ as $i$ and $j$ are not allowed to be in the same covariate group. If there are in fact multiple identical individuals within each household, Lemma~\ref{lemm:siveclust} would not hold. In this case, we would have to remove the off-diagonal elements from the matrix $A$ for people in the same covariance group and the same cluster.

We also have the following result.
\begin{lemma}
Under Assumption~\ref{ass:IV}, \ref{ass:group_size}, \ref{ass:relevance}, \ref{ass:large_sample} and \ref{ass:cluster} with $\frac{n}{\sqrt{G}}\text{FS}\rightarrow_{p}\infty$, it holds that $\hat{\beta}-\beta\rightarrow_{p} 0$.
\end{lemma}
\textit{Proof}: With the notation as in Appendix~\ref{subsec:consistency}, under Assumption~\ref{ass:cluster} it still holds that $\mathbb{E}[S_{n}|Q,X]=0$. For the variance of $S_{n}$, we have
\begin{equation}
\begin{split}
    \mathbb{E}[S_{n}^2|Q,X]
& = r_{n}^{-1}(\zeta'Z'M_{W}, \pi'Z'M_{W})\Sigma(\zeta'Z'M_{W}, \pi'Z'M_{W})'\\
&\quad + r_{n}^{-1}\sum_{k\neq l}\text{tr}\bigg(\Sigma_{(k)}\left[\begin{array}{cc}
0 & A_{(kl)}\\
A_{(kl)}&0\end{array}\right]\Sigma_{(l)}\left[\begin{array}{cc}
0 & A_{(lk)}\\
A_{(lk)}&0\end{array}\right]\bigg)\\
&\leq cr_{n}^{-1}(\zeta'Z'M_{W}Z\zeta + \pi'Z'M_W Z\pi) + cr_{n}^{-1}\text{tr}(A^2)\leq c,\quad a.s.
    \end{split}
\end{equation}
The second line is obtained as in \citet{ligtenberg2023inference}, the first inequality uses Assumption~\ref{ass:cluster} and the final inequality is the same as in Appendix~\ref{subsec:consistency}. Likewise, we obtain that $\mathbb{E}[D_{n}|Q,X]=r_{n}^{1/2}H_{n}$ and $\mathbb{V}(D_n|Q,X)\leq c$ $a.s.$ The proof is completed as in Appendix~\ref{subsec:consistency}.

For inference, we adapt \eqref{eq:Vhat} using the generalization of the HRK variance estimators to the clustered setting considered in \citet{boot2023unbiased}.
\begin{equation}\label{eq:Vhat_clust}
    \hat{\mathbb{V}}[\hat{\beta}|Q,X] = \frac{(Y-T\hat{\beta})'A \hat{\Sigma}_{u} A( Y- T\hat{\beta}) +  T' A \hat{\Sigma}_{v} A T + 2( Y- T\hat{\beta})' A \hat{\Sigma}_{uv} A T}{(T' A T)^2}.
\end{equation}
Let $F$ be an $n^2\times \sum_{k=1}^{K}n_{k}^2$ matrix that collects the nonzero terms of $\text{vec}(\Sigma_{u})=Fh$. Define
\begin{align}
    \text{vec}(\hat{\Sigma}_{u})&=F(F'(M_{W,Z}\otimes M_{W,Z})F)^{-1}F'(M_{W,Z}T\otimes M_{W,Z}T),\nonumber\\
     \text{vec}(\hat{\Sigma}_{v})&=F(F'(M_{W,Z}\otimes M_{W,Z})F)^{-1}F'(M_{W,Z}(Y-T\hat{\beta})\otimes M_{W,Z}(Y-T\hat{\beta})),\nonumber\\
      \text{vec}(\hat{\Sigma}_{uv})&=F(F'(M_{W,Z}\otimes M_{W,Z})F)^{-1}F'(M_{W,Z}T\otimes M_{W,Z}(Y-T\hat{\beta})).
  \end{align}
Consider $\text{vec}(\hat{\Sigma}_{u})$. It is straightforward to show that this is unbiased, 
\begin{align}
    \mathbb{E}[\text{vec}(\hat{\Sigma}_{u})|Q,X]&=F(F'(M_{W,Z}\otimes M_{W,Z})F)^{-1}F'(M_{W,Z}\otimes M_{W,Z})\mathbb{E}[u\otimes u|Q,X]\nonumber\\
    &=F(F'(M_{W,Z}\otimes M_{W,Z})F)^{-1}F'(M_{W,Z}\otimes M_{W,Z})\mathbb{E}[\text{vec}(uu')|Q,X]\nonumber\\
     &=F(F'(M_{W,Z}\otimes M_{W,Z})F)^{-1}F'(M_{W,Z}\otimes M_{W,Z})\text{vec}(\Sigma_{u})\nonumber\\
      &=F(F'(M_{W,Z}\otimes M_{W,Z})F)^{-1}F'(M_{W,Z}\otimes M_{W,Z})Fh\nonumber\\
      &=Fh = \text{vec}(\Sigma_{u}).
    \end{align}

\section{Jackknife instrumental variable estimation}\label{A:jive}
The following result gives the expression for the JIVE estimand from \eqref{eq:jive}. It shows that while removing the diagonal elements has removed the many instrument bias, 
the estimand can be substantially different from $\tau$ and is difficult to interpret.
\begin{lemma}\label{lemm:jive}
Under Assumption~\ref{ass:IV} and \ref{ass:indep} it holds that
\begin{align}
    \beta^{\emph{JIVE}}=\frac{\sum\nolimits_g  \tilde{\emph{P}}[X_i=x_g]\pi(x_g)^2 \tilde{\emph{V}}[Q_i|X_i=x_g](1-\frac{1}{m_{1,g}})\tau(x_g) -B_Y}{\sum\nolimits_g \tilde{\emph{P}}[X_i=x_g] \pi(x_g)^2 \tilde{\emph{V}}[Q_i|X_i=x_g](1-\frac{1}{m_{1,g}}) -B_T},
\end{align}
where
\begin{align}
    B_{Y} =& \sum_g  \tilde{\emph{P}}[X_i=x_g]\pi(x_g)(\phi_g + \tau(x_g)\psi_g) \tilde{\emph{V}}[Q_i|X_i=x_g]\frac{1}{m_{1,g}} +\frac{1}{n} \psi_g\phi_g,\nonumber \\
    B_T =& 2\sum_g \tilde{\emph{P}}[X_i=x_g]\pi(x_g)\psi_g \tilde{\emph{V}}[Q_i|X_i=x_g]\frac{1}{m_{1,g}} +\frac{1}{n} \psi_g^2,
\end{align}
with $\psi_g=\mathbb{E}[T_i|Z_{ig}=0,W_{ig}=1]$, and $\phi_g=\mathbb{E}[Y_i|Z_{ig}=0,W_{ig}=1]$.
\end{lemma}
\textit{Proof}: The JIVE estimand is
\begin{equation}\begin{split}
    \beta^{\text{JIVE}}=\frac{\mathbb{E}[T'(P-D_P) Y|Q,X]}{\mathbb{E}[T'(P-D_P)T|Q,X]}
    =\frac{\mathbb{E}[T'PY|Q,X]-\mathbb{E}[T'D_P Y|Q,X]}{\mathbb{E}[T'PT|Q,X]-\mathbb{E}[T'D_P T|Q,X]}.
\end{split}\end{equation}
For the second term in the numerator, using \eqref{eq:2slseq}, 
\begin{equation}
\begin{split}
   &\mathbb{E}[T'D_PY|Q,X]
    \\
    &=\pi'Z'D_P Z\theta+\pi'Z'D_P W \phi+\psi'W'D_P Z\theta +\psi'W'D_P W \phi+\mathbb{E}[u'D_P\varepsilon|Q,X]\\
    &= \sum_g\big(\tau(x_g) \pi(x_g)^2 +\pi(x_g)\phi_g + \theta(x_g)\psi_g\big) \frac{m_{0,g}}{n_{g}} + \psi_g\phi_g + \sum_i \mathbb{E}[u_i \varepsilon_i|Q,X] P_{ii},
\end{split}
\end{equation}
where $W'D_PW=I_G$ as $\sum_i W_{ig}^2 P_{ii}=1$, and $Z'D_PZ=W'D_PZ$ is a $G \times G$ diagonal matrix with elements $\sum_i Z_{ig}^2 P_{ii}=\frac{m_{0,g}}{n_{g}} $. Similarly, we get
\begin{equation}\begin{split}
    \mathbb{E}[T'D_PT|Q,X]&= \sum_g \big(\pi(x_g)^2 +2\pi(x_g)\psi_g\big) \frac{m_{0,g}}{n_{g}}  + \psi_g^2+ \sum_i \mathbb{E}[u_i \varepsilon_i|Q,X] P_{ii}.
\end{split}\end{equation}
Using the results in Appendix~\ref{A:2sls}, we arrive at the expression for $\beta^{\text{JIVE}}$ in Lemma~\ref{lemm:jive}.

The following result shows that IJIVE defined in \eqref{eq:jive2} reintroduces the many instrument bias.
\begin{lemma}\label{lemm:IJIVE}
Under Assumption~\ref{ass:IV} and \ref{ass:indep} it holds that
\begin{equation}
    \beta^{\emph{IJIVE}}= \frac{\sum\nolimits_g\tilde{\emph{P}}[X_i=x_g]\pi(x_g)^2\tilde{\emph{V}}[Q_{i}|X_{i}=x_g]\tau(x_g) + B_{Y,1}+B_{Y,2}}{\sum\nolimits_{g}\tilde{\emph{P}}[X_i=x_g]\pi(x_g)^2\tilde{\emph{V}}[Q_{i}|X_{i}=x_g] + B_{T,1}+B_{T,2}},
\end{equation}
where $B_{Y,1} = -\frac{1}{n}\sum_{g=1}^{G}\pi(x_g)^2\big[1-3\tilde{\emph{V}}[Q_{i}|X_{i}=x_g]\big]\tau(x_g)$,\\ $B_{Y,2} = \frac{1}{n}\sum_{i=1}^{n}\mathbb{E}[u_{i}\varepsilon_{i}|X_{i},Q_{i}]\big(\frac{2}{n_{g}}P_{ii}-\frac{1}{n_{g}^2}\big)$, $B_{T,1} = -\frac{1}{n}\sum_{g=1}^{G}\pi(x_g)^2\big[1-3\tilde{\emph{V}}[Q_{i}|X_{i}=x_g]\big]$, and $B_{T,2} = \frac{1}{n}\sum_{i=1}^{n}\mathbb{E}[u_{i}^2|X_{i},Q_{i}]\big(\frac{2}{n_{g}}P_{ii}-\frac{1}{n_{g}^2} \big).$
\end{lemma}
The proof is similar as for Lemma~\ref{lemm:jive} and omitted. The causal estimand again is different from TSLS, but under Assumption~\ref{ass:group_size} the weights on $\tau(x_g)$ will lie between 0 and the TSLS weights. To quantify the many instrument bias, consider a homoskedastic setting where $\mathbb{E}[u_{i}\varepsilon_{i}|X, Q] = \sigma_{u\varepsilon}$. We see that $B_{Y,2}=\frac{\sigma_{u\varepsilon}}{n}\sum_{g=1}^{G}\frac{1}{n_{g}}$ compared to the bias in TSLS that is $\frac{\sigma_{u\varepsilon}}{n}\sum_{g=1}^{G}1$. We conclude that IJIVE offers a substantial bias reduction, especially when $n_{g}$ is large. However, with rich support for the covariates, $n_{g}$ is fixed, and the bias is of the same order as for TSLS. 

\section{Additional results on variance estimation}\label{A:addvarest}
\subsection{Results on the matrix A used for consistency}\label{A:Aresults}
To show consistency of the variance estimator, we need to upper bound terms that are of the form $
S( A, L) = \sum_{i_{1},\ldots,i_{16}}|A_{i_{1}i_{2}}||A_{i_{3}i_{4}}||A_{i_{5}i_{6}}||A_{i_{7}i_{8}}|L_{i_{9}i_{10}}L_{i_{11}i_{12}}L_{i_{13}i_{14}}L_{i_{15}i_{16}}$, 
with some restrictions on the indices over which we sum. We will first find a matrix $ B$ such that $S( A, L)\leq S( B, L)$. To save on notation, define
\begin{equation}
\begin{split}
 H_{g}& = \left(\begin{array}{cc}
    \iota_{m_{1,g}}&  0_{m_{1,g}}\\
     0_{m_{0,g}} & \iota_{m_{0,g}} 
    \end{array}\right),\quad 
 E_{g}  = \left(\begin{array}{cc}
    \frac{1}{m_{1,g}}& 0\\
    0 & \frac{1}{m_{0,g}}\end{array}\right).
    \end{split}
\end{equation}
We then have
\begin{align}
    {A}_{(g)} 
& = \frac{m_{1,g}m_{0,g}}{n_{g}}\left(\begin{array}{cc}
\frac{1}{m_{1,g}(m_{1,g}-1)}(\iota_{m_{1,g}}\iota_{m_{1,g}}'-I_{m_{1,g}}) & -\frac{1}{m_{1,g}m_{0,g}}\iota_{m_{1,g}}\iota_{m_{0,g}}'\\
-\frac{1}{m_{1,g}m_{0,g}}\iota_{m_{0,g}}\iota_{m_{1,g}}' &\frac{1}{m_{0,g}(m_{0,g}-1)}(\iota_{m_{0,g}}\iota_{m_{0,g}}'-I_{m_{0,g}})    \end{array}\right)\nonumber\\
&\leq 2\frac{m_{1,g}m_{0,g}}{n_{g}} H_{g}E_g\iota_{2}\iota_{2}' E_{g} H_{g}'\equiv B_{(g)},\label{eq:Atilde}
    \end{align}
where $\leq$ indicates an elementwise inequality, so that $S( A, L)\leq S( B, L)$. 

We can now establish the following results
\begin{equation}\label{eq:useful2}
\begin{array}{crlcrl}
   (R1) & \text{tr}( B_{(g)}) & \leq c, &  (R7)&\iota_{n_{g}}'( B_{(g)}\odot L_{(g)}) B_{(g)}\iota_{n_{g}}&\leq c,\\
    (R2) &\lambda_{\max}( B_{(g)})&\leq c,& (R8) &  \iota_{n_{g}}' B_{(g)}( L_{(g)}\odot L_{(g)}) B_{(g)}\iota_{n_{g}} &\leq c,\\
    (R3)&  B_{(g)}^2 &=2 B_{(g)},&(R9)&\iota_{n_{g}}'(B_{(g)}\odot B_{(g)})^2\iota_{n_g} &\leq c,\\
    (R4) & B_{(g)} L_{(g)} &=  B_{(g)},&(R10) &\iota_{n_{g}}'B_{(g)}(B_{g}\odot B_{(g)})B_{(g)}\iota_{n_{g}}&\leq c,\\
    (R5)& \iota_{n_{g}}'( B_{(g)}\odot L_{(g)})\iota_{n_{g}}&\leq c,&(R11)&(\iota_{n_{g}}'(B_{(g)}\odot B_{(g)})((B_{(g)}\iota_{n_{g}})\odot(B_{(g)}\iota_{n_{g}}))&\leq c,\\ 
     (R6)& \iota_{n_{g}}'( B_{(g)}\odot L_{(g)})^2\iota_{n_{g}}&\leq c.
     \end{array}
 \end{equation}

\textit{Proof:} (R1) follows from the fact that $ H_{g}' H_{g} =  E_{g}^{-1}$ 
and $\iota_{2}' E_{g}\iota_{2} = \frac{n_{g}}{m_{1,g}m_{0,g}}$. Since $\text{rank}( B_{(g)})=1$, (R2) follows from (R1). (R3) follows from the fact that $ H_{g}' H_{g} =  E_{g}^{-1}$. For (R4)--(R8), note that 
   $ L_{(g)} =  H_{g} E_{g} H_{g}'$. Using this result and $ H_{g}' H_{g} =  E_{g}^{-1}$ yields (R4). 
   For (R5), we note that $ \iota_{n_{g}}'( B_{(g)}\odot L_{(g)})\iota_{n_{g}} = \text{tr}( B_{(g)} L_{(g)}) = \text{tr}( B_{(g)})$ with the last equality by (R4). For (R6) and (R7) we calculate the elementwise product,
\begin{equation}
\begin{split}
     B_{(g)}\odot L_{(g)} & = 2\frac{m_{1,g}m_{0,g}}{n_{g}} H_{g} E_{g}^3 H_{g}'.
\end{split}
\end{equation}
We can now explicitly calculate bounds for (R6) and (R7),
\begin{equation}
\begin{split}
  \iota_{n_{g}}'( B_{(g)}\odot L_{(g)})^2\iota_{n_{g}} &= 4\frac{m_{1,g}^2m_{0,g}^2}{n_{g}^2}\left(\frac{1}{m_{1,g}^{3}}+\frac{1}{m_{0,g}^{3}}\right) \\
 &=4\frac{n_{g}^3-3n_{g}m_{1,g}^2-3n_{g}^2m_{1,g}}{n_{g}^3m_{0,g}}\leq 4\frac{1}{n_{g}}\leq 2,\\
 \iota_{n_{g}}'( B_{(g)}\odot L_{(g)}) B_{(g)}\iota_{n_{g}}& = 4\frac{m_{1,g}^2m_{0,g}^2}{n_{g}^2}\left(\frac{1}{m_{1,g}^{2}}+\frac{1}{m_{0,g}^{2}}\right)\leq 8.
 \end{split}
 \end{equation}
 For (R8), we have a similar result as,
 \begin{equation}
\begin{split}
 \iota_{n_{g}}' B_{(g)}( L_{(g)}\odot L_{(g)}) B_{(g)}\iota_{n_{g}}  &= 4\frac{m_{1,g}^2m_{0,g}^2}{n_{g}^2}\left(\frac{1}{m_{1,g}^{2}}+\frac{1}{m_{0,g}^{2}}\right)\leq 8.
    \end{split}
\end{equation}
This completes the proof for $(R1)-(R8)$.

To prove $(R9)-(R11)$, using the definition of $B_{(g)}$ in \eqref{eq:Atilde},
\begin{align}
(B_{(g)}\odot B_{(g)})\iota_{n_g}&=4\frac{m_{1,g}m_{0,g}}{n_{g}}H_{g}E_{g}^2\iota_{2},\label{eq:iv}\\
B_{(g)}\odot B_{(g)} & = 4\frac{m_{1,g}^2m_{0,g}^2}{n_{g}^2}H_{g}E_{g}^2\iota_{2}\iota_{2}'E_{g}^2H_{g}',\label{eq:i}\\
B_{(g)}\iota_{n_{g}}&=4\frac{m_{1,g}m_{0,g}}{n_{g}}H_{g}E_g\iota_{2},\label{eq:ii}\\
(B_{(g)}\iota_{n_{g}})\odot(B_{(g)}\iota_{n_{g}})&= 16\frac{m_{1,g}^2m_{0,g}^2}{n_{g}^2}H_{g}E_{g}^2\iota_{2}.\label{eq:iii}
\end{align}
Using \eqref{eq:iv} and the fact that $H_{g}'H_{g} = E_{g}^{-1}$, $(R9)$ follows from
\begin{equation}
\begin{split}
    \frac{m_{1,g}^2m_{0,g}^2}{n_{g}^2}\iota_{2}'E_{g}^3\iota_{2}& =     \frac{m_{1,g}^2m_{0,g}^2}{n_{g}^2}\frac{m_{1,g}^3 + m_{0,g}^3}{m_{1,g}^3m_{0,g}^3}
    \leq \frac{n_{g}^3-3n_{g}^2m_{1,g}-3n_{g}m_{1,g}^2}{n_{g}^3}\leq 1,
    \end{split}
\end{equation}
as $m_{1,g}m_{0,g}\geq 2(n_{g}-2)\geq 2(n_{g}-n_{g}/2) = n_{g}$ since $n_{g}\geq 4$ by Assumption~\ref{ass:group_size}.

For $(R10)$, using first \eqref{eq:i} and then \eqref{eq:ii},
\begin{equation}
    \begin{split}
        \iota_{n_{g}}'B_{(g)}(B_{(g)}\odot B_{(g)})B_{(g)}\iota_{n_{g}}& = 64\frac{m_{1,g}^4m_{0,g}^4}{n_{g}^4}(\iota_{2}'E_{g}^2\iota_{2})^2=64 \frac{(m_{1,g}^2+m_{0,g}^2)^2}{n_{g}^4}\leq 256.
    \end{split}
\end{equation}
 For $(R11)$, using \eqref{eq:iv} and \eqref{eq:iii}, 
 \begin{equation*}
     \begin{split}
         (\iota_{n_{g}}'(B_{(g)}\odot B_{(g)})((B_{(g)}\iota_{n_{g}})\odot(B_{(g)}\iota_{n_{g}}))& = 64\frac{m_{1,g}^3m_{0,g}^3}{n_{g}^3}\iota_{2}'E_{g}^3\iota_{2}=64\frac{m_{1,g}^3+m_{0,g}^3}{n_{g}^3}\leq 128.
     \end{split}
 \end{equation*}

\subsection{Variance bounds}\label{A:varbounds}
For $\mathbb{V}(B.2a.1|Q,X)$ in \eqref{eq:IIa1}, we need to bound terms of type
\begin{equation}
r_{n}^{-2}\sum_{i_{1},\ldots,i_{12}}|A_{i_{1}i_{2}}||A_{i_{3}i_{4}}||A_{i_{5}i_{6}}||A_{i_{7}i_{8}}|L_{i_{9}i_{10}}L_{i_{11}i_{12}}.
\end{equation}
We list these terms in Table~\ref{tab:termsv1} and bound them by $cr_{n}^{-2}G$.

For $\mathbb{V}(B.2a.2|Q,X)$ in \eqref{eq:IIa2}, we need to bound terms of type
\begin{equation}    r_{n}^{-2}\sum_{i_{1},\ldots,i_{16}}|A_{i_{1}i_{2}}||A_{i_{3}i_{4}}||A_{i_{5}i_{6}}||A_{i_{7}i_{8}}|L_{i_{9}i_{10}}L_{i_{11}i_{12}}L_{i_{13}i_{14}}L_{i_{15}i_{16}}.
\end{equation}
We list these terms in Table~\ref{tab:termsv2} and \ref{tab:termsv3} and bound them by $cr_{n}^{-2}G$.

The variance in \eqref{eq:varZ2a} converges to zero almost surely by the bounds:
\begin{equation}
    \begin{split}
     &   \sum_{i,j,k}B_{ij}^2L_{ij}L_{jk}\leq \sum_{i,j}B_{ij}^2  =\text{tr}( B^2)\leq cG,\\
     &   \sum_{i,j,k}B_{ij}B_{ik}L_{ik}L_{jk}\leq c\sum_{i,j,k}B_{ij}L_{ik}L_{jk}=c\sum_{i,j}B_{ij}L_{ij} = c\iota'( B\odot L)\iota\leq cG,\\
      &  \sum_{i,j,k}B_{ij}^2L_{jk}^2 \leq \sum_{i,j}B_{ij}^2=\text{tr}( B^2)\leq cG,\\
      &  \sum_{i,j,k}B_{ij}B_{ik}L_{jk}^2\leq \text{tr}( B L B)\leq \text{tr}( B^2)\leq cG,\\
       & \sum_{i,j,k}B_{ij}B_{jk}L_{ik}L_{jk}\leq c\sum_{i,j,k}B_{ij}L_{ik}L_{jk}=c\iota'( B\odot L) \iota\leq cG,\\
       & \sum_{i,j,k}B_{ik}B_{jk}L_{ik}L_{jk} = \iota'( B\odot L)^2\iota\leq cG.
    \end{split}
\end{equation}

Likewise, the variance in \eqref{eq:varZ2b} converges to zero almost surely by the bounds:
\begin{align*}
&\sum_{i_{1},\ldots,i_{5}}
B_{i_{1}i_{4}}B_{i_{1}i_{5}}L_{i_{2}i_{4}}L_{i_{2}i_{5}}L_{i_{3}i_{4}}L_{i_{3}i_{5}}=\sum_{i_{1},i_4,i_{5}}
B_{i_{1}i_{4}}B_{i_{1}i_{5}}L_{i_{4}i_{5}}^2\leq \text{tr}( BL B)\leq c\text{tr}( B^2)\leq cG,\\
&\sum_{i_{1},\ldots,i_{4}}B_{i_{1}i_{3}}B_{i_{2}i_{4}}L_{i_{1}i_{3}}L_{i_{1}i_{4}}L_{i_{2}i_{3}}L_{i_{2}i_{4}}=\text{tr}(( B\odot L)L( B\odot L)L)\leq \iota'( B\odot L)^2\iota\leq cG,\\
&\sum_{i_{1},\ldots,i_{5}}B_{i_{1}i_{4}}B_{i_{2}i_{5}}L_{i_{1}i_{4}}L_{i_{2}i_{5}}L_{i_{3}i_{4}}L_{i_{3}i_{5}}
\leq \iota'( B\odot L)L( B\odot L)\iota\leq \iota'( B\odot L)^2\iota\leq cG,\\
&\sum_{i_{1},\ldots,i_{5}}B_{i_{1}i_{4}}B_{i_{2}i_{5}}L_{i_{1}i_{5}}L_{i_{2}i_{4}}L_{i_{3}i_{4}}L_{i_{3}i_{5}}\leq 
\text{tr}( BL BL)\leq \text{tr}( B^2)\leq cG.
\end{align*}


\begin{table}[ht]
\fontsize{10}{7.2}\selectfont
\centering
\caption{Terms needed to bound $\mathbb{V}(B.2a.1|Q,X)$ in \eqref{eq:IIa1}}\label{tab:termsv1}
\begin{tabular}{L@{}LL@{}L@{}L@{}L}
\hline
 && \text{Expression} && \text{Bound}\\
\hline
1&&\sum_{i_{1}..,i_{3}} B_{i_{1}i_{2}}^{3}B_{i_{1}i_{3}}L_{i_{1}i_{2}}L_{i_{1}i_{3}}&\leq c\sum_{i_{1}..i_{3}} B_{i_{1}i_{2}}B_{i_{1}i_{3}}L_{i_{1}i_{2}}L_{i_{1}i_{3}} &\leq c\iota'( B\odot L)^2\iota,\\
2&&\sum_{i_{1}..i_{3}}B_{i_{1}i_{2}}^{2}B_{i_{1}i_{3}}^{2}L_{i_{1}i_{2}}L_{i_{1}i_{3}}&\leq \sum_{i_{1}..i_{3}}B_{i_{1}i_{2}}B_{i_{1}i_{3}}L_{i_{1}i_{2}}L_{i_{1}i_{3}}&\leq c\iota'( B\odot  L)^2\iota,\\
3&&\sum_{i_{1}..i_{3}}B_{i_{1}i_{2}}B_{i_{1}i_{3}}^{3}L_{i_{1}i_{2}}^{2}&\leq c\sum_{i_{1}..i_{3}}B_{i_{1}i_{3}}^{2}L_{i_{1}i_{2}}^{2}&\leq c\text{tr}( B^2),\\
4&&\sum_{i_{1}..i_{4}}B_{i_{1}i_{2}}B_{i_{1}i_{3}}B_{i_{1}i_{4}}^{2}L_{i_{1}i_{2}}L_{i_{1}i_{3}}&\leq c\sum_{i_{1}..i_{3}}B_{i_{1}i_{2}}B_{i_{1}i_{3}} e_{i_{1}}' B^2 e_{i_{1}}L_{i_{1}i_{2}}L_{i_{1}i_{3}}
&\leq c\iota'( B\odot L)^2 \iota,\\
5&&\sum_{i_{1}..i_{4}}B_{i_{1}i_{3}}^{2}B_{i_{1}i_{4}}^{2}L_{i_{1}i_{2}}^{2}&\leq c\sum_{i_{1},i_3,i_4}B_{i_{1}i_{3}}^{2}B_{i_{1}i_{4}}^{2}&\leq c\iota'(B\odot B)^2\iota,\\
6&&\sum_{i_{1}..i_{3}}B_{i_{1}i_{2}}^{3}B_{i_{1}i_{3}}L_{i_{1}i_{2}}L_{i_{2}i_{3}}&\leq c\sum_{i_{1}..i_{3}}B_{i_{1}i_{3}}L_{i_{1}i_{2}}L_{i_{2}i_{3}}&\leq c  \iota'( B\odot L)\iota,\\
7&& \sum_{i_{1}..i_{3}}B_{i_{1}i_{2}}^{2}B_{i_{1}i_{3}}^{2}L_{i_{1}i_{2}}L_{i_{2}i_{3}}&\leq \sum_{i_{1}..i_{3}}B_{i_{1}i_{3}}L_{i_{1}i_{2}}L_{i_{2}i_{3}}&\leq c \iota'( B\odot L)\iota,\\
8&& \sum_{i_{1}..i_{3}}B_{i_{1}i_{2}}B_{i_{1}i_{3}}^{3}L_{i_{1}i_{2}}L_{i_{2}i_{3}}&\leq c\sum_{i_{1}..i_{3}}B_{i_{1}i_{3}}L_{i_{1}i_{2}}L_{i_{2}i_{3}}&\leq c \iota'( B\odot L)\iota,\\
9&& \sum_{i_{1}..i_{3}}B_{i_{1}i_{3}}^{4}L_{i_{1}i_{2}}L_{i_{2}i_{3}}&\leq \sum_{i_{1}..i_{3}}B_{i_{1}i_{3}}L_{i_{1}i_{2}}L_{i_{2}i_{3}}&\leq c  \iota'( B\odot L)\iota,\\
10&&\sum_{i_{1}..i_{3}}B_{i_{1}i_{2}}^{2}B_{i_{1}i_{3}}B_{i_{2}i_{3}}L_{i_{1}i_{2}}^{2}&\leq \sum_{i_{1}..i_{3}}B_{i_{1}i_{3}}B_{i_{2}i_{3}}L_{i_{1}i_{2}}&\leq c\iota' ( B^2\odot  L)\iota.\\
11&&\sum_{i_{1}..i_{3}}B_{i_{1}i_{2}}B_{i_{1}i_{3}}^{2}B_{i_{2}i_{3}}L_{i_{1}i_{2}}^{2}&\leq \sum_{i_{1}..i_{3}}B_{i_{1}i_{3}}B_{i_{2}i_{3}}L_{i_{1}i_{2}}&\leq c\iota' ( B^2\odot  L)\iota,\\
12&&\sum_{i_{1}..i_{3}}B_{i_{1}i_{2}}B_{i_{1}i_{3}}^{2}B_{i_{2}i_{3}}L_{i_{1}i_{2}}L_{i_{1}i_{3}}&\leq \sum_{i_{1}..i_{3}}B_{i_{1}i_{3}}B_{i_{2}i_{3}}L_{i_{1}i_{2}}&\leq c\iota' ( B^2\odot  L)\iota,\\
13&&\sum_{i_{1}..i_{3}}B_{i_{1}i_{3}}^{3}B_{i_{2}i_{3}}L_{i_{1}i_{2}}L_{i_{1}i_{3}}&\leq \sum_{i_{1}..i_{3}}B_{i_{1}i_{3}}B_{i_{2}i_{3}}L_{i_{1}i_{2}}&\leq c\iota' ( B^2\odot  L)\iota,\\
14&&\sum_{i_{1}..i_{3}}B_{i_{1}i_{2}}^{2}B_{i_{1}i_{3}}B_{i_{2}i_{3}}L_{i_{1}i_{2}}L_{i_{2}i_{3}}&\leq \sum_{i_{1}..i_{3}}B_{i_{1}i_{2}}B_{i_{2}i_{3}}L_{i_{1}i_{2}}L_{i_{2}i_{3}}&\leq c\iota'( B\odot L)^2\iota,\\
15&&\sum_{i_{1}..i_{4}}B_{i_{1}i_{2}}B_{i_{1}i_{3}}B_{i_{1}i_{4}}B_{i_{2}i_{4}}L_{i_{1}i_{2}}L_{i_{2}i_{3}}&=\sum_{i_{1}..i_{3}}B_{i_{1}i_{2}}B_{i_{1}i_{3}} e_{i_{1}}'( B^2\odot L) e_{i_2}L_{i_{2}i_{3}}& \leq c\iota'( B^2\odot L)\iota.\\
16&&\sum_{i_{1}..i_{4}}B_{i_{1}i_{2}}B_{i_{1}i_{4}}^{2}B_{i_{2}i_{3}}L_{i_{1}i_{2}}L_{i_{2}i_{3}}&\leq \sum_{i_{1}..i_{3}}B_{i_{1}i_{2}} e_{i_{1}}' B^2 e_{i_{1}}B_{i_{2}i_{3}}L_{i_{1}i_{2}}L_{i_{2}i_{3}}&\leq c\iota'( B\odot L)^2\iota\\
17&&\sum_{i_{1}..i_{3}}B_{i_{1}i_{3}}^{3}B_{i_{2}i_{3}}L_{i_{1}i_{2}}L_{i_{2}i_{3}}&\leq \sum_{i_{1}..i_{3}}B_{i_{1}i_{3}}B_{i_{2}i_{3}}L_{i_{1}i_{2}}&\leq c\iota'( B^2\odot  L)\iota. \\
18&&\sum_{i_{1}..i_{4}}B_{i_{1}i_{3}}^{2}B_{i_{1}i_{4}}B_{i_{2}i_{3}}L_{i_{1}i_{2}}L_{i_{3}i_{4}}&\leq\sum_{i_{1}..i_{3}}B_{i_{1}i_{3}}B_{i_{2}i_{3}}L_{i_{1}i_{2}} e_{i_{1}}'( B L) e_{i_{3}}&\leq c\iota'( B^2\odot  L)\iota. \\
19&&\sum_{i_{1}..i_{4}}B_{i_{1}i_{3}}^{2}B_{i_{1}i_{4}}B_{i_{3}i_{4}}L_{i_{1}i_{2}}L_{i_{2}i_{3}} &\leq \sum_{i_{1},i_{3},i_{4}}B_{i_{1}i_{4}}B_{i_{3}i_{4}}L_{i_{1}i_{3}}& \leq c\iota'( B^2\odot  L)\iota.\\
20&&\sum_{i_{1}..i_{4}}B_{i_{1}i_{2}}B_{i_{1}i_{3}}^{2}B_{i_{3}i_{4}}L_{i_{1}i_{2}}L_{i_{3}i_{4}}&= \iota '( B\odot L) B( B\odot L)\iota&\leq c\iota'( B\odot L)^2\iota,\\
21&&\sum_{i_{1}..i_{3}}B_{i_{1}i_{2}}B_{i_{1}i_{3}}B_{i_{2}i_{3}}^{2}L_{i_{1}i_{2}}^{2}&\leq \sum_{i_{1}..i_{3}}B_{i_{2}i_{3}}^{2}L_{i_{1}i_{2}}^{2}&\leq c\text{tr}( B^2),\\
22&&\sum_{i_{1}..i_{3}}B_{i_{1}i_{2}}B_{i_{1}i_{3}}B_{i_{2}i_{3}}^{2}L_{i_{1}i_{2}}L_{i_{1}i_{3}}&\leq \sum_{i_{1}..i_{3}}B_{i_{2}i_{3}}^{2}L_{i_{1}i_{2}}L_{i_{1}i_{3}}&\leq c\text{tr}( B^2),&\\
23&&\sum_{i_{1}..i_{3}}B_{i_{1}i_{3}}^{2}B_{i_{2}i_{3}}^{2}L_{i_{1}i_{2}}L_{i_{1}i_{3}}&\leq c\sum_{i_{1}..i_{3}}B_{i_{2}i_{3}}^{2}L_{i_{1}i_{2}}L_{i_{1}i_{3}}&\leq c\text{tr}( B^2),\\
24&&\sum_{i_{1}..i_{4}}B_{i_{1}i_{3}}^{2}B_{i_{2}i_{4}}^{2}L_{i_{1}i_{2}}^{2}& = \sum_{i_{1},i_{2}} e_{i_1}' B^2 e_{i_1} e_{i_2}' B^2  e_{i_{2}}L_{i_{1}i_{2}}^2&\leq c\text{tr}( B^2),\\
25&&\sum_{i_{1}..i_{4}}B_{i_{1}i_{3}}B_{i_{1}i_{4}}B_{i_{2}i_{3}}B_{i_{2}i_{4}}L_{i_{1}i_{2}}^{2}&\leq \sum_{i_{1},i_{2}}( e_{i_{1}}' B^2 e_{i_{2}})^2L_{i_{1}i_{2}}^2 & \leq c\iota'( B^2\odot L)\iota.\\
26&&\sum_{i_{1}..i_{4}}B_{i_{1}i_{3}}B_{i_{1}i_{4}}B_{i_{2}i_{3}}B_{i_{3}i_{4}}L_{i_{1}i_{2}}L_{i_{1}i_{3}}
&\leq \sum_{i_{1}..i_{3}}B_{i_{2}i_{3}}L_{i_{1}i_{2}}L_{i_{1}i_{3}}&\leq c\iota'( B\odot L)\iota,\\
27&&\sum_{i_{1}..i_{4}}B_{i_{1}i_{2}}B_{i_{1}i_{3}}B_{i_{3}i_{4}}^{2}L_{i_{1}i_{2}}L_{i_{1}i_{3}}&\leq c\sum_{i_{1}..i_{3}}B_{i_{1}i_{2}}B_{i_{1}i_{3}}L_{i_{1}i_{2}}L_{i_{1}i_{3}}&\leq c\iota'( B\odot L)^2\iota.\vspace{0.1cm}\\
\hline
\end{tabular}
\end{table}


\begin{table}
\fontsize{10}{7.2}\selectfont
\centering
\caption{Terms needed to bound $\mathbb{V}(B.2a.2|Q,X)$ in \eqref{eq:IIa2}}\label{tab:termsv2}
\rotatebox{90}{\begin{tabular}{L@{}L@{}L@{}L@{}L@{}L}
\hline
& \text{Expression}&&\text{Bound}\\
\hline
1		&\sum_{i_{1}..i_{4}}B_{i_{1}i_{2}}B_{i_{1}i_{3}}B_{i_{2}i_{4}}^{2}L_{i_{1}i_{2}}L_{i_{1}i_{3}}L_{i_{2}i_{4}}L_{i_{3}i_{4}}&	\leq c\sum_{i_{1}..i_{4}}B_{i_{1}i_{2}}B_{i_{2}i_{4}}L_{i_{1}i_{2}}L_{i_{1}i_{3}}L_{i_{2}i_{4}}L_{i_{3}i_{4}}&\leq c\iota'( B\odot L)^2\iota,\\
2&\sum_{i_{1}..i_{5}}B_{i_{1}i_{4}}^{2}B_{i_{2}i_{5}}^{2}L_{i_{1}i_{2}}L_{i_{1}i_{3}}L_{i_{2}i_{5}}L_{i_{3}i_{5}}	&\leq \sum_{i_{1},i_{2},i_{4},i_{5}}B_{i_{1}i_{4}}^{2}B_{i_{2}i_{5}}^{2}L_{i_{1}i_{2}}L_{i_{1}i_{5}}L_{i_{2}i_{5}}&\leq c\text{tr}( B^2),\\
3	&\sum_{i_{1}..i_{5}}B_{i_{1}i_{3}}B_{i_{1}i_{4}}B_{i_{2}i_{5}}^{2}L_{i_{1}i_{2}}L_{i_{1}i_{3}}L_{i_{2}i_{5}}L_{i_{4}i_{5}}&\leq c\sum_{i_{1}..i_{5}}B_{i_{1}i_{3}}B_{i_{1}i_{4}}L_{i_{1}i_{2}}L_{i_{1}i_{3}}L_{i_{2}i_{5}}L_{i_{4}i_{5}}
&\leq c\iota'( B\odot L)^2\iota,\\
4&\sum_{i_{1}..i_{5}}B_{i_{1}i_{3}}B_{i_{1}i_{4}}B_{i_{2}i_{5}}^{2}L_{i_{1}i_{2}}L_{i_{1}i_{3}}L_{i_{3}i_{5}}L_{i_{4}i_{5}}&\leq c\sum_{i_{1}..i_{5}}B_{i_{1}i_{3}}B_{i_{1}i_{4}}B_{i_{2}i_{5}}L_{i_{1}i_{2}}L_{i_{1}i_{3}}L_{i_{3}i_{5}}L_{i_{4}i_{5}}&\leq c\text{tr}(B),\\
5&\sum_{i_{1}..i_{5}}B_{i_{1}i_{4}}^{2}B_{i_{2}i_{5}}^{2}L_{i_{1}i_{2}}L_{i_{1}i_{3}}L_{i_{3}i_{5}}L_{i_{4}i_{5}}&=\sum_{i_{1},i_{2},i_{4},i_{5}}B_{i_{1}i_{4}}^{2}B_{i_{2}i_{5}}^{2}L_{i_{1}i_{2}}L_{i_{1}i_{5}}L_{i_{4}i_{5}}&\leq c\iota'( B\odot L)\iota,	\\
6	&\sum_{i_{1}..i_{5}}B_{i_{1}i_{4}}^{2}B_{i_{2}i_{5}}B_{i_{3}i_{5}}L_{i_{1}i_{2}}L_{i_{1}i_{3}}L_{i_{2}i_{5}}L_{i_{3}i_{5}}	&\leq \sum_{i_{1}..i_{3},i_5} e_{i_{1}}' B^2 e_{i_{1}}B_{i_{2}i_{5}}B_{i_{3}i_{5}}L_{i_{1}i_{2}}L_{i_{1}i_{3}}L_{i_{2}i_{5}}L_{i_{3}i_{5}} &\leq c\iota'( B\odot L)^2\iota,\\
7	&\sum_{i_{1}..i_{5}}B_{i_{1}i_{3}}B_{i_{1}i_{4}}B_{i_{2}i_{5}}B_{i_{3}i_{5}}L_{i_{1}i_{2}}L_{i_{1}i_{3}}L_{i_{2}i_{5}}L_{i_{4}i_{5}}&	\leq c\sum_{i_{1}..i_{5}}B_{i_{1}i_{3}}B_{i_{1}i_{4}}L_{i_{1}i_{2}}L_{i_{1}i_{3}}L_{i_{2}i_{5}}L_{i_{4}i_{5}}&\leq c\iota'( B\odot L)^2\iota\\
8&\sum_{i_{1}..i_{5}}B_{i_{1}i_{4}}^{2}B_{i_{2}i_{5}}B_{i_{3}i_{5}}L_{i_{1}i_{2}}L_{i_{1}i_{3}}L_{i_{2}i_{5}}L_{i_{4}i_{5}}&\leq c	\sum_{i_{1},i_{2},i_5}( e_{i_{1}}'( B L) e_{i_{5}})^2 B_{i_{2}i_{5}}L_{i_{1}i_{2}}L_{i_{2}i_{5}}&\leq c\text{tr}( B^2), \\
9&\sum_{i_{1}..i_{5}}B_{i_{1}i_{2}}B_{i_{1}i_{4}}B_{i_{2}i_{5}}B_{i_{3}i_{5}}L_{i_{1}i_{2}}L_{i_{1}i_{3}}L_{i_{3}i_{5}}L_{i_{4}i_{5}}&\leq c \sum_{i_{1}..i_{3},i_5}B_{i_{1}i_{2}}B_{i_{2}i_{5}}B_{i_{3}i_{5}}L_{i_{1}i_{2}}L_{i_{1}i_{3}}L_{i_{3}i_{5}}	&\leq c\iota'( B\odot L)\iota,\\
10&\sum_{i_{1}..i_{5}}B_{i_{1}i_{4}}^{2}B_{i_{2}i_{5}}B_{i_{3}i_{5}}L_{i_{1}i_{2}}L_{i_{1}i_{3}}L_{i_{3}i_{5}}L_{i_{4}i_{5}}&\leq \sum_{i_{2},i_5}B_{i_{2}i_{5}}L_{i_{2}i_{5}}&\leq c\iota'( B\odot L)\iota,	\\
11&\sum_{i_{1}..i_{6}}B_{i_{1}i_{4}}B_{i_{1}i_{5}}B_{i_{2}i_{6}}B_{i_{3}i_{6}}L_{i_{1}i_{2}}L_{i_{1}i_{3}}L_{i_{4}i_{6}}L_{i_{5}i_{6}}&\leq \sum_{i_{1},i_6}( e_{i_{1}}' B L e_{i_{6}})^4 &\leq c\text{tr}( B^2),	\\
12	&\sum_{i_{1}..i_{5}}B_{i_{1}i_{3}}B_{i_{1}i_{4}}B_{i_{2}i_{5}}B_{i_{4}i_{5}}L_{i_{1}i_{2}}L_{i_{1}i_{3}}L_{i_{2}i_{5}}L_{i_{3}i_{5}}&\leq c\sum_{i_{1}..i_{3},i_5}B_{i_{1}i_{3}}B_{i_{2}i_{5}}L_{i_{1}i_{2}}L_{i_{1}i_{3}}L_{i_{2}i_{5}}L_{i_{3}i_{5}}	&\leq c\iota'( B\odot L)^2\iota,\\
13&\sum_{i_{1}..i_{5}}B_{i_{1}i_{4}}^{2}B_{i_{2}i_{5}}B_{i_{4}i_{5}}L_{i_{1}i_{2}}L_{i_{1}i_{3}}L_{i_{2}i_{5}}L_{i_{3}i_{5}}	&\leq c\sum_{i_{1}..i_{3},i_5}B_{i_{2}i_{5}}L_{i_{1}i_{2}}L_{i_{1}i_{3}}L_{i_{2}i_{5}}L_{i_{3}i_{5}}&\leq c\iota'( B\odot L) \iota,\\
14	&\sum_{i_{1}..i_{5}}B_{i_{1}i_{3}}^{2}B_{i_{2}i_{4}}B_{i_{4}i_{5}}L_{i_{1}i_{2}}L_{i_{1}i_{3}}L_{i_{2}i_{4}}L_{i_{4}i_{5}}&\leq c\sum_{i_{1}..i_{3},i_5} e_{i_{2}}'( B\odot L)^2 e_{i_{5}}L_{i_{1}i_{2}}L_{i_{1}i_{3}}	& \leq c\iota' ( B\odot L)^2\iota \\
15	&	\sum_{i_{1}..i_{5}}B_{i_{1}i_{3}}B_{i_{1}i_{4}}B_{i_{2}i_{5}}B_{i_{4}i_{5}}L_{i_{1}i_{2}}L_{i_{1}i_{3}}L_{i_{2}i_{5}}L_{i_{4}i_{5}}&\leq c\sum_{i_{2},i_{3},i_{4},i_{5}} e_{i_2}'( B\odot L) e_{i_{5}} e_{i_{5}}'( B\odot L) e_{i_{4}} L_{i_{2}i_{3}}&\leq c\iota( B\odot L)^2\iota,\\
16	&\sum_{i_{1}..i_{5}}B_{i_{1}i_{2}}^{2}B_{i_{2}i_{4}}B_{i_{4}i_{5}}L_{i_{1}i_{2}}L_{i_{1}i_{3}}L_{i_{3}i_{4}}L_{i_{4}i_{5}}&\leq c\sum_{i_{2},i_{4},i_{5}}B_{i_{2}i_{4}}L_{i_{2}i_{4}}B_{i_{4}i_{5}}L_{i_{4}i_{5}}&\leq c\iota ( B\odot L)^2\iota,	\\
17&\sum_{i_{1}..i_{5}}B_{i_{1}i_{2}}B_{i_{1}i_{3}}B_{i_{2}i_{4}}B_{i_{4}i_{5}}L_{i_{1}i_{2}}L_{i_{1}i_{3}}L_{i_{3}i_{4}}L_{i_{4}i_{5}}	&\leq c\sum_{i_{2},i_{4},i_{5}}B_{i_{2}i_{4}}L_{i_{2}i_{4}}B_{i_{4}i_{5}}L_{i_{4}i_{5}}&\leq c\iota ( B\odot L)^2\iota, \\
18&	\sum_{i_{1}..i_{5}}B_{i_{1}i_{2}}B_{i_{1}i_{4}}B_{i_{2}i_{5}}B_{i_{4}i_{5}}L_{i_{1}i_{2}}L_{i_{1}i_{3}}L_{i_{3}i_{5}}L_{i_{4}i_{5}}&\leq c\sum_{i_{2},i_{4},i_{5}}B_{i_{2}i_{4}}L_{i_{2}i_{4}}B_{i_{4}i_{5}}L_{i_{4}i_{5}}&\leq c\iota ( B\odot L)^2\iota,	\\	
19	&	\sum_{i_{1}..i_{5}}B_{i_{1}i_{3}}B_{i_{1}i_{4}}B_{i_{2}i_{5}}B_{i_{4}i_{5}}L_{i_{1}i_{2}}L_{i_{1}i_{3}}L_{i_{3}i_{5}}L_{i_{4}i_{5}}&\leq c\sum_{i_{2},i_{4},i_{5}}B_{i_{2}i_{4}}L_{i_{2}i_{4}}B_{i_{4}i_{5}}L_{i_{4}i_{5}}&\leq c\iota ( B\odot L)^2\iota,	\\	
20&\sum_{i_{1}..i_{6}}B_{i_{1}i_{4}}B_{i_{1}i_{5}}B_{i_{2}i_{6}}B_{i_{4}i_{6}}L_{i_{1}i_{2}}L_{i_{1}i_{3}}L_{i_{3}i_{6}}L_{i_{5}i_{6}}&\leq c \sum_{i_{2},i_{4},i_{5},i_{6}}B_{i_{2}i_{6}}B_{i_{4}i_{6}}L_{i_{2}i_{6}}L_{i_{5}i_{6}}&\leq  C\iota'( B\odot  L) B\iota,\\
21	&	\sum_{i_{1}..i_{6}}B_{i_{1}i_{4}}B_{i_{1}i_{5}}B_{i_{2}i_{6}}B_{i_{5}i_{6}}L_{i_{1}i_{2}}L_{i_{1}i_{3}}L_{i_{3}i_{6}}L_{i_{4}i_{6}}&\leq c \sum_{i_{2},i_{4},i_{5},i_{6}}B_{i_{2}i_{6}}B_{i_{4}i_{6}}L_{i_{2}i_{6}}L_{i_{5}i_{6}}& \leq  C\iota'( B\odot  L) B\iota, \\
22&\sum_{i_{1}..i_{6}}B_{i_{1}i_{4}}^{2}B_{i_{2}i_{5}}B_{i_{5}i_{6}}L_{i_{1}i_{2}}L_{i_{1}i_{3}}L_{i_{3}i_{5}}L_{i_{5}i_{6}}&	\leq c \sum_{i_{2},i_{5},i_{6}}B_{i_{2}i_{5}}B_{i_{5}i_{6}}L_{i_{2}i_{5}}L_{i_{5}i_{6}}&\leq  C\iota'( B\odot  L)^2\iota\\
23&\sum_{i_{1}..i_{6}}B_{i_{1}i_{3}}B_{i_{1}i_{4}}B_{i_{2}i_{5}}B_{i_{5}i_{6}}L_{i_{1}i_{2}}L_{i_{1}i_{3}}L_{i_{4}i_{5}}L_{i_{5}i_{6}}&\leq c\sum_{i_{2},i_{3},i_{4},i_{5},i_{6}}B_{i_{2}i_{5}}B_{i_{5}i_{6}}L_{i_{2}i_{3}}L_{i_{4}i_{5}}L_{i_{5}i_{6}}	&\leq c\iota' B( B\odot L)\iota,\\
24&\sum_{i_{1}..i_{5}}B_{i_{1}i_{4}}^{2}B_{i_{3}i_{5}}^{2}L_{i_{1}i_{2}}L_{i_{1}i_{3}}L_{i_{2}i_{5}}L_{i_{3}i_{5}}&\leq c\sum_{i_{1},i_{3},i_{4},i_{5}}B_{i_{1}i_{4}}B_{i_{3}i_{5}}L_{i_{1}i_{5}}L_{i_{3}i_{5}}&\leq c\iota' B( B\odot L) \iota,	\\
25	&\sum_{i_{1}..i_{5}}B_{i_{1}i_{4}}^{2}B_{i_{3}i_{5}}^{2}L_{i_{1}i_{2}}L_{i_{1}i_{3}}L_{i_{2}i_{5}}L_{i_{4}i_{5}}&\leq \sum_{i_{1},i_{3},i_{4},i_{5}}B_{i_{1}i_{4}}^{2}B_{i_{3}i_{5}}^{2}L_{i_{1}i_{5}}L_{i_{1}i_{3}}L_{i_{4}i_{5}} 	&	\leq c\text{tr}( B^2),\\
26	&	\sum_{i_{1}..i_{5}}B_{i_{1}i_{2}}B_{i_{1}i_{4}}B_{i_{3}i_{5}}^{2}L_{i_{1}i_{2}}L_{i_{1}i_{3}}L_{i_{3}i_{5}}L_{i_{4}i_{5}}&\leq c\sum_{i_{1},i_{3},i_{4},i_{5}}B_{i_{1}i_{4}}B_{i_{3}i_{5}}L_{i_{1}i_{3}}L_{i_{3}i_{5}}L_{i_{4}i_{5}}
&\leq c\text{tr}( B^2).\vspace{0.1cm}	\\
\hline
\end{tabular}}
\end{table}

\begin{table}
\fontsize{10}{7.2}\selectfont
\centering
\caption{Terms needed to bound $\mathbb{V}(B.2a.2|Q,X)$ in \eqref{eq:IIa2} -- continued from Table~\ref{tab:termsv2}}\label{tab:termsv3}
\rotatebox{90}{\begin{tabular}{LLLLLL}
   \hline
& \text{Expression}&&\text{Bound}\\
\hline
27&\sum_{i_{1}..i_{5}}B_{i_{1}i_{2}}B_{i_{1}i_{4}}B_{i_{3}i_{5}}B_{i_{4}i_{5}}L_{i_{1}i_{2}}L_{i_{1}i_{3}}L_{i_{2}i_{5}}L_{i_{3}i_{5}}&\leq c\sum_{i_{1},i_{2},i_{5}}B_{i_{1}i_{2}} e_{i_{1}}' B^2 e_{i_{5}}L_{i_{1}i_{2}}L_{i_{1}i_{5}}^2&\leq \leq c\iota'( B\odot L) B\iota,	\\
28&\sum_{i_{1}..i_{5}}B_{i_{1}i_{4}}^{2}B_{i_{3}i_{5}}B_{i_{4}i_{5}}L_{i_{1}i_{2}}L_{i_{1}i_{3}}L_{i_{2}i_{5}}L_{i_{3}i_{5}}&\leq 	C\sum_{i_{1},i_{3},i_{5}}B_{i_{1}i_{5}}B_{i_{3}i_{5}}L_{i_{1}i_{5}}L_{i_{3}i_{5}}&\leq c\iota'( B\odot  L)^2\iota,\\
29&\sum_{i_{1}..i_{5}}B_{i_{1}i_{2}}B_{i_{1}i_{3}}B_{i_{3}i_{4}}B_{i_{4}i_{5}}L_{i_{1}i_{2}}L_{i_{1}i_{3}}L_{i_{2}i_{4}}L_{i_{4}i_{5}}&\leq \sum_{i_{1}..i_{5}}B_{i_{3}i_{4}}B_{i_{4}i_{5}}L_{i_{1}i_{2}}L_{i_{1}i_{3}}L_{i_{2}i_{4}}L_{i_{4}i_{5}}&\leq \iota'(B\odot L)^2\iota\\
30	&	\sum_{i_{1}..i_{5}}B_{i_{1}i_{2}}B_{i_{1}i_{4}}B_{i_{3}i_{5}}B_{i_{4}i_{5}}L_{i_{1}i_{2}}L_{i_{1}i_{3}}L_{i_{2}i_{5}}L_{i_{4}i_{5}}&\leq \sum_{i_{1}..i_{5}}B_{i_{3}i_{5}}B_{i_{4}i_{5}}L_{i_{1}i_{2}}L_{i_{1}i_{3}}L_{i_{2}i_{5}}L_{i_{4}i_{5}}&\leq \iota'(B\odot L)^2\iota\\	
31	&\sum_{i_{1}..i_{5}}B_{i_{1}i_{3}}B_{i_{1}i_{4}}B_{i_{3}i_{5}}B_{i_{4}i_{5}}L_{i_{1}i_{2}}L_{i_{1}i_{3}}L_{i_{2}i_{5}}L_{i_{4}i_{5}}	&\leq \sum_{i_{1}..i_{5}}B_{i_{3}i_{5}}B_{i_{4}i_{5}}L_{i_{1}i_{2}}L_{i_{1}i_{3}}L_{i_{2}i_{5}}L_{i_{4}i_{5}}	&\leq \iota'(B\odot L)^2\iota,\\
32	&	\sum_{i_{1}..i_{6}}B_{i_{1}i_{4}}B_{i_{1}i_{5}}B_{i_{3}i_{6}}B_{i_{4}i_{6}}L_{i_{1}i_{2}}L_{i_{1}i_{3}}L_{i_{2}i_{6}}L_{i_{5}i_{6}}	&\leq c\sum_{i_{1}..i_{6}}B_{i_{3}i_{6}}B_{i_{4}i_{6}}L_{i_{1}i_{2}}L_{i_{1}i_{3}}L_{i_{2}i_{6}}L_{i_{5}i_{6}}& \leq c\iota'( B\odot L) B\iota\\
33&\sum_{i_{1}..i_{5}}B_{i_{1}i_{2}}^{2}B_{i_{3}i_{4}}B_{i_{4}i_{5}}L_{i_{1}i_{2}}L_{i_{1}i_{3}}L_{i_{3}i_{4}}L_{i_{4}i_{5}}&\leq 	c\sum_{i_{1}..i_{5}}B_{i_{3}i_{4}}B_{i_{4}i_{5}}L_{i_{1}i_{2}}L_{i_{1}i_{3}}L_{i_{3}i_{4}}L_{i_{4}i_{5}}&\leq c\iota'( B\odot L)^2\iota,\\
34	&	\sum_{i_{1}..i_{5}}B_{i_{1}i_{2}}B_{i_{1}i_{4}}B_{i_{3}i_{5}}B_{i_{4}i_{5}}L_{i_{1}i_{2}}L_{i_{1}i_{3}}L_{i_{3}i_{5}}L_{i_{4}i_{5}}&\leq c\sum_{i_{1}..i_{5}}B_{i_{3}i_{5}}B_{i_{4}i_{5}}L_{i_{1}i_{2}}L_{i_{1}i_{3}}L_{i_{3}i_{5}}L_{i_{4}i_{5}}	& \leq c\iota'(B\odot L)^2\iota,\\
35	&\sum_{i_{1}..i_{6}}B_{i_{1}i_{4}}B_{i_{1}i_{5}}B_{i_{3}i_{6}}B_{i_{5}i_{6}}L_{i_{1}i_{2}}L_{i_{1}i_{3}}L_{i_{2}i_{6}}L_{i_{4}i_{6}}&\leq \sum_{i_{1},i_{3},i_{4},i_{5},i_6}B_{i_{1}i_{4}}B_{i_{1}i_{5}}B_{i_{3}i_{6}}B_{i_{5}i_{6}}L_{i_{1}i_{6}}L_{i_{1}i_{3}}L_{i_{4}i_{6}}	& \leq \iota'( B\odot L)\iota,\\
36&\sum_{i_{1}..i_{6}}B_{i_{1}i_{4}}^{2}B_{i_{3}i_{5}}B_{i_{5}i_{6}}L_{i_{1}i_{2}}L_{i_{1}i_{3}}L_{i_{2}i_{5}}L_{i_{5}i_{6}}&	\leq c\sum_{i_{1},i_{3},i_{4},i_{5},i_6}B_{i_{1}i_{4}}B_{i_{3}i_{5}}B_{i_{5}i_{6}}L_{i_{1}i_{3}}L_{i_{1}i_{5}}L_{i_{5}i_{6}}&\leq c\iota' B ( B\odot L)\iota,\\
37	&	\sum_{i_{1}..i_{6}}B_{i_{1}i_{2}}B_{i_{1}i_{4}}B_{i_{3}i_{5}}B_{i_{5}i_{6}}L_{i_{1}i_{2}}L_{i_{1}i_{3}}L_{i_{4}i_{5}}L_{i_{5}i_{6}}&\leq \sum_{i_{2},i_{3},i_{4},i_{5},i_6}B_{i_{3}i_{5}}B_{i_{5}i_{6}}L_{i_{2}i_{3}}L_{i_{4}i_{5}}L_{i_{5}i_{6}}
&\leq c\iota' B( B\odot L)\iota,\\
38&\sum_{i_{1}..i_{5}}B_{i_{1}i_{2}}^{2}B_{i_{4}i_{5}}^{2}L_{i_{1}i_{2}}L_{i_{1}i_{3}}L_{i_{2}i_{4}}L_{i_{3}i_{4}}&\leq c\sum_{i_{1},i_{2},i_{4},i_{5}}B_{i_{1}i_{2}}B_{i_{4}i_{5}}L_{i_{1}i_{2}}L_{i_{1}i_{4}}L_{i_{2}i_{4}}		& \leq c\iota'( B( B\odot L)\iota\\
39&\sum_{i_{1}..i_{5}}B_{i_{1}i_{2}}B_{i_{1}i_{3}}B_{i_{4}i_{5}}^{2}L_{i_{1}i_{2}}L_{i_{1}i_{3}}L_{i_{2}i_{4}}L_{i_{3}i_{4}}&	\leq c\sum_{i_{1}..i_{5}}B_{i_{1}i_{2}}B_{i_{1}i_{3}}B_{i_{4}i_{5}}L_{i_{1}i_{2}}L_{i_{1}i_{3}}L_{i_{2}i_{4}}
&\leq c\iota' B( B\odot L)\iota\\
40&\sum_{i_{1}..i_{5}}B_{i_{1}i_{2}}B_{i_{1}i_{4}}B_{i_{4}i_{5}}^{2}L_{i_{1}i_{2}}L_{i_{1}i_{3}}L_{i_{2}i_{5}}L_{i_{3}i_{5}}&\leq c\sum_{i_{1},i_{2},i_{4},i_{5}}B_{i_{1}i_{2}}B_{i_{1}i_{4}}B_{i_{4}i_{5}}^{2}L_{i_{1}i_{2}}L_{i_{1}i_{5}}L_{i_{2}i_{5}}	
&\leq c\iota' B( B\odot L)\iota,	\\
41&\sum_{i_{1}..i_{5}}B_{i_{1}i_{3}}B_{i_{1}i_{4}}B_{i_{4}i_{5}}^{2}L_{i_{1}i_{2}}L_{i_{1}i_{3}}L_{i_{2}i_{5}}L_{i_{3}i_{5}}&\leq c\sum_{i_{1},i_{3},i_{4},i_{5}}B_{i_{1}i_{3}}B_{i_{1}i_{4}}B_{i_{4}i_{5}}^{2}L_{i_{1}i_{5}}L_{i_{1}i_{3}}L_{i_{3}i_{5}}&\leq c\text{tr}(( B\odot L) L( B\odot L)),\\
42&\sum_{i_{1}..i_{5}}B_{i_{1}i_{2}}B_{i_{1}i_{3}}B_{i_{4}i_{5}}^{2}L_{i_{1}i_{2}}L_{i_{1}i_{3}}L_{i_{2}i_{4}}L_{i_{4}i_{5}}&	\leq c\sum_{i_{1}..i_{5}}B_{i_{1}i_{2}}B_{i_{1}i_{3}}B_{i_{4}i_{5}}L_{i_{1}i_{2}}L_{i_{1}i_{3}}L_{i_{2}i_{4}}L_{i_{4}i_{5}}& \leq c\iota'( B\odot L)^2\iota,\\
43	&	\sum_{i_{1}..i_{5}}B_{i_{1}i_{3}}^{2}B_{i_{4}i_{5}}^{2}L_{i_{1}i_{2}}L_{i_{1}i_{3}}L_{i_{2}i_{4}}L_{i_{4}i_{5}}&\leq c\sum_{i_{1},i_{3},i_{4},i_{5}}B_{i_{1}i_{3}}B_{i_{4}i_{5}}L_{i_{1}i_{4}}L_{i_{1}i_{3}}L_{i_{4}i_{5}}		&\leq c\iota'( B\odot L) L( B\odot L)\iota,\\
44&\sum_{i_{1}..i_{5}}B_{i_{1}i_{3}}B_{i_{1}i_{4}}B_{i_{4}i_{5}}^{2}L_{i_{1}i_{2}}L_{i_{1}i_{3}}L_{i_{2}i_{5}}L_{i_{4}i_{5}}&	\leq c\sum_{i_{1},i_{3},i_{4},i_{5}}B_{i_{1}i_{3}}B_{i_{1}i_{4}}B_{i_{4}i_{5}}L_{i_{1}i_{5}}L_{i_{1}i_{3}}L_{i_{4}i_{5}}&\leq c\iota'( B\odot  L) L( B\odot  L)\iota, \\
45&\sum_{i_{1}..i_{5}}B_{i_{1}i_{2}}^{2}B_{i_{4}i_{5}}^{2}L_{i_{1}i_{2}}L_{i_{1}i_{3}}L_{i_{3}i_{4}}L_{i_{4}i_{5}}&\leq \sum_{i_{1},i_{2},i_{4},i_{5}}B_{i_{1}i_{2}}B_{i_{4}i_{5}}L_{i_{1}i_{2}}L_{i_{1}i_{4}}L_{i_{4}i_{5}}	&\leq c\iota'( B\odot  L) L( B\odot  L)\iota,\\
46&\sum_{i_{1}..i_{5}}B_{i_{1}i_{2}}B_{i_{1}i_{3}}B_{i_{4}i_{5}}^{2}L_{i_{1}i_{2}}L_{i_{1}i_{3}}L_{i_{3}i_{4}}L_{i_{4}i_{5}}&\leq c\sum_{i_{1}..i_{3},i_{5}}B_{i_{1}i_{2}}B_{i_{1}i_{3}}B_{i_{3}i_{5}}L_{i_{1}i_{2}}L_{i_{1}i_{3}}&\leq c\iota' B( B\odot L)\iota,\\
47	&\sum_{i_{1}..i_{5}}B_{i_{1}i_{2}}B_{i_{1}i_{4}}B_{i_{4}i_{5}}^{2}L_{i_{1}i_{2}}L_{i_{1}i_{3}}L_{i_{3}i_{5}}L_{i_{4}i_{5}}&\leq c\sum_{i_{1},i_{2},i_{4},i_{5}}B_{i_{1}i_{2}}B_{i_{1}i_{4}}B_{i_{4}i_{5}}L_{i_{1}i_{2}}L_{i_{1}i_{5}}L_{i_{4}i_{5}}	& \leq c\iota' B( B\odot L)\iota,\\
48&\sum_{i_{1}..i_{6}}B_{i_{1}i_{4}}B_{i_{1}i_{5}}B_{i_{4}i_{6}}B_{i_{5}i_{6}}L_{i_{1}i_{2}}L_{i_{1}i_{3}}L_{i_{2}i_{6}}L_{i_{3}i_{6}}&\leq c\sum_{i_{1},i_{2},i_{6}}B_{i_{1}i_{6}}^3L_{i_{1}i_{2}}L_{i_{1}i_{6}}L_{i_{2}i_{6}}	&\leq c\iota'( B\odot L)\iota,\\
49&\sum_{i_{1}..i_{6}}B_{i_{1}i_{3}}B_{i_{1}i_{4}}B_{i_{4}i_{5}}B_{i_{5}i_{6}}L_{i_{1}i_{2}}L_{i_{1}i_{3}}L_{i_{2}i_{5}}L_{i_{5}i_{6}}&\leq c\sum_{i_{1},i_{3},i_{5},i_{6}}B_{i_{1}i_{3}}B_{i_{1}i_{5}}B_{i_{5}i_{6}}L_{i_{1}i_{5}}L_{i_{1}i_{3}}L_{i_{5}i_{6}}	& \leq c\iota'( B\odot L)^2\iota,\\
50&\sum_{i_{1}..i_{6}}B_{i_{1}i_{2}}B_{i_{1}i_{4}}B_{i_{4}i_{5}}B_{i_{5}i_{6}}L_{i_{1}i_{2}}L_{i_{1}i_{3}}L_{i_{3}i_{5}}L_{i_{5}i_{6}}&\leq \sum_{i_{1},i_{2},i_{5}}B_{i_{1}i_{2}}B_{i_{1}i_{5}}L_{i_{1}i_{2}}L_{i_{1}i_{5}}	& \leq c\iota'( B\odot L)\iota,\\
51&\sum_{i_{1}..i_{6}}B_{i_{1}i_{4}}^{2}B_{i_{5}i_{6}}^{2}L_{i_{1}i_{2}}L_{i_{1}i_{3}}L_{i_{2}i_{5}}L_{i_{3}i_{5}}&\leq \sum_{i_{1},i_{4},i_{5},i_{6}}B_{i_{1}i_{4}}B_{i_{5}i_{6}}L_{i_{1}i_{5}}^2&\leq c\iota' B( L\odot  L) B\iota,\\
52&\sum_{i_{1}..i_{6}}B_{i_{1}i_{3}}B_{i_{1}i_{4}}B_{i_{5}i_{6}}^{2}L_{i_{1}i_{2}}L_{i_{1}i_{3}}L_{i_{2}i_{5}}L_{i_{4}i_{5}}&\leq c \sum_{i_{1},i_{3},i_{4}}B_{i_{1}i_{3}}B_{i_{1}i_{4}}L_{i_{1}i_{4}}L_{i_{1}i_{3}}&\leq c\iota'( B\odot L)^2\iota,\\
53&\sum_{i_{1}..i_{6}}B_{i_{1}i_{2}}B_{i_{1}i_{4}}B_{i_{5}i_{6}}^{2}L_{i_{1}i_{2}}L_{i_{1}i_{3}}L_{i_{3}i_{5}}L_{i_{4}i_{5}}&\leq c\sum_{i_{1},i_{2},i_{4}}B_{i_{1}i_{2}}B_{i_{1}i_{4}}L_{i_{1}i_{2}}L_{i_{1}i_{4}}&  \leq c\iota'( B\odot L)^2\iota.
\end{tabular}}
\end{table}


\subsection{Consistency of the variance bias correction}\label{supp:consistencycorrection}
Define 
\begin{align}
    \hat{\xi}_{11,ij}&=\frac{\alpha_{1,g}^2}{(\alpha_{1,g}+1)(\alpha_{1,g}-2)}\hat{\sigma}_{u,i}^2\hat{\sigma}_{v,j}^2 + \frac{\alpha_{1,g}(\alpha_{1,g}+2)}{(\alpha_{1,g}+1)(\alpha_{1,g}-2)}\hat{\sigma}_{uv,i}\hat{\sigma}_{uv,j}-u_{i}^2v_{j}^2-u_{i}v_{i}u_{j}v_{j},\notag\\
    \hat{\xi}_{00,ij}&=\frac{\alpha_{0,g}^2}{(\alpha_{0,g}+1)(\alpha_{0,g}-2)}\hat{\sigma}_{u,i}^2\hat{\sigma}_{v,j}^2 + \frac{\alpha_{0,g}(\alpha_{0,g}+2)}{(\alpha_{0,g}+1)(\alpha_{0,g}-2)}\hat{\sigma}_{uv,i}\hat{\sigma}_{uv,j}-u_{i}^2v_{j}^2-u_{i}v_{i}u_{j}v_{j},\notag\\
\hat{\xi}_{10,ij}&=\hat{\sigma}_{u,i}^2\hat{\sigma}_{v,j}^2 + \hat{\sigma}_{uv,i}\hat{\sigma}_{uv,j}-u_{i}^2v_{j}^2-u_{i}v_{i}u_{j}v_{j}. \label{eq:xi}
\end{align}
For consistency, we need to show that the following converges to zero. 
\begin{align}\label{eq:toshowconsistency}
&\mathbb{V}(r_{n}^{-1}(\hat{\Delta}_{b}-\Delta_{b})|Q,X)= \frac{1}{r_{n}^2}\sum_{g=1}^{G}(A_{11,g}^2+2C_{11,g})^2\mathbb{E}\bigg[\bigg(\sum_{i}\sum_{j\neq i}\hat{\xi}_{11,ij}Z_{ig}Z_{jg}\bigg)^2\big|Q,X\bigg] \notag\\
&\qquad+ (A_{00,g}^2+2C_{00,g})^2\mathbb{E}\bigg[\bigg(\sum_{i}\sum_{j\neq i}\hat{\xi}_{00,ij}Z_{ig}^{c}Z_{jg}^{c}W_{ig}W_{jg}\bigg)^2\big|Q,X\bigg]\notag\\
& \qquad + A_{10,g}^2\mathbb{E}\bigg[\bigg(\sum_{i}\sum_{j\neq i}\hat{\xi}_{10,ij}Z_{ig}Z_{jg}^{c}W_{jg}\bigg)^2\big|Q,X\bigg] \notag\\&\qquad +A_{01,g}^2\mathbb{E}\bigg[\bigg(\sum_{i}\sum_{j\neq i}\hat{\xi}_{10,ji}W_{ig}Z_{ig}^{c}Z_{jg}\bigg)^2\big|Q,X\bigg].
\end{align}
We will focus on the first term, where $\hat{\xi}_{11,ij}$ is mean zero by construction. Consider
\begin{align*}
&\sum_{i}\sum_{j\neq i}Z_{ig}Z_{jg}\hat{\sigma}_{uv,i}\hat{\sigma}_{uv,j} =\sum_{i}\sum_{j\neq i}Z_{ig}Z_{jg}
\bigg(\underbrace{u_{i}v_{i}}_{(a)} +\underbrace{ \frac{1}{m_{1,g}-1}u_{i}\sum_{j\neq i}Z_{jg}v_{j}}_{(b)}\label{eq:shatshat}\\& + \frac{1}{m_{1,g}-1}v_{i}\sum_{j\neq i}Z_{jg}u_{j}  + \underbrace{\frac{1}{(m_{1,g}-1)(m_{1,g}-2)}\sum_{l}\sum_{k\neq l}Z_{jg}Z_{kg}u_{j}v_{k})}_{(c)}\bigg)\notag\\
&\times \bigg(u_{j}v_{j} + \underbrace{\frac{1}{m_{1,g}-1}u_{j}\sum_{l\neq j}Z_{lg}v_{l}}_{(d)} + \frac{1}{m_{1,g}-1}v_{j}\sum_{l\neq j}Z_{lg}u_{l} + \underbrace{\frac{\sum\nolimits_{l}\sum\nolimits_{k\neq l}Z_{lg}Z_{kg}u_{l}v_{k}}{(m_{1,g}-1)(m_{1,g}-2)}}_{(e)}\bigg). \notag
\end{align*}
 We now subtract all terms with nonzero expectation and consider the following five cross-products, from which the others follow: $(I)\colon (a)\times (d)$, $(II)\colon (b)\times (d)$, $(III)\colon (a)\times (e)$, $(IV)\colon (b)\times (e)$ and $(V)\colon (c)\times (e)$.

Let $y_{ki}$ denote a random variable independent over $i=1,\ldots,N$. Let $k=1,\ldots,4$. For example, we could choose $y_{1i}=u_{i}$, $y_{2i}=u_i$, $y_{3i}=v_{i}$, $y_{4i}=v_i$, and $N=m_g$. After subtracting terms with nonzero expectation, we obtain terms of the following general structure.
\begin{equation}
\begin{split}
(I)&=\frac{1}{N}\sum_{i}\sum_{l\neq i}\sum_{j\neq l}y_{1i}y_{2i}y_{3l}y_{4j},\\
(II)&=\frac{1}{N^2}\sum_{i}\sum_{l\neq i}\sum_{k\neq i}\sum_{j\neq l}y_{1i}y_{2l}y_{3k}y_{4j}-\frac{1}{N^2}\sum_{i}\sum_{l\neq i}y_{1i}y_{2l}y_{3l}y_{4i},\\
(III)&=\frac{1}{N^2}\sum_{i}\sum_{l\neq i}\sum_j \sum_{k\neq j}y_{1i}y_{2i}y_{3j}y_{4k},\\
(IV)&=\frac{1}{N^3}\sum_{i}\sum_{l\neq i}\sum_{j\neq i}\sum_{k}\sum_{m\neq k}y_{1i}y_{2j}y_{3k}y_{4m}\\
&\quad -\frac{N-1}{N^3}\sum_{i}\sum_{j\neq i}(y_{1i}y_{2j}y_{3i}y_{4j}   +y_{1i}y_{2j}y_{3j}y_{4i}),\\
(V)&=\frac{1}{N^4}\sum_{i}\sum_{l\neq i}\sum_{j}\sum_{k\neq j}\sum_{m}\sum_{n\neq m}y_{1j}y_{2k}y_{3m}y_{4n}\\
&\quad -\frac{N-1}{N^3}\sum_{j}\sum_{k\neq j}(y_{1j}y_{2k}y_{3k}y_{4j}+y_{1j}y_{2k}y_{3j}y_{4j}).
\end{split}
\end{equation}
We will show that the variance of these terms is at most of order $N^2$. Starting with $(I)$,
\begin{equation}
\begin{split}
(I)
&= \underbrace{\frac{1}{N}\sum_{i}\sum_{l\neq i}\sum_{j\neq \{l,i\}}(y_{1i}y_{2i}-\mathbb{E}[y_{1i}y_{2i}])y_{3l}y_{4j}}_{(I.a)} + \underbrace{\frac{1}{N}\sum_{i}\sum_{l\neq i}(y_{1i}y_{2i}y_{4i}-\mathbb{E}[y_{1i}y_{2i}y_{4i}])y_{3l}}_{(I.b)}\\
&+  \underbrace{\frac{1}{N}\sum_{i}\sum_{l\neq i}\sum_{j\neq \{l,i\}}\mathbb{E}[y_{1i}y_{2i}]y_{3l}y_{4j} }_{(I.c)}+ \underbrace{\frac{1}{N}\sum_{i}\sum_{l\neq i}\mathbb{E}[y_{1i}y_{2i}y_{4i}]y_{3l}}_{(I.d)}.
\end{split}
\end{equation}
Since by Assumption~\ref{ass:large_sample} we can assume that $y_{kj}$ has bounded moments up to order 8, it holds that $
\mathbb{V}(I.a)\leq cN$, $\mathbb{V}(I.b)\leq c$, $\mathbb{V}(I.c)\leq cN^2$, $\mathbb{V}(I.d)\leq cN$.

Considering $(II)$, 
\begin{equation}
\begin{split}
(II)&=\underbrace{\frac{1}{N^2}\sum_{i}\sum_{l\neq i}\sum_{k\neq \{l,i\}}\sum_{j\neq \{i,l,k\}}y_{1i}y_{2l}y_{3k}y_{4j}}_{(II.a)} \\
&\quad +\underbrace{\frac{1}{N^2}\sum_{i}\sum_{l\neq i}\sum_{k\neq \{l,i\}}(y_{1i}y_{2l}y_{3k}y_{4k}+y_{1i}y_{2l}y_{3k}y_{4i}+y_{1i}y_{2l}y_{3l}y_{4k})}_{(II.b)}.
\end{split}
\end{equation}
The variance of $(II.a)$ and $(II.b)$ can be bounded as $\mathbb{V}(II.a)\leq c$, $\mathbb{V}(II.b)\leq c.$

Considering $(III)$,
\begin{equation}
\begin{split}
(III)
&=\underbrace{\frac{N-1}{N^2}\sum_{i}\sum_{j\neq i}( y_{1i}y_{2i}y_{3j}y_{4i}+y_{1i}y_{2i}y_{3i}y_{4j})}_{(III.a)}+\underbrace{\frac{N-1}{N^2}\sum_{i}\sum_{j\neq i} \sum_{k\neq \{i,j\}}y_{1i}y_{2i}y_{3j}y_{4k}}_{(III.b)}.
\end{split}
\end{equation}
The variance of $(III.a)$ and $(III.b)$ can be bounded as $\mathbb{V}(III.a)\leq c$, $\mathbb{V}(III.b)\leq cN^2$.

Considering $(IV)$,
\begin{equation}
\begin{split}
    (IV)
    &=\underbrace{\frac{N-1}{N^3}\sum_{i}\sum_{j\neq i}\sum_{k\neq \{i,j\}}\sum_{m\neq \{i,j,k\}}y_{1i}y_{2j}y_{3k}y_{4m}}_{(IV.a)}\\
  &  +\underbrace{\frac{N-1}{N^3}\sum_{i}\sum_{j\neq i}\sum_{k\neq \{i,j\}}(y_{1i}y_{2j}y_{3k}y_{4j}+y_{1i}y_{2j}y_{3k}y_{4i}+y_{1i}y_{2j}y_{3i}y_{4k}+y_{1i}y_{2j}y_{3j}y_{4k})}_{(IV.b)}.
    \end{split}
\end{equation}
The variance of $(IV.a)$ and $(IV.b)$ can be bounded as $
\mathbb{V}(IV.a)\leq c$, $\mathbb{V}(IV.b)\leq c$.

Finally, considering $(V)$:
\begin{equation}
\begin{split}
& \underbrace{\frac{N-1}{N^3}\sum_{j}\sum_{k\neq j}\sum_{m\neq \{j,k\}}(y_{1j}y_{2k}y_{3k}y_{4m}+y_{1j}y_{2k}y_{3m}y_{4k}+y_{1j}y_{2k}y_{3j}y_{4m}+y_{1j}y_{2k}y_{3m}y_{4j})}_{(V.a)}\\ 
&+\underbrace{\frac{N-1}{N^3}\sum_{j}\sum_{k\neq j}\sum_{m\neq \{j,k\}}\sum_{n\neq \{j,m,k\}}y_{1j}y_{2k}y_{3m}y_{4n}}_{(V.b)}.
    \end{split}
\end{equation}
The variance of $(V.a)$ and $(V.b)$ can be bounded as $\mathbb{V}(V.a)\leq cN$, $\mathbb{V}(V.b)\leq cN^2$.

Taking $N=m_{1,g}$, these results show that
\begin{equation}
\begin{split}
\frac{1}{r_{n}^2}\sum_{g=1}^{G}(A_{11,g}^2+2C_{11,g})^2\mathbb{E}\big[\big(\sum_{i}\sum_{j\neq i}\hat{\xi}_{11,ij}Z_{ig}Z_{jg}\big)^2\big|Q,X\big] &\leq \frac{c}{r_{n}^2}\sum_{g=1}^{G}A_{11,g}^4m_{1,g}^2
\leq c\frac{G}{r_{n}^2}\rightarrow_{a.s.} {0}.
\end{split}
\end{equation}
Analogous calculations show that the remaining terms in \eqref{eq:toshowconsistency} converge to zero.

\subsection{Small groups}\label{sec:smallgroups}
Consider a group with $m_{1,g}=2$ or $m_{0,g}=2$. To avoid singularity in the variance estimators of Appendix~\ref{subsec:unbvarest}, we can use the following rescaled version of the conventional variance estimator for the observations that only share their instrument status with one other observation in the group,
\begin{equation}\label{eq:biased1}
\begin{split}
    \tilde{\sigma}_{u,i}^2&=4 e_{i}'( M_{W,Z} u\odot  M_{W,Z} u)= u_{i}^2-4u_{i}\sum_{j\neq i}L_{ij}u_{j} + 4\sum_{j\neq i}L_{ij}^2 u_{j}^2.
\end{split}
\end{equation}
Notice that here $L_{ij}=1/2$. Similarly, 
\begin{equation}\label{eq:biased2}
\begin{split}
    \tilde{\sigma}_{v,i}^2(\beta)& =4e_{i}'( M_{W,Z} v\odot  M_{W,Z} v)=u_{i}^2-4v_{i}\sum_{j\neq i}L_{ij}v_{j} + 4\sum_{j\neq i}L_{ij}^2 v_{j}^2,\\
    \tilde{\sigma}_{uv,i}(\beta)& =4 e_{i}'( M_{W,Z} u\odot  M_{W,Z} v)= u_{i}v_{i}-2u_{i}\sum_{j\neq i}L_{ij}v_{j}-2v_{i}\sum_{j\neq i}L_{ij}u_{j}  + 4\sum_{j\neq i}L_{ij}^2 v_{j}u_{j}.
\end{split}
\end{equation}
The first term in the expressions coincides with the first terms in \eqref{eq:unbvarest_decomp}. The same holds up to a scaling constant for the second term. Those terms are therefore covered by the proof in Appendix~\ref{app:consistentvar}. The only new terms are the last terms of the respective expressions that yield a positive bias in the variance estimator. Substituting these terms into \eqref{eq:Vhat}, we see that the bias contribution by observation $i$ if observation $j(i)$ is the only observation in the same covariate group with the same instrument status, is
\begin{equation}\label{eq:manyterms}
\begin{split}
   &[e_{i}'A(Y-T\beta)]^2u_{j(i)}^2 
   +[e_{i}'AT]^2v_{j(i)}^2
   +2[e_{i}'A(Y-T\beta)e_{i}'AT]u_{j(i)}v_{j(i)} 
 \\
 &\qquad + R(|\hat{\beta}-\beta|)\\
&=\big(e_{i}'A(Y-T\beta)u_{j(i)}+e_{i}'AT v_{j(i)}\big)^2 +  R(|\hat{\beta}-\beta|),
    \end{split}
\end{equation}
where $ R(|\hat{\beta}-\beta|)\rightarrow_{p} 0$. This shows that the bias is positive up to an $o_{p}(1)$ term. 

\section{Additional Monte Carlo experiments}\label{A:mc}
This appendix discusses the results from a set of additional Monte Carlo experiments. First, we consider experiments similar to the ones in Section~\ref{sec:mc} extended with treatment effect heterogeneity. Second, we run experiments with additional variance estimators to the ones used in Section~\ref{sec:mc}.

\begin{figure}[t]
\centering
\caption{Absolute median bias and test size with treatment effect heterogeneity}\label{mc:add_hetero}
     \includegraphics[width=1\textwidth,trim=1.8cm 0 2cm 0,clip]{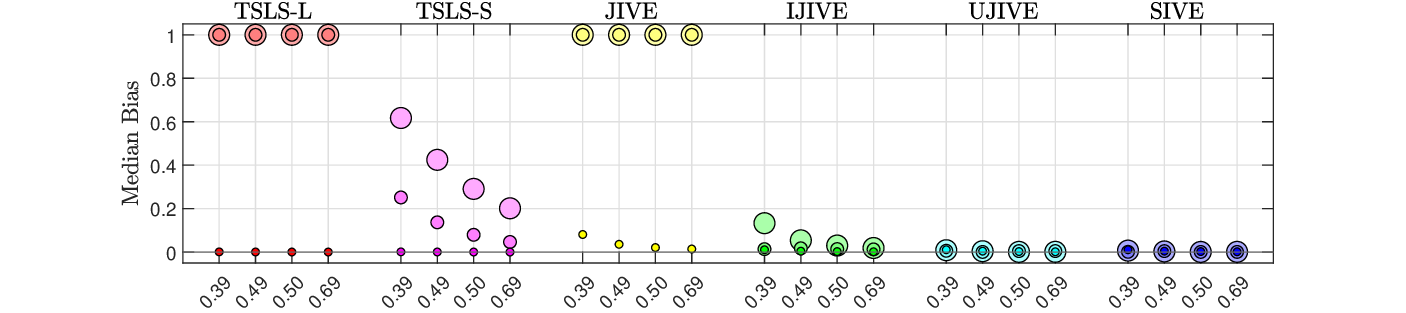}
     \includegraphics[width=1\textwidth,trim=1.8cm 0 2cm 0,clip]{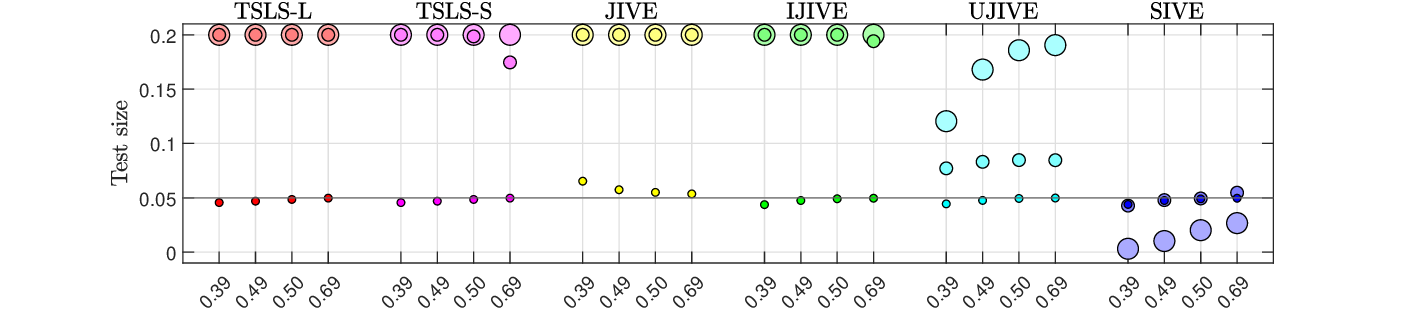}
     \vspace{0.2cm}
     \begin{minipage}{0.9\textwidth}
        \footnotesize{
\textit{Note}: see notes Figure~\ref{mc:bias} and Figure~\ref{mc:size}}
\end{minipage}
\end{figure}

\textbf{Treatment effect heterogeneity:}
Consider the setup in Section~\ref{sec:mc}, with $\gamma_i=5\times1.2$ for observations in covariate groups with less observations than the median sized covariate group, and $\gamma_i=1.2$ for the remaining observations. Figure~\ref{mc:add_hetero} shows the absolute median bias and test size as a function of instrument strength. We find thet treatment effect heterogeneity does not affect the bias, but the test size further deteriorates for all methods but SIVE when the number of covariate groups increases. 

\textbf{Alternative variance estimator SIVE:}
For SIVE, we use the variance estimator given in \eqref{eq:Vhat}, which is robust to treatment effect heterogeneity. We now assess the effect of treatment effect heterogeneity on tests that rely on the variance estimator proposed by \citet{chao2023jackknife}:
\begin{equation}\label{var:chao}
    V_{c} = (T'AD_{1}AT + (\hat{\varepsilon}\odot\hat{u})'J(A\odot A)J(\hat{\varepsilon}\odot\hat{u}))/(T'AT)^2,
\end{equation}
where $J = (M_{W}\odot M_{W})^{-1}$, $[D_{1}]_{i} = [J(\hat{\varepsilon}\odot\hat{\varepsilon})]_{i}$, $\hat{\varepsilon} = M_{W,Z}(y-T\hat{\beta}^{\text{SIVE}})=M_{W,Z}(\varepsilon-u\hat{\beta}^{\text{SIVE}})$ and $\hat{u} = M_{W,Z}T=M_{W,Z}u$. This variance estimator is proposed in the context of a linear panel data model with homogeneous slope coefficient.

\begin{figure}[t]
\centering
\caption{Test size of \citet{chao2023jackknife} with treatment effect heterogeneity}\label{mc:chao}
 \includegraphics[width=1\textwidth,trim=1.8cm 0 2cm 0,clip]{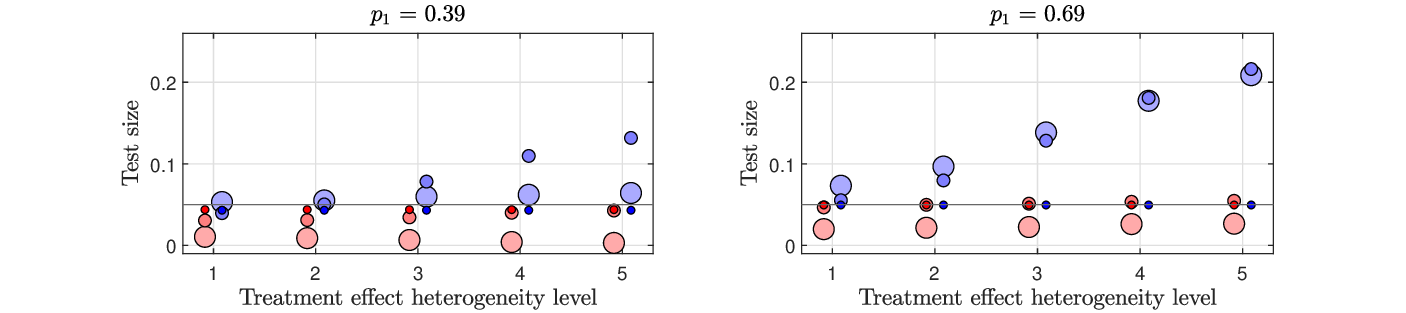}
    \begin{minipage}{0.95\textwidth}
    \vspace{0.2cm}
        \footnotesize{
\textit{Note}: 
the figure shows the size of testing $H_{0}\colon \beta=\tau$ at a nominal level of $5\%$ in a setting with treatment effect heterogeneity. The red circles correspond to SIVE with the variance estimator in \eqref{eq:Vhat}, and the green circles to SIVE with variance estimator in \eqref{var:chao}. The size of the circles indicates the number of covariate groups with the small circle corresponding to $G=1$, the medium circle corresponding to $G=50$ and the large circle corresponding to $G=200$. The $x$-axis is treatment effect heterogeneity level.}
\end{minipage}
\end{figure}

Consider the setup in Section~\ref{sec:mc}, with $\gamma_i=h\times1.2$ for observations in covariate groups with less observations than the median sized covariate group, and $\gamma_i=1.2$ for the remaining observations. The treatment effect heterogeneity level $h$ ranges from 1 to 5. Figure~\ref{mc:chao} shows the size of the test of $H_{0}\colon \beta=\tau$ in a setting with weak instruments ($p(2)=0.39$, left panel) and with strong instruments ($p(2)=0.69$, right panel). On the $x$-axis we vary the treatment effect heterogeneity level. 

The red circles in the left panel confirm that our proposed method offers a conservative test under weak instruments. As expected based on the theory, the level of treatment effect heterogeneity has no effect on the size of the test. For the green circles representing the alternative variance estimator, we see that increasing the level of treatment effect heterogeneity leads to a oversized test. The fact that treatment effect heterogeneity has only a mild effect in this case is due to the fact that it is flooded by the uncertainty introduced by the presence of many weak instruments. When the instruments are strong, as in the right panel of Figure~\ref{mc:chao}, accounting for treatment effect heterogeneity becomes more important. Again, SIVE shows no dependence on the level of treatment effect heterogeneity. The alternative variance estimator now becomes progressively more oversized as the level of heterogeneity increases for all values of the number of covariate groups.

\textbf{Alternative variance estimator TSLS:}
\citet{lee2018consistent} proposes a variance estimator for TSLS that is valid in overidentified specifications where each instrument identifies a different LATE. It therefore allows for treatment effect heterogeneity. However, the analysis in \citet{lee2018consistent} assumes that the number of instruments is fixed relative to the sample size. 

\begin{figure}[t]
\centering
\caption{Test size of \citet{lee2018consistent} with treatment effect heterogeneity}\label{mc:lee}
 \includegraphics[width=1\textwidth,trim=1.8cm 0 2cm 0,clip]{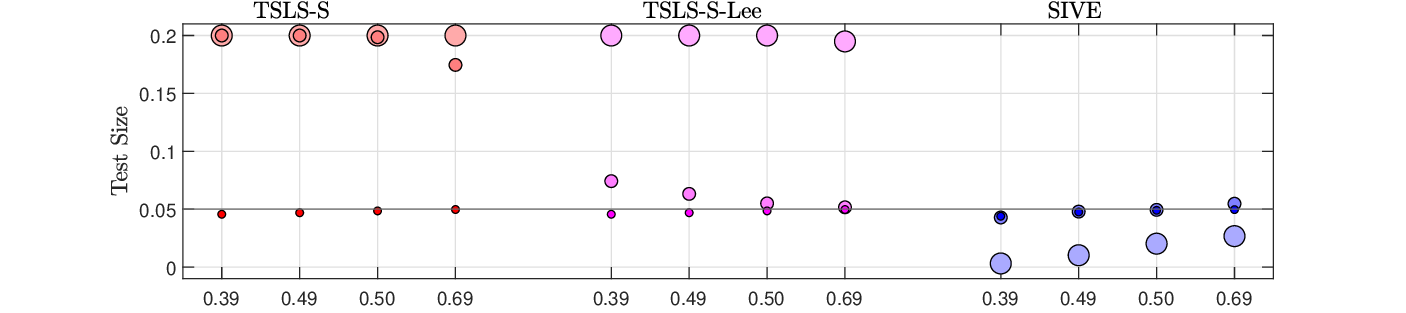}
    \begin{minipage}{0.95\textwidth}
    \vspace{0.2cm}
        \footnotesize{
\textit{Note}: 
the figure shows the size of testing $H_{0}\colon \beta=\tau$ at a nominal level of $5\%$ in a setting with treatment effect heterogeneity. See notes Figure~\ref{mc:size} for details.}
\end{minipage}
\end{figure}

Consider the setup in Section~\ref{sec:mc}, with $\gamma_i=5\times1.2$ for observations in covariate groups with less observations than the median sized covariate group, and $\gamma_i=1.2$ for the remaining observations. Figure~\ref{mc:lee} shows that TSLS-S with the variance estimator proposed by \citet{lee2018consistent} (TSLS-S-Lee) shows excellent size control when the number of covariate groups equals one. Also with $G=50$, TSLS-S-Lee shows a substantial improvement over the standard Eicker-Huber-White variance estimator used by TSLS-S. However, when the number of covariate groups increases, the many instrument bias manifests itself and the size increases above its nominal value. 

\section{Additional empirical applications}\label{A:application}
This appendix discusses additional empirical applications to the LATEs of schooling, health insurance, and an educational program. 

\subsection{Returns to schooling}\label{sec:card}
First, we discuss the LATE of years of schooling on log wage by \citet{card1995using}. The linear TSLS estimate for this LATE has been investigated by \citet{blandhol2022tsls} due to concerns about this estimator in this setting.
%
The sample consists of the National Longitudinal Survey of Young Men (NLSYM) respondents who reported educational attainment and wages in the 1976 interview. The survey started in 1966 with 14-24 year old men. 
The outcome variable is log hourly wage and the treatment variable is years of education. The binary instrument equals one if there was an accredited four-year college in the local labor market of the respondent when the respondent was 14 years old.
The analysis includes fifteen controls: years of potential experience itself and squared, a binary race indicator for Black, binary indicators for living in the South and in a standard metropolitan statistical area (SMSA), a set of binary indicators for region of residence in 1966, and a binary indicator for residence in an SMSA in 1966. The sample size is 3010. 

The TSLS estimator linearly includes these controls (TSLS-L), which in this case does not coincide with a saturated specification. Since years of experience consists of only 24 unique values, and the remaining controls are binary, the set of covariates can be saturated. After saturation, we (1) remove the covariate groups without either an active or nonactive instrument, and (2) remove the observations without a covariate group. The resulting number of covariate groups and number of observations in the saturated specification (TSLS-S) equals 264 and 1864, respectively. Since SIVE can only consider covariate groups with at least two active and two nonactive instruments, the number of covariate groups and observations is further reduced to 111 and 1229, respectively. 

The first three columns of Table~\ref{tab:card} show the TSLS-L, TSLS-S, and SIVE estimates, with their corresponding standard errors. The estimate in the first column is identical to the estimates in \citet[ Table 3A, Column (5)]{card1995using} and \citet[Table 7, Column (2)]{blandhol2022tsls}, and the estimate in the second column is identical to \citet[Table 7, Column (4)]{blandhol2022tsls}. TSLS-L relies on the parametric assumption of a linear relation between the controls and the outcome and treatment, and a strong monotonicity assumption on the instrument. The estimator is inconsistent if either of those assumptions is violated. The TSLS-S estimate suffers from a many and potentially weak instrument bias. These biases are also pointed out by \citet{blandhol2022tsls}.

\begin{table}[t]
  \centering
  \caption{Return to schooling and impact of health insurance estimates}
    \begin{tabular}{lrrrrrr}
          & \multicolumn{3}{c}{\citet{card1995using}} & \multicolumn{3}{c}{\citet{finkelstein2016effect}} \\
          \cline{2-4}  \cline{5-7}
          & \multicolumn{1}{l}{TSLS-L} & \multicolumn{1}{l}{TSLS-S} & \multicolumn{1}{l}{SIVE} & \multicolumn{1}{l}{TSLS-L} & \multicolumn{1}{l}{TSLS-S} & \multicolumn{1}{l}{SIVE} \\
          \hline
    Estimate & 0.132 & 0.072 & 0.079 & 0.088 & 0.077 & 0.077 \\
    Std. Error & 0.054 & 0.011 & 0.324 & 0.018 & 0.019 & 0.019  \\ \hline
    Variables & 15    & 528   & 222   & 12    & 68    & 66 \\
    Observations & 3010  & 1864  & 1229  & 24646 & 24622 & 24618 \\
    \hline
    \end{tabular}\label{tab:card}
   	\begin{minipage}{0.95\textwidth}
  	\vspace{0.2cm}
  	\footnotesize{\textit{Note:} the total number of variables includes the controls and instrument in TSLS-L, and the covariate groups and instrument interactions with all covariate groups in TSLS-S and SIVE. Due to lack of variation in the instrument for some covariate groups, the final two estimators may have less observations, denoted in the fourth row. TSLS standard errors are heteroskedasticity-robust for Card (1995) and standard errors are clustered at the household level for Finkelstein et al. (2016). }
    \end{minipage}
\end{table}%

The SIVE estimate is similar to the TSLS-S estimate, and albeit substantially smaller than the TSLS-L estimate, it resides in the 95\% confidence intervals of both estimates. However, there is a substantial increase in the magnitude of its standard error, which accounts both for many controls, many and weak instruments, and treatment effect heterogeneity. We conclude that based on this sample, we cannot reject that the LATE of returns to schooling differs from zero, in contrast to what is suggested by the TSLS estimates. 

We conclude that our proposed method is able to correct for common issues in LATE estimation in a widely studied example. The SIVE estimates for the returns to schooling are in line with the literature, and correct for the upward bias in TSLS estimates caused by violations of the linearity assumption on the controls or the strong monotonicity assumption on the instruments. In contrast to existing estimators robust to these violations, SIVE does not suffer from many or weak instrument bias, and its standard errors allow for treatment effect heterogeneity. Hence, our inference method is robust against violations of the (implicit) assumptions made by the researcher in an empirically relevant setting.

\subsection{The impact of health insurance}
\label{sec:ohie_bank}
Second, we consider a carefully conducted field experiment for which we have no a priori indications that the TSLS estimates are problematic. \cite{finkelstein2012oregon} use the randomized opportunity to apply for Medicaid in the Oregon Health Insurance Experiment (OHIE) to study the effects of Medicaid enrollment. In a follow-up paper, \cite{finkelstein2016effect} study the impact on emergency department use in a sample of 24,646 individuals with Portland-area zip codes. The binary instrument equals one if an individual was randomly given the opportunity to apply for Medicaid. The binary outcome variable equals one if an individual visited an emergency department within 180 days of the lottery notification, and the binary treatment variable  equals one if an individual was enrolled in Medicaid during the same 180-day period. The IV analysis includes twelve controls: three household size dummies, eight lottery draw dummies, and the pre-lottery value of the binary outcome variable. Standard errors are clustered at the household level, and our sample includes 22,188 different households.

Table~\ref{tab:card} shows the TSLS-L, TSLS-S, and SIVE estimates for the LATE of Medicaid, with their corresponding standard errors. The estimate by TSLS-L is identical to the estimate in \citet[Appendix Table 6, Column (6)]{finkelstein2016effect}. The estimators using a saturated specification result in a slightly smaller LATE with very similar standard errors, leading to the same conclusions as with TSLS-L.

\subsection{The effect of a technology-led instructional program on math test scores}\label{sec:tech}
Although the conditional validity of instrument variables is most commonly argued on the basis of discrete controls/strata, a valid question is what happens in the presence of continuous controls. In this case, covariate groups have to be merged before the methods in this paper can be applied. The results we derive hold exactly when the merged covariate groups $g$ and $g'$ satisfy the following conditions. For $q\in\{0,1\}$,
\begin{equation}\label{eq:homo_ass2}
\begin{split}
    \mathbb{P}[T_i(q)|X_i=x_g] &= \mathbb{P}[T_i(q)|X_i=x_{g'}] ,\quad
    \mathbb{E}[Y_i(q)|X_i=x_g] = \mathbb{E}[Y_i(q)|X_i=x_{g'}].
    \end{split}
\end{equation}
Although restrictions on treatment effect heterogeneity are often considered to be strong, the assumptions in \eqref{eq:homo_ass2} are considerably weaker than the necessary conditions required for a causal interpretation of the widely used linear TSLS estimand with covariates. As an alternative to \eqref{eq:homo_ass2}, one can impose a smoothness assumption on the conditional expectations and exploit the fact that we can finely discretize the control as the number of control dummies can be of the same order as the sample size. In this case, results from the matching literature \citep{abadie2006large} show that the resulting bias is small when the number of continuous controls is small. 

This appendix illustrates the effect of discretizing a continuous control in an empirical application. \citet{muralidharan2019disrupting} use a randomized opportunity to receive free access to a technology-led instructional program (called Mindspark) to estimate the LATE of attending the Mindspark centers on math and language test scores. The study considers students from public middle schools in Delhi, who are tested before and after the 4.5-month-long intervention.
We focus on the outcome variable that equals the endline math test score, which is standardized to have a mean of zero and standard deviation of one in the baseline test before the intervention. The treatment variable is the number of days a student logged into the Mindspark system. The binary instrument equals one for students who were randomly selected for the Mindspark voucher offer. The controls include 18 randomization strata fixed effects and the baseline math test score. The sample size is 535 students, and we report Huber-White standard errors.

\begin{figure}[t]
\centering
\caption{Estimated effect of an educational program on math scores}\label{fig:Xc}
\includegraphics[width=\textwidth,trim=.6cm 0.5 1.2cm .5cm,clip]{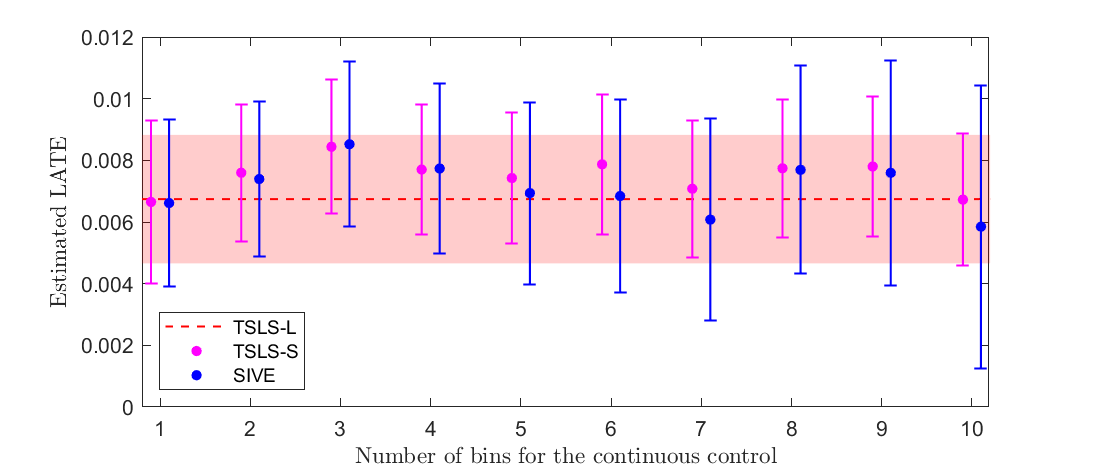} 
    \begin{minipage}{0.95\textwidth}
    \vspace{0.2cm}
        \footnotesize{
\textit{Note}: 
the figure shows the confidence intervals for the estimated LATE of the educational program in \citet{muralidharan2019disrupting} on math scores, as a function of the level of granularity the continuous control is accounted for.}
\end{minipage}
\end{figure}
The TSLS-L estimator linearly includes the controls, and due to the continuous baseline math test score this does not coincide with a saturated specification. The baseline score consists of 528 unique values, which makes saturation in a sample of size 535 observations infeasible. However, instead of discretizing the continuous control to its most granular level, we can include a set of dummies representing equally sized intervals of the sorted values in the control. Setting the number of dummies equal to one is identical to including an intercept, and the continuous control is not taken into account, while ten mutually exclusive dummies approximate the variation in the control. 

Figure~\ref{fig:Xc} shows the TSLS-L estimate (black line) and the corresponding 95\% confidence interval (gray shaded area). These estimates (Estimated LATE of 0.0067 with a standard error equal to 0.0011) are identical to the estimate in \citet[Table 9, Column (1)]{muralidharan2019disrupting}. The red solid line represents the TSLS-S estimates for an increasing number of dummies representing the continuous control, and the dashed red lines represent the lower and upper bounds of the corresponding confidence intervals. Similarly, the solid and dashed blue lines represent the SIVE estimates and corresponding confidence intervals. For any level of granularity of the continuous control, the TSLS-S and SIVE estimate lie in the TSLS-L confidence interval. Saturation only slightly inflates the confidence intervals, even with ten dummies accounting for the continuous control. 

We conclude that saturation with a discretized continuous control does not necessarily result in a substantial loss of statistical power. In case the continuous control affects the validity of the instrument, instead of governing the precision of the LATE estimate in the application here, discretizing the continuous control may effect the validity of the instrument. However, the conditional validity of an instrument almost always relies on categorical controls, representing different groups or strata. 

\end{document}